\documentclass{pasj00}
\twocolumn

\begin{document}
\SetRunningHead{Y. Shimoda et al.}{Metals in the Intracluster Medium of MS~1512.4+3647}
\Received{2012/11/26}
\Accepted{2013/07/17}

\title{Metals in the Intracluster Medium of MS~1512.4+3647 Observed
with Suzaku: Implications for the Metal Enrichment History}

 \author{
   Yuya \textsc{shimoda},\altaffilmark{1}
   Madoka \textsc{kawaharada},\altaffilmark{2}
   Kosuke \textsc{sato},\altaffilmark{3}
   Takaya \textsc{ohashi},\altaffilmark{4}
   Yoshitaka \textsc{ishisaki},\altaffilmark{4}
   Ikuyuki \textsc{mitsuishi},\altaffilmark{4}
   Hiroki \textsc{akamatsu},\altaffilmark{5}
   and
   Makoto S. \textsc{tashiro}\altaffilmark{1}
   }
   \altaffiltext{1}{Department of Physics, Saitama University, 255
   Shimo-Okubo, Sakura, Saitama 338-8570}
   \email{shimoda@heal.phy.saitama-u.ac.jp}
   \altaffiltext{2}{Institute of Space and Astronautical Science, JAXA,
   3-1-1 Yoshinodai, Chuo-ku, Sagamihara, Kanagawa 252-5210}
   \altaffiltext{3}{Department of Physics, Tokyo University of Science,
   1-3 Kagurazaka, Shinjuku-ku, Tokyo 162-8601}
   \altaffiltext{4}{Department of Physics, Tokyo Metropolitan
   University, 1-1 Minami-Osawa, Hachioji, Tokyo 192-0397}
   \altaffiltext{5}{SRON Netherlands Institute for Space Research,
   Sorbonnelaan 2, 3584, CA, Utrecht, The Netherlands}

\KeyWords{galaxies: clusters: individual (MS~1512.4+3647) -- X-rays:
galaxies: clusters -- X-rays: ICM} 

\maketitle

\begin{abstract}
The cluster of galaxies MS 1512.4+3647 ($z=0.372$) was observed with
Suzaku for 270 ks.
Besides the Fe abundance, the abundances of Mg, Si, S, and Ni are
 separately determined for the first time in a medium redshift cluster
 ($z>0.3$).
The derived abundance pattern of MS~1512.4+3647 is consistent with those
 of nearby clusters, suggesting that the system has similar
 contributions from supernovae (SNe)~Ia and SNe~II to nearby clusters.
The number ratio of SNe~II to SNe~Ia is $\sim3$.
The estimated total numbers of both SNe~II and SNe~Ia against gas mass
 indicate similar correlation with those for the nearby clusters.
The abundance results of MS~1512.4+3647 is consistent with the standard scenario
that the SN~II rate history roughly follows the star-formation history
 which has a peak at $1<z<2$ and then declines by about one order of
 magnitude toward $z\sim0$.
The similar number of SNe~Ia to the nearby clusters suggests that the
 SN~Ia rate declines steeply from $z=0.37$ to $z=0$ and/or SN~Ia
 explosions occurred predominantly at larger redshifts.
\end{abstract}

\section{Introduction}
Clusters of galaxies are the largest virialized structures in the
universe, and they gravitationally bind hot thin-thermal plasma
(intracluster medium; ICM).
The high temperature reaching up to several $10^{7}$ K make the ICM to
produce X-rays via thermal bremsstrahlung.
The ICM is enriched with metals having been synthesized in stars and
supernova (SN) explosions and ejected to the intra-galactic space by SN
explosions (e.g. \cite{key-4}; Renzini et al. 1993). 
The metals in galaxies are thought to be exported to the ICM
via galaxy wind (Mathews \& Baker 1971) and/or ram pressure stripping
(Gunn \& Gott 1972).
These metals are highly ionized in collisional ionization equilibrium
and excited metal ions emit atomic lines in the X-ray range.

The majority of supernovae are classified into type Ia (SN~Ia) and type
II (SN~II).
The former is the explosion of a white dwarf with gas accretion from
a companion star, in which nuclides are in the thermal equilibrium
and provides a lot of iron family elements (Iwamoto et al. 1999).
On the other hand, in the latter the iron core of a massive star are
photodisintegrated into helium nuclei and mainly alpha elements are synthesized
in the explosion (Nomoto et al. 2006).
Thus the emission lines observed in the X-ray spectrum reflect
integrated activities of both types of supernovae in the galaxies, and enable
us to study chemical enrichment history of the ICM, which is a
major component of the known baryon in the universe.

Observational studies of metals in the ICM were greatly advanced by
ASCA. In particular, it allowed to measure spatial distributions of
Si and Fe in the ICM (Fukazawa et al. 1998, 2000; Finoguenv et al. 2000,
2001; Ezawa et al. 1997).
Recently, XMM-Newton and Chandra, thanks to their large
effective areas and high angular resolutions, have measured ``metal
abundance'' of the ICM in high-redshift clusters,
which is a representative abundance for various elements,
mainly determined by using Fe-L and Fe-K lines.
Using XMM-Newton and Chandra data of 56 clusters at $0.3\leq z\leq1.3$,
Balestra et al. (2007) measured the iron abundance $Z_{\rm Fe}$ within
a spatial region of 0.15 $r_{\rm 200}$ -- 0.3 $r_{\rm 200}$.
They found that the clusters at $z>0.5$ have a constant average
abundance $Z_{\rm Fe}\approx0.25$ solar,
while clusters at $z<0.5$ exhibit a significantly larger abundance
of $Z_{\rm Fe}\approx0.4$ solar.
Maughan et al. (2008) also reported the same result using 115
clusters at $0.1<z<1.3$ observed with Chandra.
They found significant evolution in the metal abundance which drops by
50\% from $z=0.1$ to $z\sim1$.
Anderson et al. (2009) found the similar drop in the metal abundance
from $z=0.1$ to $z\sim1$ using 29 clusters observed with XMM-Newton,
together with the 115 clusters of Maughan et al. (2008) and 70
clusters at $z<0.3$ in Snowden et al. (2008).
Baldi et al. (2012) tried to study dependence of the metal
evolution on distance from the center by spatially resolved
analysis, but they obtained no statistically significant conclusion
about the different evolutionary path that the different regions
of the clusters may have traversed.

After the launch of Suzaku (Mitsuda et al. 2007), studies of alpha
elements have made progress.
X-ray Imaging Spectrometer (XIS: Koyama et al. 2007)
onboard Suzaku has good sensitivity for lines in lower energies thanks
to good line spread function and low particle background level,
especially in the earlier phases after launch for the less contamination
affected the effective area in lower energies (Koyama et al. 2007).
The metal abundances synthesized mainly in SNe~II, such as O and Mg,
in the ICM outside cool-core regions up to $\sim0.3 r_{200}$ have
been measured for several nearby clusters and groups ($z\sim0.001$)
with Suzaku (Matsushita et al. 2007; Komiyama et al. 2009; Sato et al.
2007a, 2008, 2009a, 2010; Sakuma et al. 2011).
Combining the Suzaku results with SNe nucleosynthesis model,
Sato et al. (2007b) calculated integrated numbers of
SNe~Ia and SNe~II explosions in nearby clusters.
They showed that the number ratio of SNe~II to Ia is $\sim3.5$
and Fe has been synthesized predominantly by SNe~Ia.
A similar result is reported in de Plaa et al. (2007)
using data of XMM-Newton.

In order to conclude the metal enrichment history of the ICM
observationally, it is crucial to measure metal abundances of individual
elements in clusters at high redshifts.
This is beyond the ability of X-ray observatories currently in orbit,
and would be one of major sciences in future X-ray missions with
huge effective areas.
However, Suzaku has the ability to pilot that survey for bright sources
at medium redshifts.

We selected a target for Suzaku to perform the study
of chemical evolution in the ICM. The criteria for the cluster
are
1) distant but bright enough,
2) metal rich, and
3) cool (below 4 keV).
In general, a luminous cluster has high temperature.
However, in the high temperature gas, most alpha elements are fully
ionized and hence no (or very weak) emission lines are expected.
This is why we set the third criterion.
As a result, we chose MS~1512.4+3647 at $z=0.372$  (Stocke et al. 1991).
Using ASCA data of MS~1512.4+3647, Ota (2000) reported 
its flux of
$3.9 \times 10^{-13}$ erg s$^{-1}$ cm$^{-2}$ in 2.0--10.0 keV,
luminosity of
$3.6^{+0.2}_{-0.2}\times10^{44}$ erg s$^{-1}$,
metal abundance of
$1.05^{+1.34}_{-0.67}$ solar (Anders and Grevesse 1989) and
temperature of
$2.85^{+0.91}_{-0.49}$ keV.
In this paper, we report results of our Suzaku long observation of
 the cluster of galaxies MS~1512.4+3647.
Throughout this paper, cosmological parameters of
$H_{0}=71$ km s$^{-1}$ Mpc$^{-1}$,
$\Omega_{\rm M}=0.27$, and
$\Omega_{\Lambda}=0.73$
are adopted.
At $z=0.372$, $1'$ corresponds to 306 kpc.
The virial radius of MS~1512.4+3647 is $r_{200}=1.22$ Mpc
by substituting the average temperature 2.85 keV (Ota 2000)
into equation 2 of Henry et al. (2009), where  $r_{200}=1.22$ Mpc
is the radius within which the average density is 200 times
the critical density at the redshift of the cluster.  
The Galactic hydrogen column density in the direction of
MS 1512.4+3647 is 
$N_{\rm H}=1.4\times 10^{20}$ cm$^{-2}$ (Dickey \& Lockman 1990).
The definition of one solar abundance is taken from Lodders (2003).
Errors are given at the 90 \% confidence level unless otherwise stated. 


\section{Observation and Data Reduction}
\subsection{Suzaku Observation}
A Suzaku observation of MS~1512.4+3647  was performed from
2007 December 29 15:37:13 UT to 2008 January 4 16:40:18 UT
with the total exposure of 268.9 ks (ObsID 802034010).
In this paper, we utilize only the XIS data, since no signals are
observed by the HXD.
Two front-illuminated (FI: XIS0 and XIS3) CCD cameras and
one back-illuminated (BI: XIS1) CCD camera were in operation.
The average pointing direction of the XIS was at ($\alpha$, 
$\delta$)$=$(\timeform{15h14m25s.4}, \timeform{+36D37'11''.3}).
All the three XIS detectors were in the normal clocking mode
(8 s exposure per frame) with the standard 5 $\times$ 5 and 3
$\times$ 3 editing modes during the data rates of
SH/H and M/L, respectively.

We reprocessed unscreened XIS event files using Suzaku software version
19 in HEAsoft 6.12 while referring to the calibration data base (CALDB)
of the XIS and the X-ray telescope (XRT: Serlemitsos et al. 2007)
released on 2012 February 10 and 2011 June 30, respectively.
The \texttt{aepipeline} tool in the HEAsoft package was used
for the reprocessing. We applied the standard screening
criteria for good time intervals (GTI):
the spacecraft is outside the South Atlantic Anomaly (SAA),
the time interval after an exit from the SAA is longer than 436 s,
the geomagnetic cutoff rigidity (COR) is higher than 8 GV,
the source elevation above the rim of bright and night Earth (ELV)
is higher than 20$^{\circ}$ and 5$^{\circ}$, respectively,
and the XIS data are free from telemetry saturation.
We selected XIS events with the ASCA grades of 0, 2, 3, 4, or 6.
These procedures yielded an effective exposure of 208.6 ks.
The resulting summed image of XIS0, XIS1, and XIS3 in the 0.5--7.0
keV energy range is shown in panel (a) of figure~\ref{f1}.
The spectral analysis was performed with XSPEC~12.7.1.

\subsection{Archival Chandra Data}
MS~1512.4+3647 was observed with Chandra (Weisskopf et al. 2000, 2002)
for 49.6 ks in 2000 (ObsID 800). The observation separated into two
intervals, from 2000 June 6 15:53:46 UT to June 7 02:49:17 UT (hereafter
OBS1), and from 2000 July 7 05:17:39 UT to July 7 09:25:15 UT (hereafter
OBS2). We utilized this Chandra data to make ancillary response files (ARFs)
of Suzaku and extract point sources. Chips S1, S2, S3, S4, I2 and I3
were used in the Advanced CCD Imaging Spectrometer (ACIS; Bautz et
al. 1998; Garmire et al. 2003).

\begin{figure*}[htbp]
 \begin{minipage}{0.5\hsize}
  \begin{center}
   \FigureFile(80mm,40mm){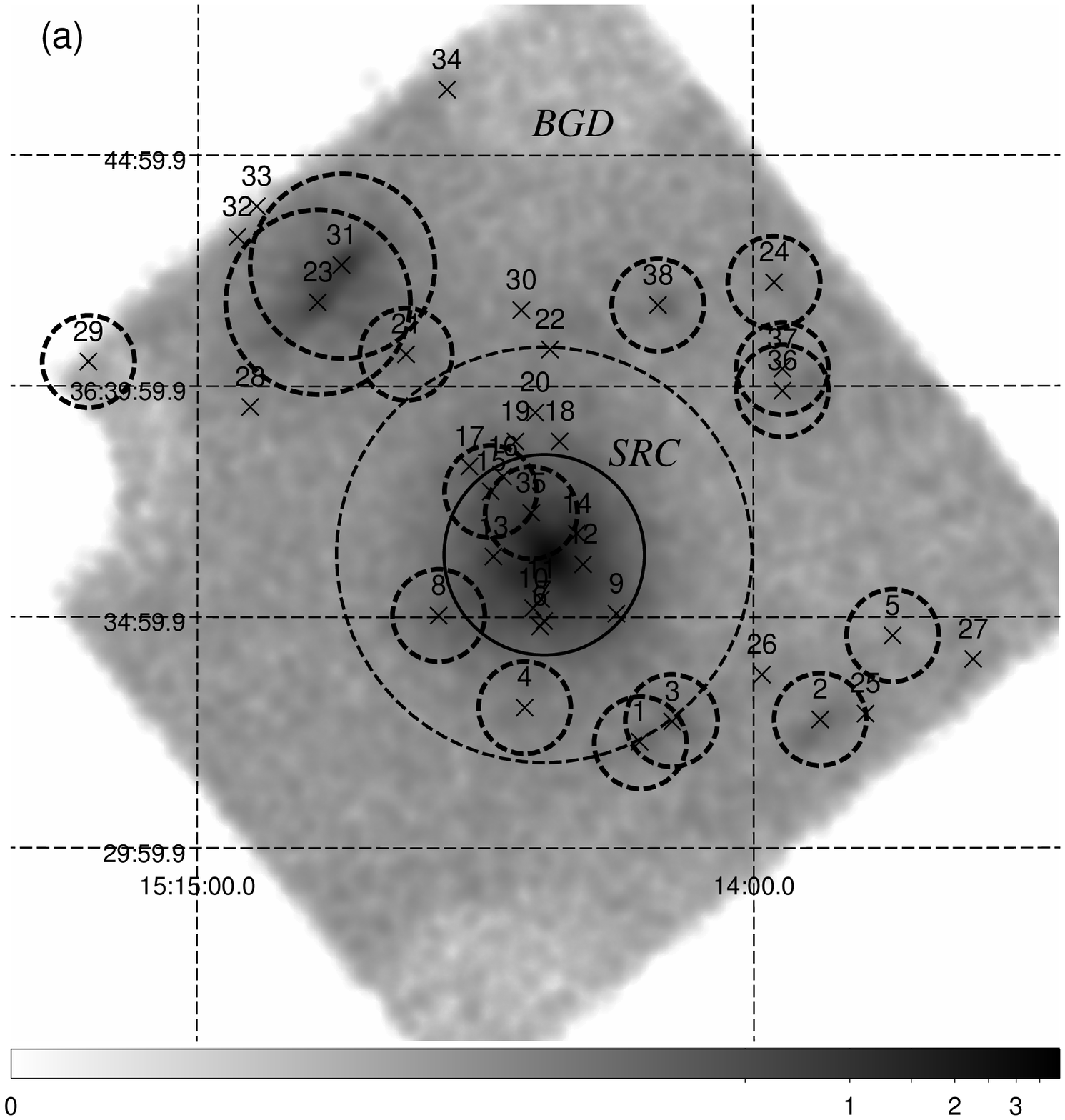}
  \end{center}
 \end{minipage}
 \begin{minipage}{0.5\hsize}
  \begin{center}
   \FigureFile(80mm,40mm){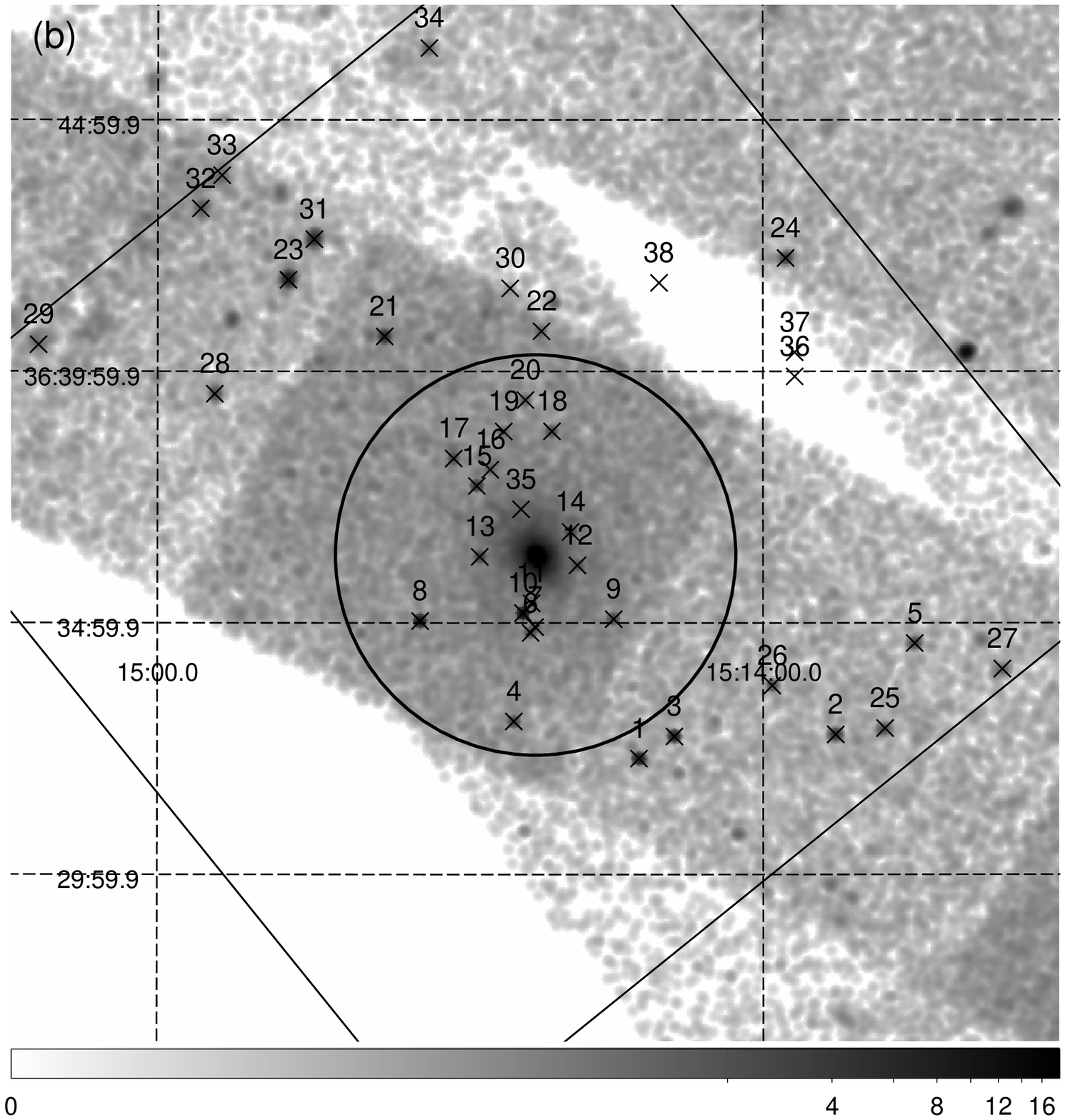}
  \end{center}
 \end{minipage}
 \caption{(a) XIS image of MS~1512.4+3647 in the 0.5--7.0 keV
 energy range, smoothed with a Gaussian kernel of $\sigma=16$ pixel
 $\simeq17''$. The scale bar indicates the X-ray counts per pixel. The
 observed XIS0, 1, 3 images were added on the sky coordinates after
 removing calibration-source regions. The NXB was not subtracted, and
 the exposure and vignetting were not corrected. The removed regions of
 point sources from spectral analysis are indicated with dashed circles.
 The SRC and the BGD regions are within the solid circle and outside of
 the dashed circle, respectively.
 (b) ACIS image in 0.5--7.0 keV energy range, smoothed with
 a Gaussian kernel of $\sigma=4$ pixel $\simeq2''$. The Suzaku FOV and
 the virial radius are shown as the square and the solid circle on the
 image, respectively. The scale bar indicates the X-ray counts per
 pixel. The OBS1 and OBS2 images were added on the sky coordinates. The
 point sources detected with \texttt{wavdetect} or extracted from the
 CXC and 2XMMi catalogs are indicated with black crosses.
 }\label{f1}
\end{figure*}

All the data were reprocessed with \texttt{chandra\_repro} software to
create new level 2 event files for the two intervals using CIAO 4.3
software package, referring to CALDB 4.4.6.1. The effective exposure
became 36.4 ks for OBS1 and 12.6 ks for OBS2. In the following analysis,
the same grade selection as that for the Suzaku XIS data (i.e., 0, 2, 3,
4, or 6) was applied. The resulting ACIS image in the 0.5--7.0 keV energy
range is shown in panel (b) of figure~\ref{f1} with the field of view
(FOV) of XIS.

\subsection{Response Files}\label{resp}
Redistribution matrix files (RMFs) of the XIS were produced by
\texttt{xisrmfgen}, and ARFs by \texttt{xissimarfgen} version 2010-11-05
(Ishisaki et al. 2007).
The effect of contaminations on the optical blocking filter of the XIS
is included in ARFs.
Two types of ARFs were generated, A$^{\rm U}$ and A$^{\rm I}$, for the
uniform background emission and ICM emission, respectively.
The input image of A$^{\rm U}$ is the uniform emission
over a circular region of 20$'$ radius, and the extract region of
A$^{\rm U}$ is the BGD region of figure~\ref{f1}(a).

In order to make A$^{\rm I}$, a background subtracted X-ray surface
brightness profile of MS~1512.4+3647 was created from an ACIS
0.5--7.0 keV image of the Chandra OBS1 data.
In this process, point sources were detected and removed from the
ACIS image, using the \texttt{wavdetect} (Freeman et al. 2002) of CIAO
with a detection threshold of 5$\sigma$.
The background was extracted from a region outside the virial
radius of MS~1512.4+3647 (4$'$).
The center of the profile, ($\alpha$,
$\delta$)$=$(\timeform{15h14m22s.6},\timeform{+36D36'21''.0}), was
determined from the position of X-ray emission centroid in the ACIS
image.
Then the profile was fitted with $\beta$-model (King 1962) of
which parameters were determined through the least chi-square
method.
The resulting parameter set of $\beta$-model is $(r_{\rm c},
\beta)=(21.7 {\rm \ kpc}, 0.53)$, where $r_{\rm c}$ is the core
radius. The input image of A$^{\rm I}$ is the two dimensional
$\beta$-model over the entire FOV of XIS, 
and the extract region of A$^{\rm I}$ is SRC region
of figure~\ref{f1}(a).

RMFs and ARFs of the ACIS were produced by \texttt{specextract} which
is included in the CIAO package, together with source and background
spectra of the ACIS.


\section{Estimation of Background}

\begin{table*}[htbp]
\begin{center}
\caption{Results of Model Fittings to XIS spectra of the BGD region}\label{t2}
\begin{tabular}{ccccccccc}
\hline
\hline
\multicolumn{3}{c}{LHB}&\multicolumn{3}{c}{MWH}&\multicolumn{2}{c}{CXB}&\\
$kT$ (keV)&$Norm^\ast$&$S^\dagger$&$kT$ (keV)&$Norm^\ast$&$S^\dagger$&$\Gamma$&$S^\ddagger$&$\chi^{2}$/dof\\
\hline
$0.12^{+0.04}_{-0.02}$&$1.60^{+1.54}_{-0.68}$&$3.91^{+3.62}_{-1.74}$&$0.26^{+0.05}_{-0.03}$&$0.32^{+0.16}_{-0.20}$&$3.70^{+1.70}_{-2.47}$&1.41
 (fix)&$5.09^{+0.12}_{-0.12}$&492.0/441\\
\hline
\end{tabular}
\begin{description}
\item[$^\ast$] Normalization of the $apec$ component divided by
           400$\pi$ which is the area of the circular region ($r=20'$ in
           radius) used for the uniform-sky ARF calculation.  
           $Norm=1/400\pi \int n_{\rm e} n_{\rm H} dV/[4\pi (1+z)^2D^2_{\rm A}]\times 10^{-20}$ cm$^{-5}$ arcmin$^{-2}$, where $D_{\rm A}$ is the angular diameter distance to the source.
\item[$^\dagger$] The 0.5--1.0 keV surface brightness in units of $10^{-9}$ erg s$^{-1}$ cm$^{-2}$ sr$^{-1}$.
\item[$^\ddagger$] The 2.0--10.0 keV surface brightness in units of $10^{-8}$ erg s$^{-1}$ cm$^{-2}$ sr$^{-1}$.
\end{description}
\end{center}
\end{table*}

Accurate estimation of background is the key to a spectral analysis of
low surface brightness X-ray emission such as ICM.
In the case of MS~1512.4+3647, the peripheral region of XIS
(outside the virial radius, for example) where no ICM emission
is expected can be used for that purpose.
However, the background level of in the BGD region is not exactly the
same as that in the SRC region.
Therefore, we estimate non-X-ray background (NXB) and X-ray background
(XRB) separately which are specific to the region to be analyzed.
XRB is composed of cosmic X-ray background (CXB) and Galactic
foreground emission (GFE).
We identified point sources with Chandra and XMM-Newton data
and excluded them.

\subsection{Non-X-Ray Background}
\begin{figure}
 \begin{center}
  \FigureFile(80mm,40mm){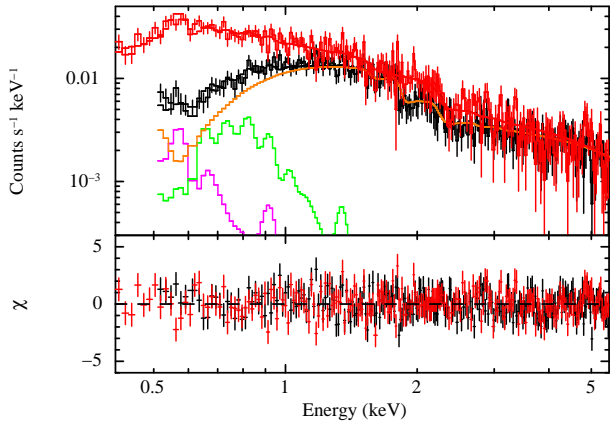}
 \end{center}
 \caption{NXB-subtracted XIS spectra of the BGD region, fitted with the X-ray
 background model described in section~\ref{bgd}. Black and red are
 0.5--5.5 keV XIS-FI (averaged over XIS0 and XIS3) and 0.4--5.5 keV
 XIS-BI spectra, respectively. The CXB component is shown in orange
 line, and the LHB and MWH emissions are indicated by magenta and green
 lines, respectively.}\label{f3}
\end{figure}

The NXB, which is the instrumental background except for the celestial
X-ray background, was generated by \texttt{xisnxbgen}, which sorts
night Earth data around an observation according to COR and makes a sum
of the night Earth spectra weighted by the 
actual COR distribution of the observation (Tawa et al. 2008).
Integrated period for the NXB is between $\pm365$ days of the Suzaku
MS~1512.4+3647 observation.
Night Earth events from the same detector area as the region to
be analyzed in the MS~1512.4+3647 XIS image were extracted
in \texttt{xisnxbgen}.
The systematic error associated with the NXB is $\pm$3\%
(Tawa et al. 2008) in 1.0--12.0 keV energy range.

\subsection{Point Sources}\label{point}
In the analysis of MS~1512.4+3647, the contribution of point sources
has to be subtracted. The XIS image shown in figure~\ref{f1}(a) is
thought to be contaminated by a number of point sources, but they
are not resolved well due to a moderate spatial resolution of X-ray
telescope (a half-power diameter of $\sim 2'$; Serlemitsos et al. 2007).
To detect point sources, we utilized the Chandra
Source Catalog (CSC; Evans et al. 2010) and the XMM-Newton 2nd
Incremental Source Catalogue (2XMMi; Watson et al. 2009).
The cataloged point sources are shown in figure~\ref{f1}
(serial numbers from 25 to 38).
Position, photon index, and energy flux of these point sources are
summarized in table~\ref{t1}.
The detection limits of these catalogs are about
$1\times10^{-14}$ erg s$^{-1}$ cm$^{-2}$ in 0.5--7.0 keV energy range
for CSC and
$1.5\times10^{-14}$ erg s$^{-1}$ cm$^{-2}$ in 2.0--12.0 keV energy range
for 2XMMi.
Because the cataloged sources are located far from MS~1512.4+3647,
we also utilized the Chandra data to detect point sources around
MS~1512.4+3647.
The CIAO tool \texttt{wavdetect} (Freeman et al. 2002) was executed
on the 0.5--7.0 keV Chandra ACIS images of OBS1 with a detection
threshold of 5$\sigma$.
As shown in black crosses of figure~\ref{f1}(b) with serial numbers
from 1 to 24, 24 point sources were detected in the ACIS FOV.
The detection limit is about
$2.0\times10^{-15}$ erg s$^{-1}$ cm$^{-2}$ in 2.0--10.0 keV energy
range.

In the case of point sources detected with \texttt{wavdetect}
in table~\ref{t1}, photon index, and energy flux were derived
from spectral model fitting to the ACIS data, extracted from a
circular region of $1''$ or $2''$ in radius depending on
the source extent.
Background was extracted from a source-free region of $1'$ 
in radius.
In the fitting, the two sets of data (OBS1 and OBS2) were added,
and absorbed power-law model, $wabs\times powerlaw$, was utilized
with free or fixed (1.7) photon index depending on statistics.
Column density was fixed to the Galactic value of
$1.4\times10^{20}$ cm$^{-2}$. The results are shown in figure~\ref{f2}.
The energy flux of the point sources ranges from
$\sim 10^{-15}$ to $\sim 10^{-13}$ erg s$^{-1}$ cm$^{-2}$.
Some sources which were detected by \texttt{wavdetect} are also
listed in CSC.
We confirmed that energy flux obtained by \texttt{wavdetect} and CSC
is consistent within errors.
In following spectral analysis, point sources with energy flux larger than
1.6$\times10^{-14}$ erg s$^{-1}$ cm$^{-2}$ in 2.0--10.0 keV
are excluded in circular regions of $1'$ or $2'$ radius depending on
source extent, as shown in figure~\ref{f1}(a).
The point sources with energy flux lower than 1.6$\times10^{-14}$ erg
s$^{-1}$ cm$^{-2}$ in 2.0--10.0 keV are included in the CXB model
(detailed in section~\ref{cxb}), corresponding to 30\% of CXB
intensity, and have roughly consistent numbers with the log$N$--log$S$
relation (Kushino et al. 2002).

\subsection{Cosmic X-Ray Background}\label{cxb}
After the removal of point sources, there remains contribution
from unresolved extragalactic sources (CXB) to XRB.
In the spectral analysis, an absorbed power-law model with a fixed
photon index of 1.41 (Kushino et al. 2002) $wabs\times powerlaw$ was
used as the CXB model.
Intensity of the CXB after the removal of point sources with flux higher than
$1.6\times10^{-14}$ erg s$^{-1}$ cm$^{-2}$ in 2.0--10.0 keV
is calculated to be
$(3.80^{+0.46}_{-1.52}\pm0.38)\times 10^{-8}$ erg s$^{-1}$
cm$^{-2}$ sr$^{-1}$ (90\% statistical and systematic errors)
from equation (6) of Kushino et al. (2002).

\subsection{Galactic Foreground Emission}
GFE typically consists of an unabsorbed plasma (LHB; representing the
local hot bubble and the solar-wind charge exchange) and an absorbed
plasma (MWH; representing the Milky Way halo).
To express the GFE, we employed two-temperature model,
$apec_{1}+wabs\times apec_{2}$, according to Tawa et al. (2008),
where the $apec$ model is a thin thermal plasma emission model
detailed in Smith et al. (2001).
The redshift and abundance of the two $apec$ components were
fixed to zero and unity, respectively.

\subsection{Spectral Analysis of Background Region}\label{bgd}

In order to estimate the XRB around MS~1512.4+3647, 
we extracted FI and BI spectra from a background region of the XIS
image (BGD region in figure~\ref{f1}(a)).
This BGD region is further than 1.1$r_{200}$ of MS~1512.4+3647,
and the contribution of ICM is 8\% of XRB there,
assuming the ICM extends beyond $r_{200}$ following the $\beta$-model
profile obtained in section 2.3 with the point spread function of XRT.
We fitted NXB-subtracted FI and BI spectra of the BGD region
simultaneously with a model of
$apec_{1}+wabs\times(apec_{2}+powerlaw)$
that represents the CXB and GFE emissions in 0.5--5.5 keV for FI
and 0.4--5.5 keV for BI.
Galactic absorption was fixed to the Galactic value of
$1.4\times10^{20}$ cm$^{-2}$.
The ARFs of A$^{\rm U}$ were used for the model fitting.
The best-fit parameters for the BGD region are summarized in
table~\ref{t2}, and the spectra are shown in figure~\ref{f3}.
The fit is moderately good with the probability that chi squared value
becomes larger than the fitting value is 0.047.

The 2.0--10.0 keV surface brightness of the CXB component is
$(5.09\pm0.12\pm0.27)\times 10^{-8}$ erg s$^{-1}$ cm$^{-2}$ sr$^{-1}$
(90\% statistical and systematic errors), where the systematic error of
5.4\% is derived by scaling the fluctuation analysis with Ginga
(Hayashida 1989) to our flux limit and FOV of
$1.6\times10^{-14}$ erg s$^{-1}$ cm$^{-2}$ and $0.06$ deg$^{2}$,
respectively (detailed in Hoshino et al. 2010; Akamatsu et al. 2011).
This CXB fluctuation is consistent with Nakazawa et al. (2009) scaled
by HEAO-1, -2 results.
The obtained surface brightness is slightly larger
($\sim25\%$) than the calculated value described in section~\ref{cxb},
although consistent within statistical and systematic errors.
In the spectral analysis of the ICM, we also tried the calculated CXB
surface brightness instead of that obtained from the BGD region, and
confirmed that the fitting results did not change significantly.

The temperature and surface brightness of LHB are
$kT=0.12^{+0.04}_{-0.02}$ keV and 
$(3.91^{+3.62}_{-1.74})\times10^{-9}$ erg s$^{-1}$ cm$^{-2}$ sr$^{-1}$
in 0.5--1.0 keV energy range,
respectively. Those of MWH are $kT=0.26^{+0.05}_{-0.03}$ keV and 
$(3.70^{+1.70}_{-2.47})\times10^{-9}$ erg s$^{-1}$ cm$^{-2}$ sr$^{-1}$
in 0.5--1.0 keV energy range,
respectively. The obtained temperatures of LHB and MWH are consistent
with typical values of these GFE components
(Lumb et al. 2002; Yoshino et al. 2009).


\begin{figure*}
\begin{minipage}{0.5\hsize}
 \begin{center}
  \FigureFile(80mm,40mm){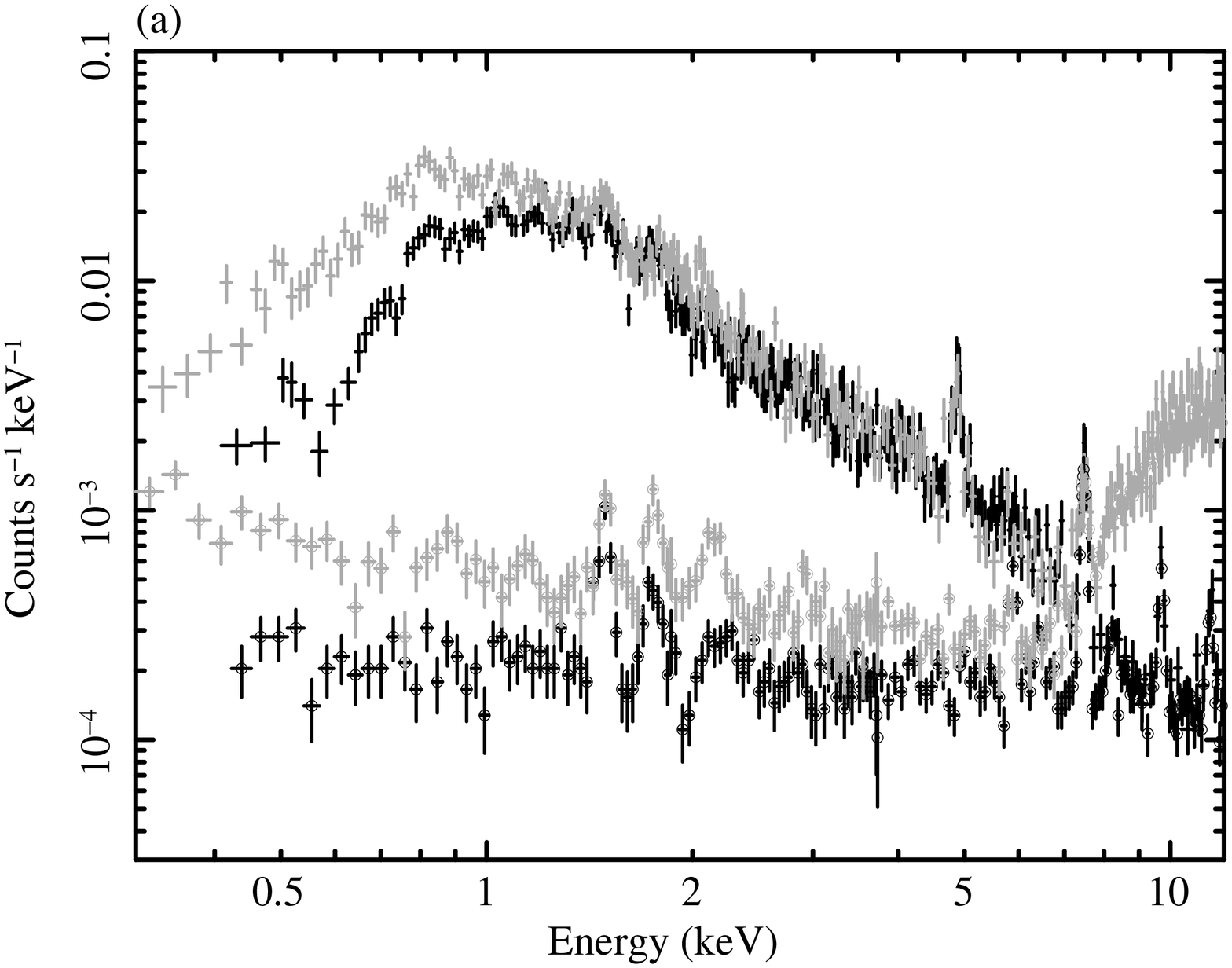}
 \end{center}
\end{minipage}
\begin{minipage}{0.5\hsize}
 \begin{center}
  \FigureFile(80mm,40mm){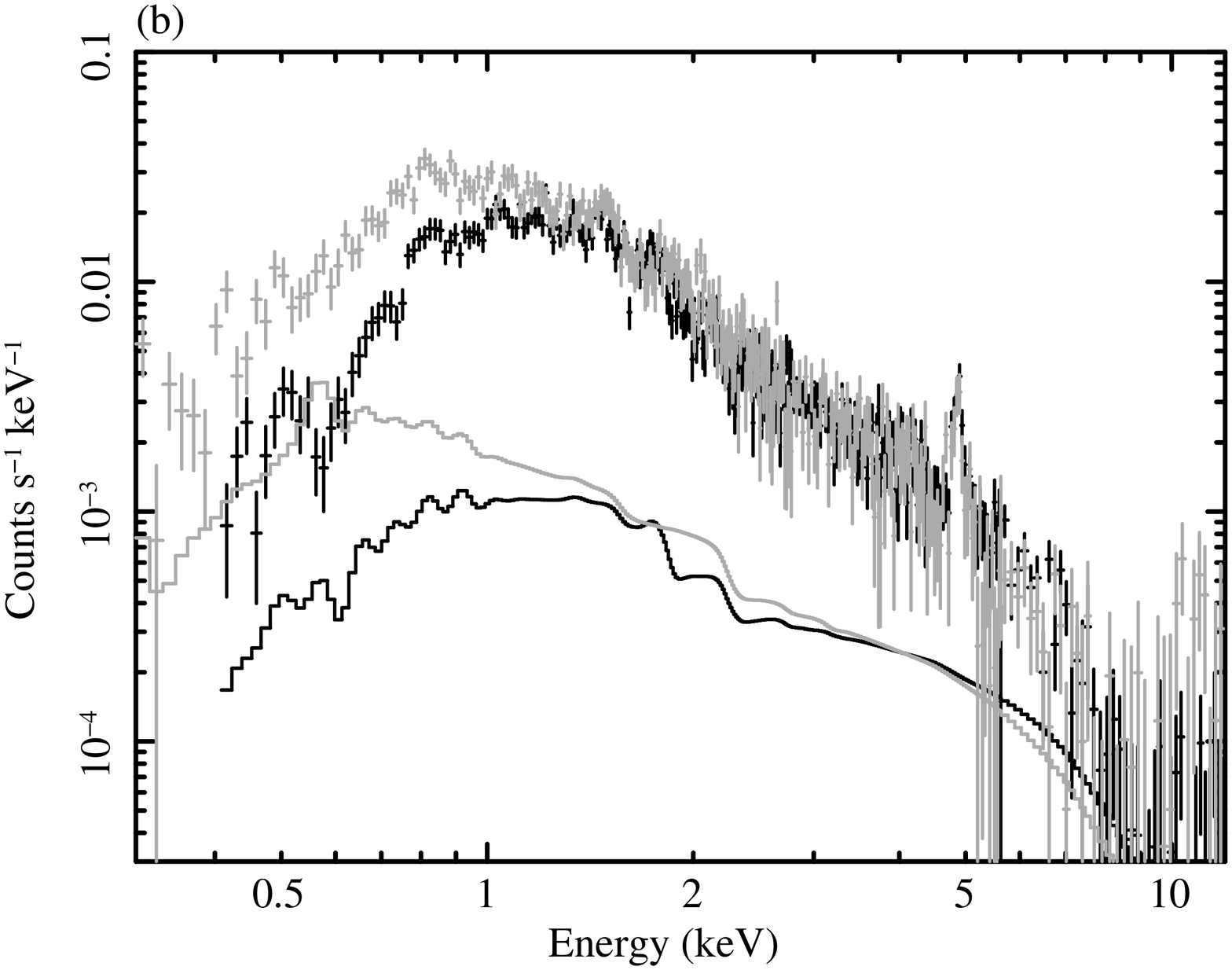}
 \end{center}
\end{minipage}
\caption{(a) Raw XIS spectra of the SRC region and NXB spectra
 extracted from the same region. Black and gray crosses are 0.3--12.0 keV XIS-FI
 (averaged over XIS0 and  XIS3) and XIS-BI raw spectra, respectively, and
 circled black and gray crosses are XIS-FI and XIS-BI NXB spectra,
 respectively. (b) NXB-subtracted spectra of the SRC region with the
 best-fit X-ray background model described in section~\ref{bgd}.}
\label{f4}
\end{figure*}

\begin{figure*}
 \begin{minipage}{0.5\hsize}
  \begin{center}
   \FigureFile(80mm,40mm){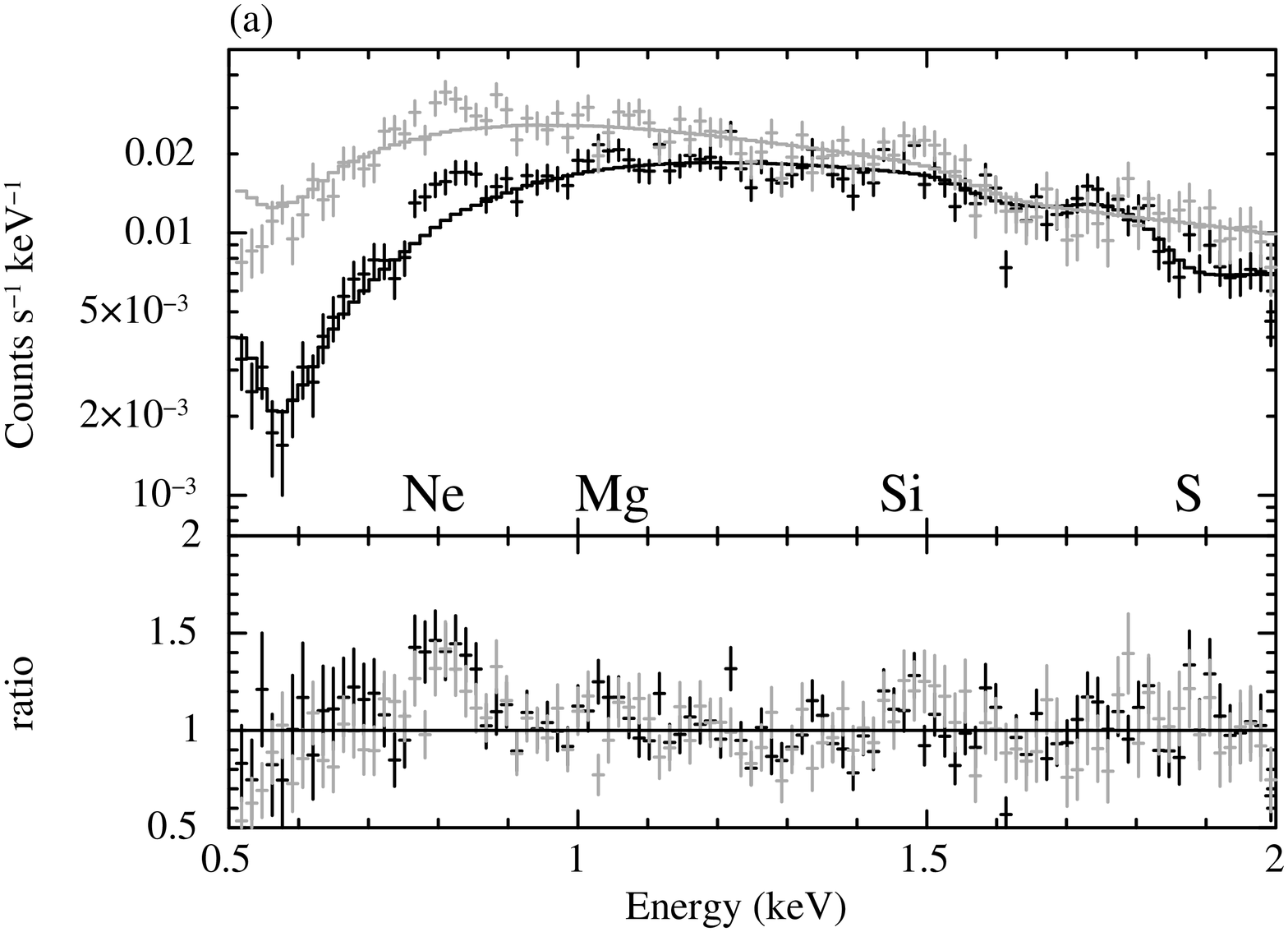}
  \end{center}
 \end{minipage}
 \begin{minipage}{0.5\hsize}
  \begin{center}
   \FigureFile(80mm,40mm){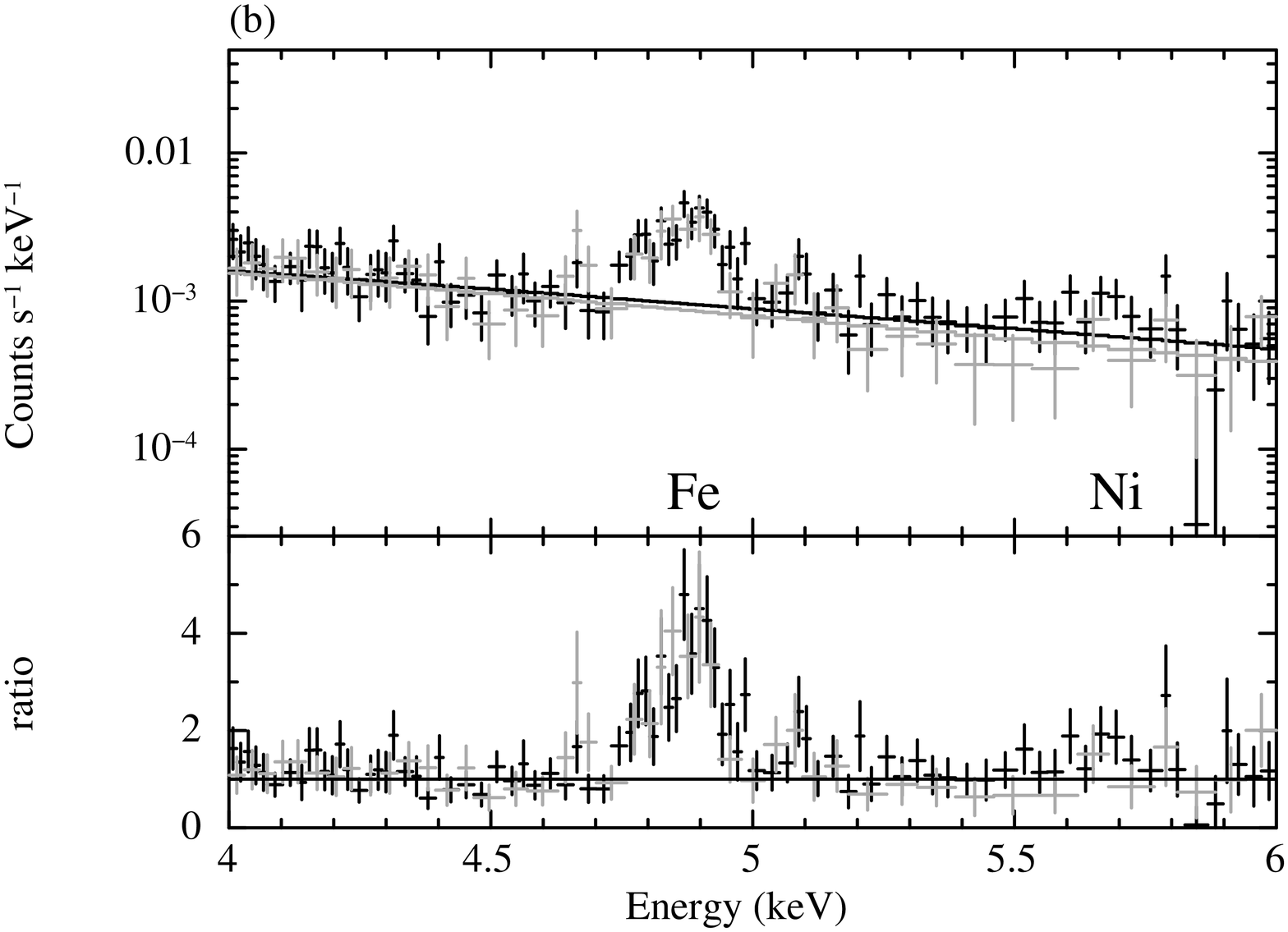}
  \end{center}
 \end{minipage}
\caption{(a) XIS-FI (black) and XIS-BI (gray) spectra around
 the Ne, Mg, S, and Si emission lines extracted from the
 SRC region, with the best-fit $wabs\times bremth$ model
 (solid lines). The positions of the redshifted metal lines are also
 indicated. The lower panel shows the ratio of the data to model.
 (b) The same as panel(a), but for spectra around the Fe and Ni
 emission lines.}\label{f8}
\end{figure*}

\section{Spectral Analysis and Results}

For the spectral analysis of the ICM in MS~1512.4+3647, 
we extracted XIS spectra within a circular region of 2.2$'$ in radius
($\sim0.6r_{\rm200}$, shown as the SRC region in figure~\ref{f1}(a)),
centered on
($\alpha$, $\delta$)$=$
(\timeform{15h14m22s.6},\timeform{+36D36'21''.0}).
The center was determined from the position of X-ray emission centroid
in the Chandra ACIS image, as described in section 2.3.
The radius of extraction was determined to maximize significance of
the ICM signal with respect to the noise, composed of the NXB, CXB, and
GFE.
The three point sources within the SRC region were masked by a circle
of 1$'$ radius (dashed circles in figure~\ref{f1}(a)).
Since MS~1512.4+3647 is rather distant, X-ray signals from core regions
and outer regions are highly mixed due to the Suzaku PSF.
Therefore, we did not study radial dependence of parameters such as
temperature, gas density, and metal abundances. 
After extracting spectra of the XIS0, XIS1, and XIS3, we created
a single XIS-FI spectra by averaging the XIS0 and XIS3 spectra.

In figure~\ref{f4}(a), we compared raw SRC spectra (black cross for
XIS-FI and gray cross for XIS-BI) with corresponding NXB spectra
(circled black cross and circled gray cross).
The raw SRC spectra have significant excess in 0.4--6.5 keV (XIS-FI)
and 0.3--6.5 keV (XIS-BI), respectively, due to the ICM and XRB emission.
In the higher energies than 6.5 keV, the raw SRC spectra overlap with
the NXB spectra, because the NXB is dominant in the energy range.
Figure~\ref{f4}(b) is NXB-subtracted SRC spectra (black cross for
XIS-FI and gray cross for XIS-BI) with the best-fit XRB model
for the BGD region described in section~\ref{bgd}
(black and gray solid line).
The signal from ICM is significantly detected in 0.5--6.5 keV
(XIS-FI) and 0.4--6.5 keV (XIS-BI), respectively.
The emission lines from ionized metals,
especially the redshifted Fe-K line,
are seen in the spectra of figure~\ref{f4}(b).
These lines are clearer as excess in each atomic-line
energy of Ne, Mg, Si, S, Fe, and Ni shown in figure~\ref{f8},
where we fitted the NXB-subtracted SRC spectra
with $wabs\times bremss+$XRB model.

\begin{figure}
 \begin{center}
  \FigureFile(80mm,40mm){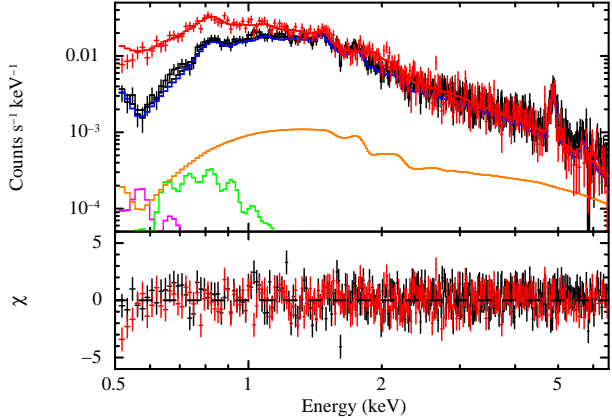}
 \end{center}
 \caption{NXB-subtracted spectra of the SRC region. Black and red are
 0.5--6.5 keV XIS-FI (averaged over XIS0 and XIS3) and XIS-BI
 spectra, respectively. These spectra are fitted simultaneously with
 the 1T model added to the XRB model (section~\ref{bgd}). Black and red
 lines are the best-fit model of the XIS-FI and XIS-BI, respectively.
 For the XIS-FI, model components of the ICM (1T), LHB, MWH, and CXB
 are shown in blue, magenta, green, and orange, respectively.
 The lower panel shows the fit residuals in units of $\sigma$.
 XRB spectra are not plotted here for clarity.}\label{f5}
\end{figure}

In order to determine metal abundances, the NXB-subtracted XIS-FI
and XIS-BI SRC spectra were fitted with an absorbed single-temperature
thin-thermal emission model (1T), $wabs \times vapec$
($vapec$: Smith et al. 2001) added to the XRB model (section~\ref{bgd}),
in 0.5--6.5 keV energy range as shown in figure~\ref{f5}.
In this fitting, the BGD spectra (figure~\ref{f3}) were simultaneously
fitted with the same XRB model to determine the XRB in MS~1512.4+3647.
For the SRC spectra, we utilized the NXB and response (RMF and ARF)
files of the SRC region (section~\ref{resp}),
while those of the BGD region were used for the BGD spectra.
The uniform sky ARF, A$^{\rm U}$, and the $\beta$-model ARF,
A$^{\rm I}$, were assigned to the XRB model and the ICM model,
respectively.
In the 12 metal abundances in $vapec$, we set He = C = N
= O = one solar abundance because the emission lines from these elements
are out of energy range. 
The emission lines from Al, Ar, and Ca are not significantly detected
(figure~\ref{f8}).
Then we set S = Ar = Ca, and set Al as free parameter.
We also examined two cases: 1) free parameters of S, Ar, and Ca, 2)
linked parameters of Mg = Al.
The errors of S and Mg abundances increased about 5\% in the case 1) and
10\% in the case 2), respectively.
The other abundances, Ne, Si, Fe, and Ni, were free.
Therefore, we had seven free metal-abundance parameters.
The hydrogen column density was fixed to the Galactic value of
$1.4\times 10^{20}$ cm$^{-2}$ and redshift was also fixed to 0.372.
In the XRB model, all parameters were free except for
the photon index of the CXB model (section~\ref{bgd}).
The fitting result is summarized in table~\ref{t3}.
The Al abundance is not determined significantly and we don't
indicate about Al abundance in the following analysis.

\begin{table*}[htbp]
\begin{center}
\small{
\caption{Results of Model Fittings to the XIS spectra of SRC region.}\label{t3}
\begin{tabular}{lccccccccc}
\hline
\hline
Model&kT1&kT2&Ne&Mg&Si&S=Ar=Ca&Fe&Ni&$\chi^{2}$/d.o.f.\\
&(keV)&(keV)&(solar)&(solar)&(solar)&(solar)&(solar)&(solar)&\\
\hline
1T&$3.28^{+0.09}_{-0.09}$&---&$<2.26$&$0.64^{+0.54}_{-0.51}$&$0.63^{+0.21}_{-0.20}$&$0.42^{+0.23}_{-0.23}$&$0.64^{+0.08}_{-0.07}$&$1.94^{+1.06}_{-1.02}$&$1087.8/1014$\\
2T&$0.79<$&$0.92<$&1 (fix)&$0.73^{+0.48}_{-0.26}$&$0.67^{+0.17}_{-0.11}$&$0.49^{+0.20}_{-0.12}$&$0.62^{+0.06}_{-0.07}$&$2.00^{+0.91}_{-0.90}$&$1085.4/1013$\\
CXB$-$6\%&$3.30^{+0.08}_{-0.09}$&---&$<2.49$&$0.65^{+0.55}_{-0.51}$&$0.64^{+0.21}_{-0.20}$&$0.42^{+0.23}_{-0.23}$&$0.65^{+0.08}_{-0.08}$&$2.02^{+1.07}_{-1.03}$&$1095.0/1015$\\
CXB$+$6\%&$3.29^{+0.09}_{-0.04}$&---&$<2.26$&$0.65^{+0.54}_{-0.52}$&$0.64^{+0.21}_{-0.23}$&$0.42^{+0.21}_{-0.21}$&$0.64^{+0.04}_{-0.07}$&$1.94^{+1.04}_{-0.76}$&$1083.3/1015$\\
NXB$\pm$3\%&$3.28^{+0.08}_{-0.09}$&---&$<2.27$&$0.64^{+0.54}_{-0.51}$&$0.63^{+0.21}_{-0.20}$&$0.42^{+0.23}_{-0.23}$&$0.64^{+0.08}_{-0.07}$&$1.94^{+1.06}_{-1.02}$&$1080.8/1014$\\
CONTAMI$-$10\%&$3.38^{+0.11}_{-0.09}$&---&$<2.01$&$0.81^{+0.58}_{-0.54}$&$0.74^{+0.23}_{-0.21}$&$0.51^{+0.25}_{-0.24}$&$0.64^{+0.08}_{-0.08}$&$1.99^{+1.09}_{-1.06}$&$1092.3/1014$\\
CONTAMI$+$10\%&$3.19^{+0.04}_{-0.09}$&---&$<2.59$&$0.51^{+0.52}_{-0.48}$&$0.55^{+0.20}_{-0.19}$&$0.35^{+0.22}_{-0.21}$&$0.64^{+0.08}_{-0.07}$&$1.93^{+1.04}_{-1.00}$&$1092.8/1014$\\
\hline
\end{tabular}
}
\end{center}
\end{table*}

The metal abundances of individual elements are determined in $3\sigma$
significance except for Ne and Mg which have large error bars.
These metal abundances of MS~1512.4+3647 are similar to those of nearby
clusters (e.g. Sato et al. 2009b).
The best-fit luminosity and temperature are
$(2.9\pm0.2)\times10^{44}$ erg s$^{-1}$ in 2.0--10.0 keV energy range
and $3.28\pm0.09$ keV, respectively.
The temperature is consistent with ASCA results (Ota 2000) within
statistical errors.
The luminosity is also consistent with the ASCA value (Ota 2000) 
when it is corrected for the difference of integrated areas
between ASCA GIS and Suzaku XIS.

The ICM in MS~1512.4+3647 is considered to have temperature structures,
especially between the core and outer regions.
If this is true, a multi-temperature model may reproduce the SRC
spectra better. 
We tried two-temperature model (2T), $wabs \times (vapec+vapec)$. 
In the 2T model, each metal abundance was tied between the two thermal
components and Ne abundance was fixed to one solar value. Therefore, in
the 2T model, only one temperature parameter was added as a new free
parameter to the 1T model.
The fitting result is summarized in table~\ref{t3}.
From the F-test, we cannot decline the null hypothesis at 5\%
significance level that adding one thermal component from 1T to 2T
does not improve the fit, because the P-value is 0.137.
In fact, the two ICM temperatures are obtained only as lower limits,
and the metal abundances are consistent with those of the 1T model
within statistical errors.
Thus, the 1T model is statistically sufficient to reproduce the data.
It suggests that the potential temperature drop in the very central
region is moderate, if any.

Finally, we tested the robustness of our results against the systematic
errors associated with the background estimation and the contamination
estimation on the blocking filter.
We repeated the spectral fit by changing the NXB intensity by $\pm 3\%$.
The error due to the CXB intensity fluctuation was also examined by
varying the CXB normalization by 6\%, which was derived by scaling the
Ginga results with the XIS sensitivity and the field of view
(section~\ref{bgd}).
The systematic error in the contamination thickness of the XIS blocking
filter is typically 10\%, and we evaluated its effect through spectral
fits by adjusting the detector response which included the contamination
thickness. The results are summarized in table~\ref{t3}.
The temperature and abundances did not change significantly by these
systematic uncertainties. We employed the systematic errors,
$\sigma_{\rm NXB}$, $\sigma_{\rm CXB}$, and $\sigma_{\rm CONTAMI}$, by
examining the change of the fitted parameters in table\ref{t3}.
As a result, $\sigma_{\rm CXB}$ and $\sigma_{\rm NXB}$ are always less
than 10\% of the statistical errors, and $\sigma_{\rm CONTAMI}$ is
comparable to the statistical error for the temperature and roughly 50\%
of that for the abundances of Mg, Si and S.
In the following analysis we used the error defined by $\sigma_{\rm
error} = (\sigma_{\rm NXB}^2+\sigma_{\rm CXB}^2 +\sigma_{\rm
CONTAMI}^2)^{1/2}$.

\begin{figure*}[tbp]
 \begin{minipage}{0.5\hsize}
  \begin{center}
   \FigureFile(80mm,40mm){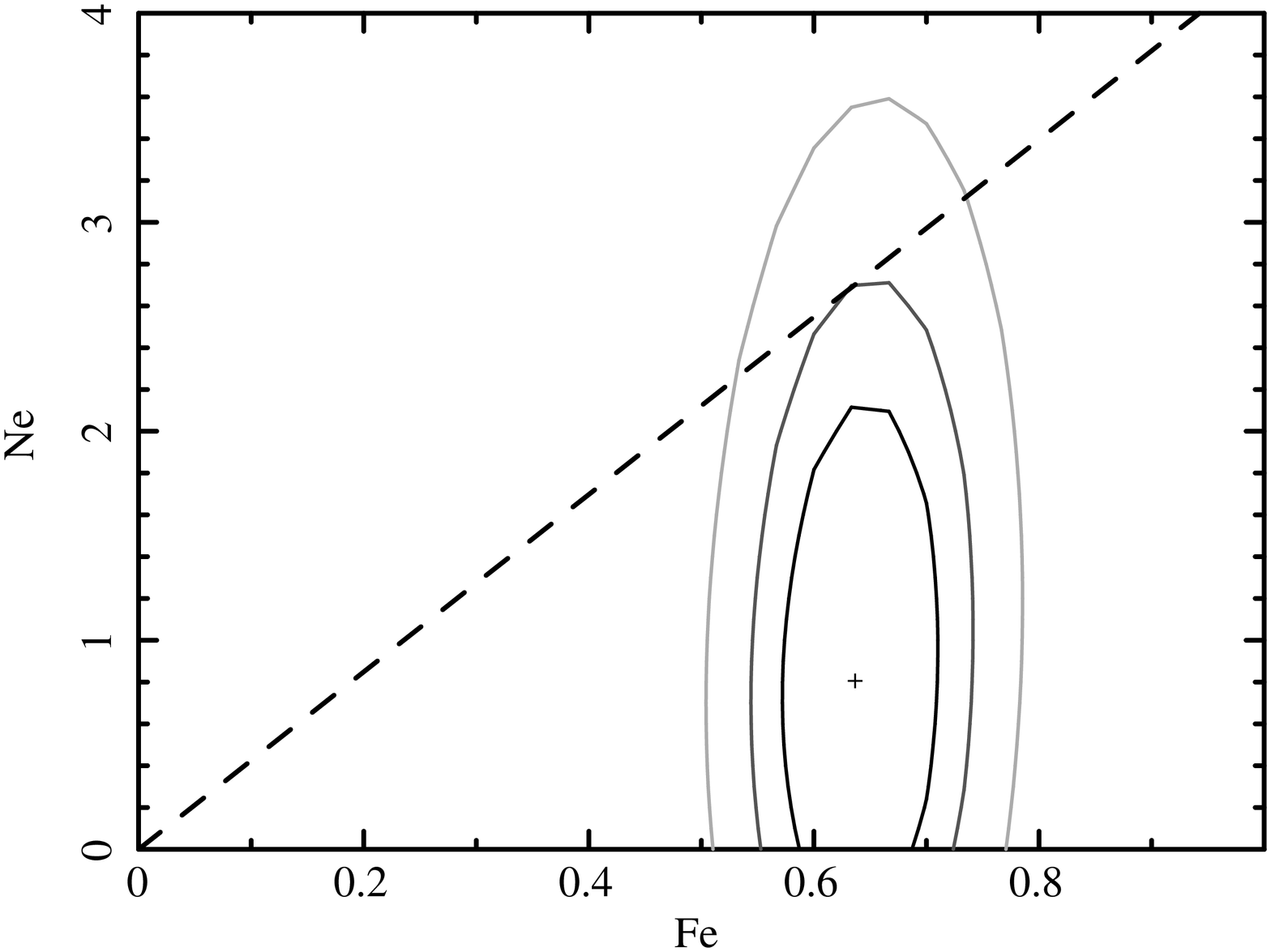}
  \end{center}
 \end{minipage}
 \begin{minipage}{0.5\hsize}
  \begin{center}
   \FigureFile(80mm,40mm){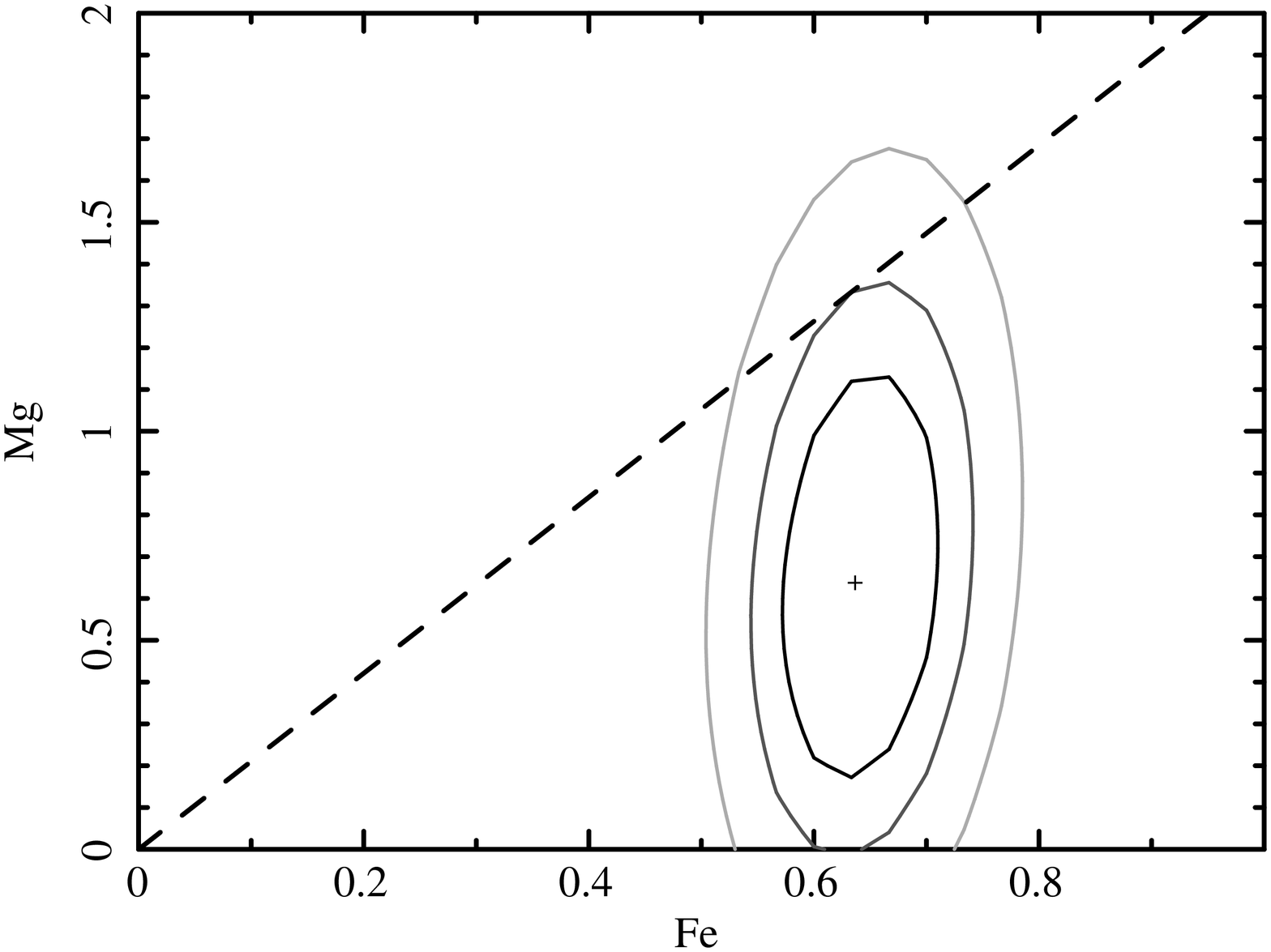}
  \end{center}
 \end{minipage}
 \begin{minipage}{0.5\hsize}
  \begin{center}
   \FigureFile(80mm,40mm){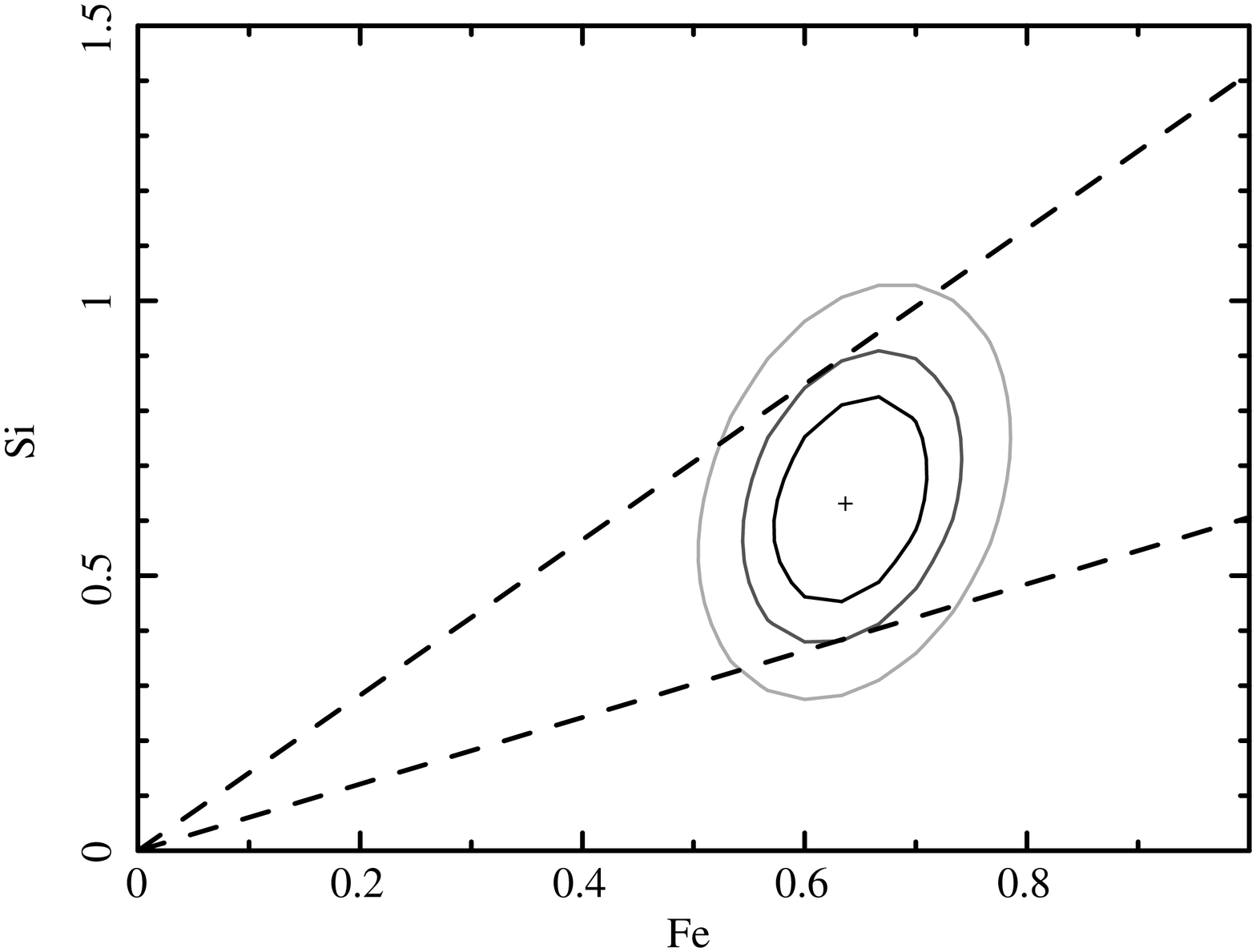}
  \end{center}
 \end{minipage}
 \begin{minipage}{0.5\hsize}
  \begin{center}
   \FigureFile(80mm,40mm){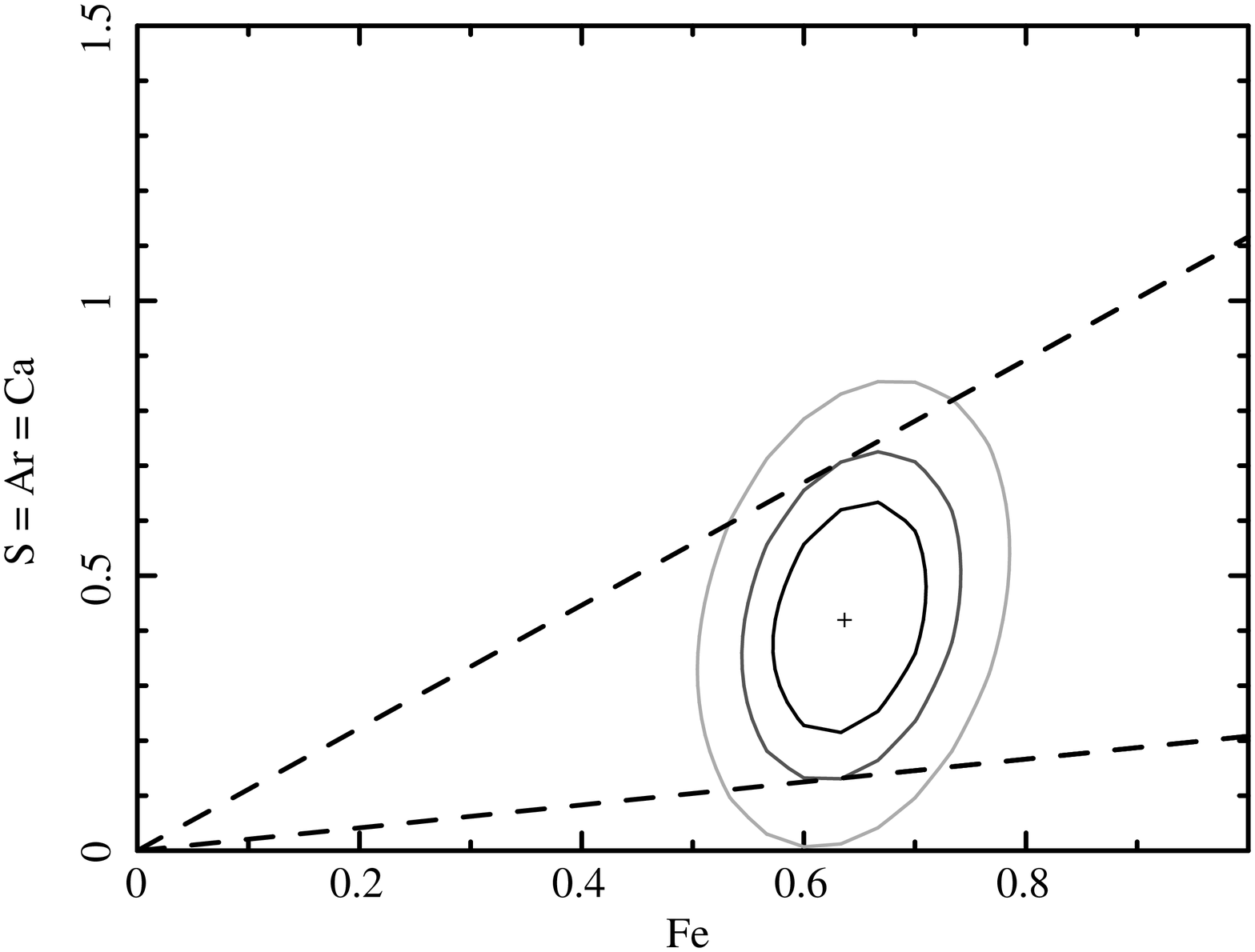}
  \end{center}
 \end{minipage}
 \begin{minipage}{0.5\hsize}
  \begin{center}
   \FigureFile(80mm,40mm){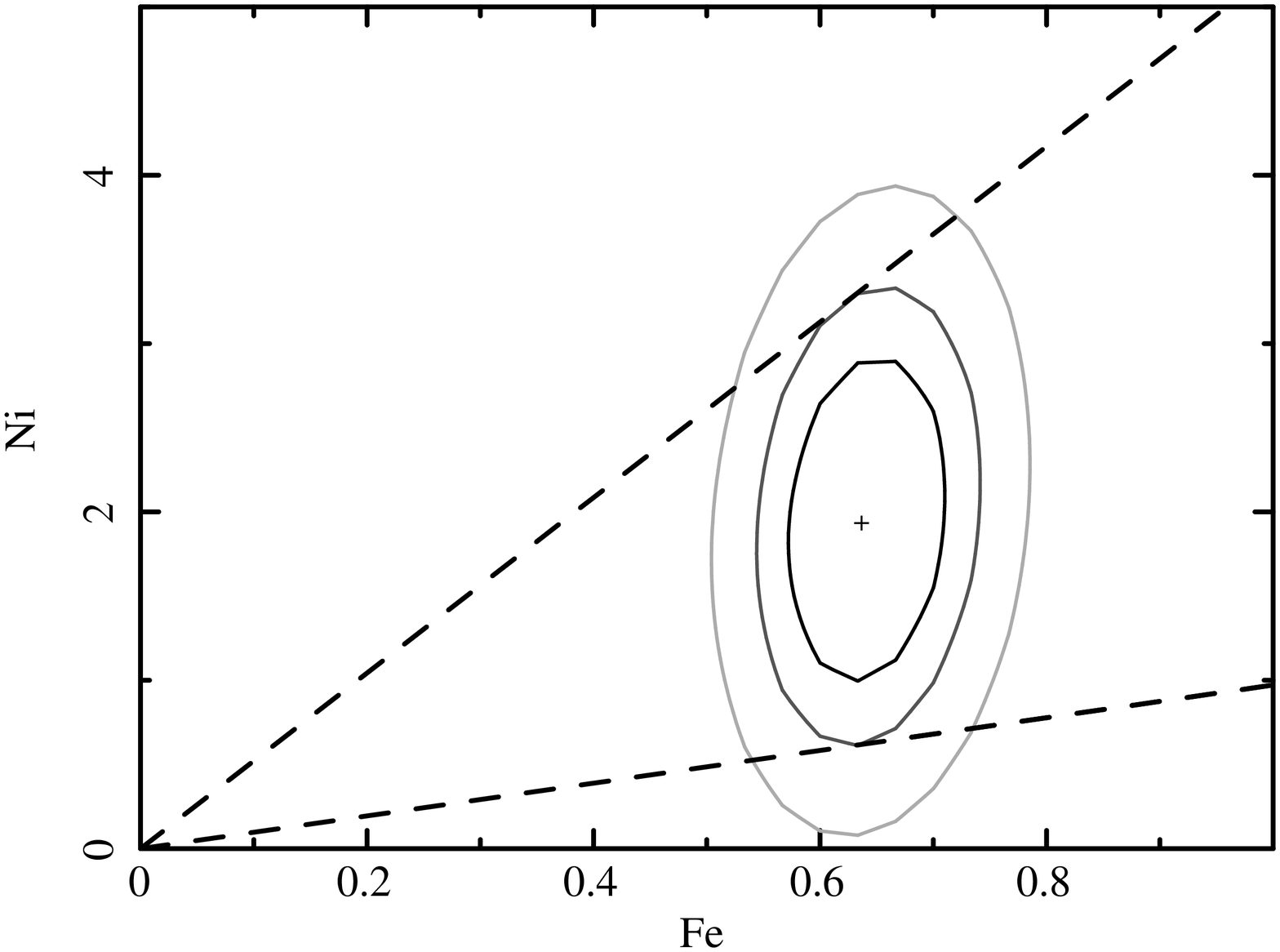}
  \end{center}
 \end{minipage}
 \caption{Confidence contours between a metal (Ne, Mg, Si, S, or Ni) and
 Fe abundances. The black, gray, and light gray contours represent 68\%, 90\%,
 and 99\% confidence regions, respectively.}\label{f7}
\end{figure*}

From the spectral analysis of Suzaku data, we obtained metal abundances
of MS~1512.4+3647, which are similar to those of nearby clusters.
To study the contributions of two types of supernovae (SNe~Ia and SNe~II),
we derived the relative abundance ratios to Fe by calculating confidence
contours between the Fe abundance and another abundance.
The results are shown in figure~\ref{f7} and table~\ref{t4}.
The relative abundance ratio (90\% confidence level) is determined
from the slope of two lines which pass the origin (zero abundance point)
and come in contact with the 90\% confidence region.
The abundance ratios are $\sim1$ solar ratio within statistical errors
(table~\ref{t4}).

\begin{figure}
 \begin{minipage}{0.5\hsize}
 \begin{center}
    \FigureFile(80mm,40mm){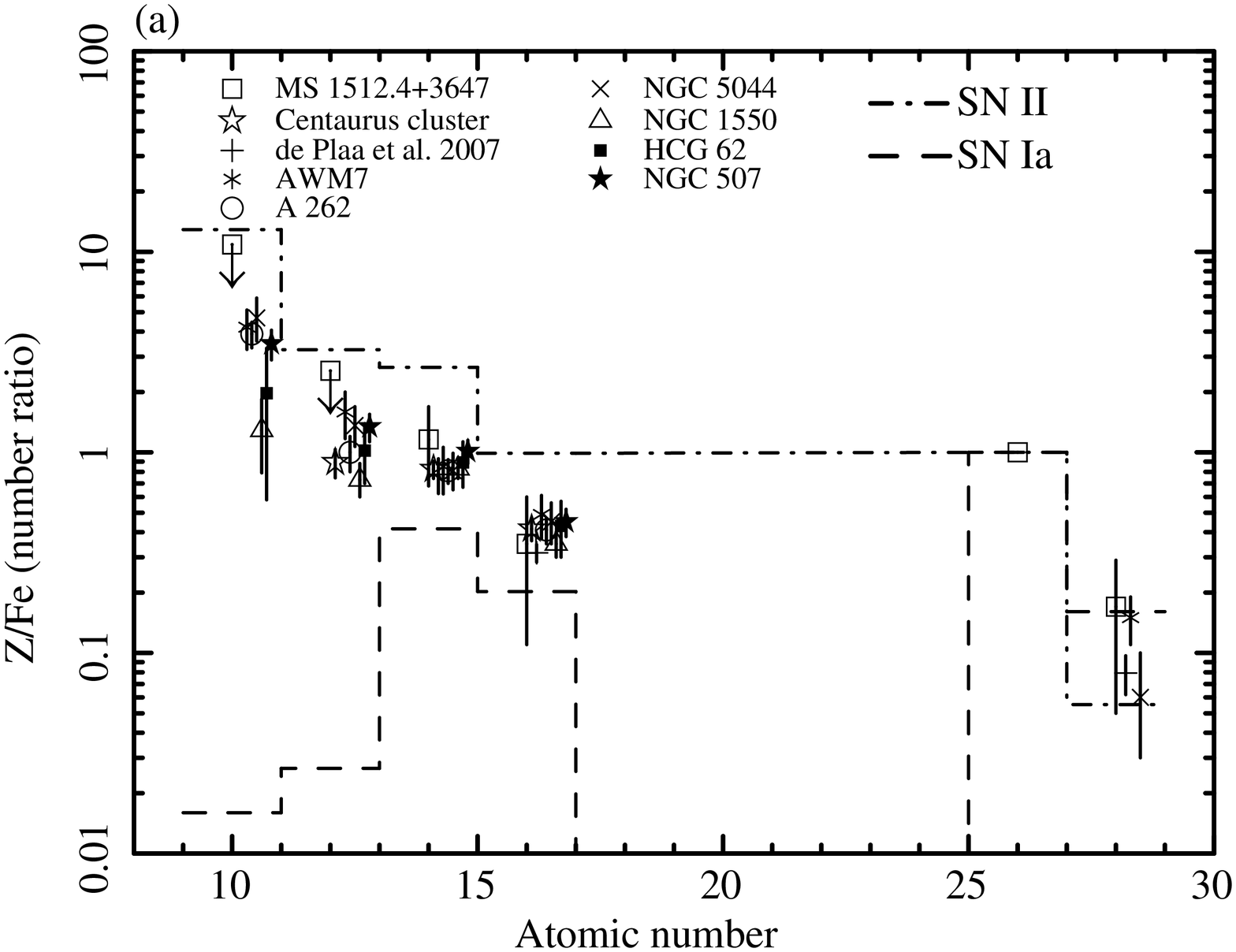}\\
 \end{center}
\end{minipage}
 \begin{minipage}{0.5\hsize}
 \begin{center}
    \FigureFile(80mm,40mm){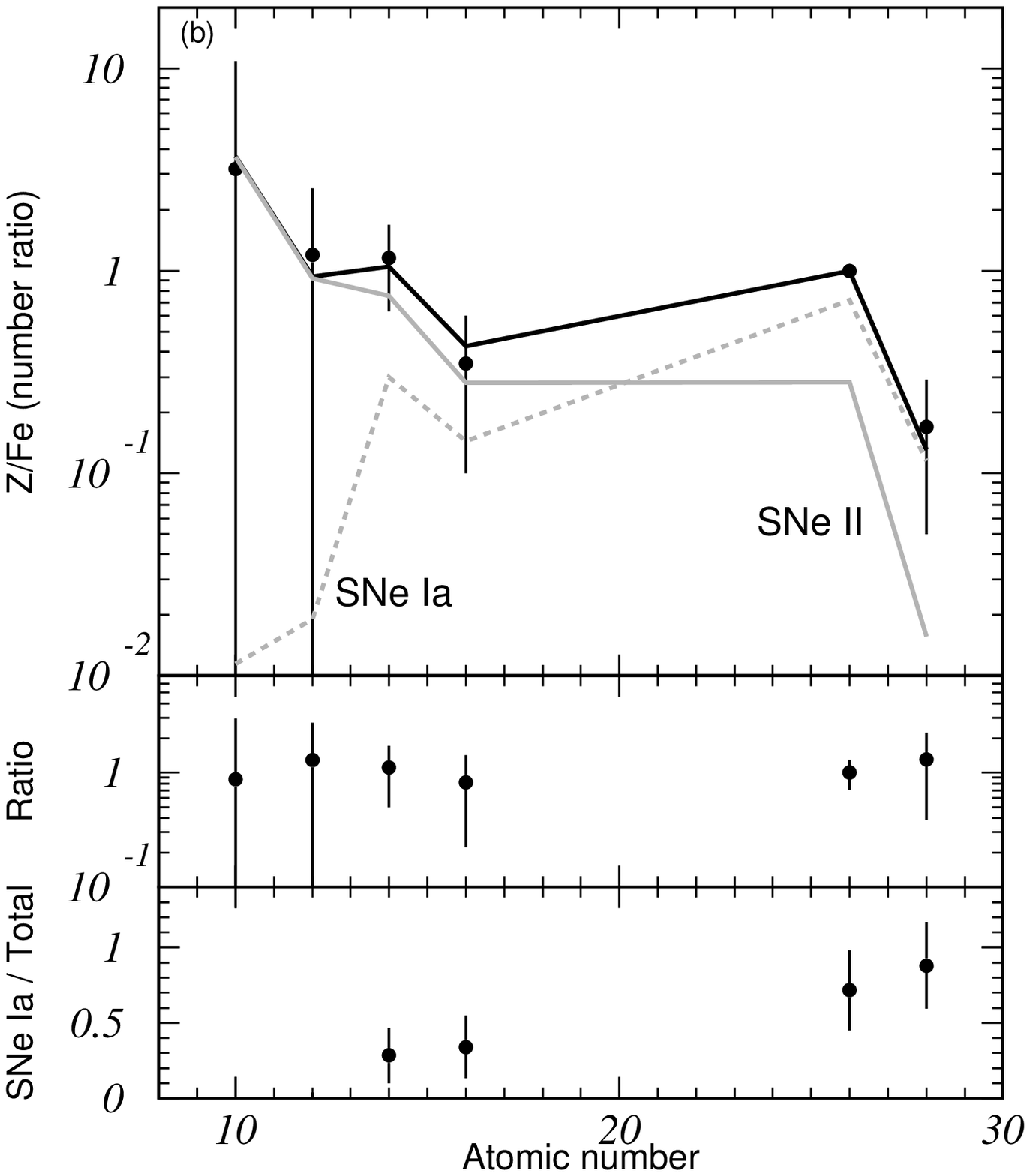}
 \end{center}
\end{minipage}
 \caption{(a) The abundance ratios of Ne, Mg, Si, S, and Ni to Fe within
 $0.6 r_{200}$.
 Those of neaby clusters are also plotted. The supernova
 yield models for SN~Ia (W7: Iwamoto et al. 1999) and SN~II ($Z=0.02$
 and Salpeter IMF: Nomoto et al. 2006) are plotted in
 black dashed and dot-dashed lines, respectively.
 (b) Top panel shows the abundance ratios within 0.6$r_{200}$ (black
 points) fitted by $N_{1}+(N_{2}/N_{1})$ (black line). 
 Gray dashed and solid lines correspond to the contributions
 of SNe~Ia and SNe~II, respectively. Middle and bottom panels indicate
 ratios of data to the best fit model, and fractions of the SNe~Ia
 contribution to total metal numbers of individual elements,
 respectively.}\label{f6}
\end{figure}

Figure~\ref{f6}(a) shows metal to Fe number ratios in MS~1512.4+3647,
obtained from the present Suzaku observation.
In comparison, the same ratios for nearby clusters observed with Suzaku
are also plotted.
The values are for the radius range of 0.05-0.1$r_{180}$, and the
clusters are 
AWM~7 ($z=0.0172$; Sato et al. 2008),
Abell~262 ($z=0.0163$; Sato et al. 2009b), 
Centaurus cluster ($z=0.0104$; Sakuma et al. 2011),
NGC~5044 ($z=0.0090$; Komiyama et al. 2009), 
NGC~1550 ($z=0.0124$; Sato et al. 2010),
HCG~62 ($z=0.0145$; Tokoi et al. 2008), 
and NGC~507 ($z=0.0165$; Sato et al. 2009a).
The figure also shows the XMM-Newton result by de Plaa et al. (2007),
indicating the average values for 22 clusters ($0.0214\leq z\leq0.1840$).

Figure~\ref{f6}(a) shows that MS~1512.4+3647 has the metal to Fe number
ratios consistent with those in nearby clusters within the statistical
and systematic errors, noting that Ne and Mg lines had insufficient
statistics and only upper limits were obtained.
In the same figure, the expected abundance patterns of SNe~Ia and II
yields, based on the calculation by Iwamoto et al. (1999) and Nomoto et
al. (2006), are also plotted. 
We assumed the W7 model for SNe~Ia, along with the
Salpeter initial mass function (IMF) for stellar masses from 10 to 50
$M_{\solar}$ with a progenitor metallicity of $Z=0.02$ for SNe~II. 
The observed number ratios for Ne, Mg, Si, S, and Ni over Fe all lie
between the SN~Ia and SN~II values.
Therefore, both types of the supernova are considered to have enriched
the ICM of the clusters shown in figure~\ref{f6}(a), including
MS~1512.4+3647.

\begin{figure*}
  \begin{minipage}{0.5\hsize}
  \begin{center}
   \FigureFile(80mm,40mm){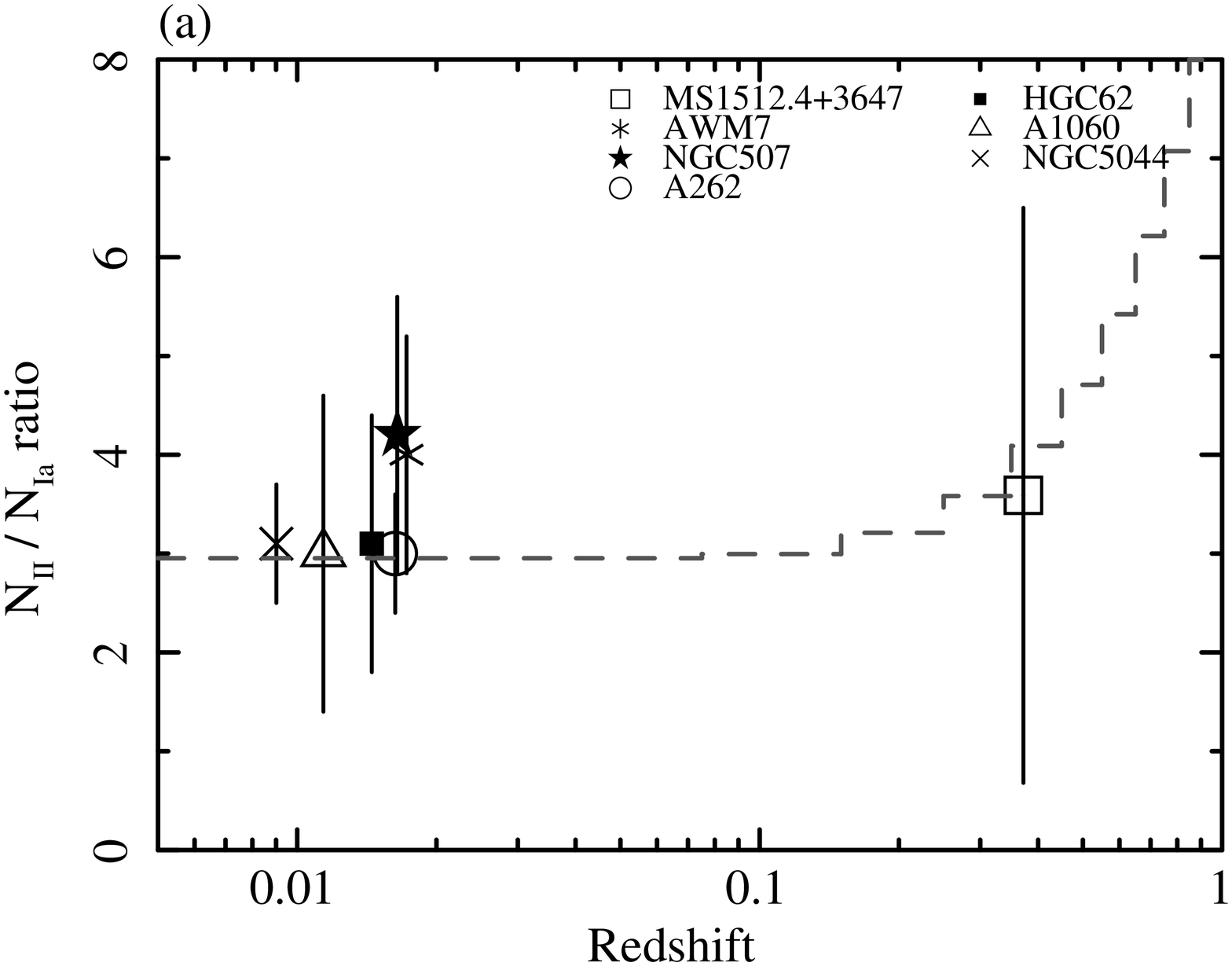}
  \end{center}
  \end{minipage}
  \begin{minipage}{0.5\hsize}
  \begin{center}
   \FigureFile(85mm,40mm){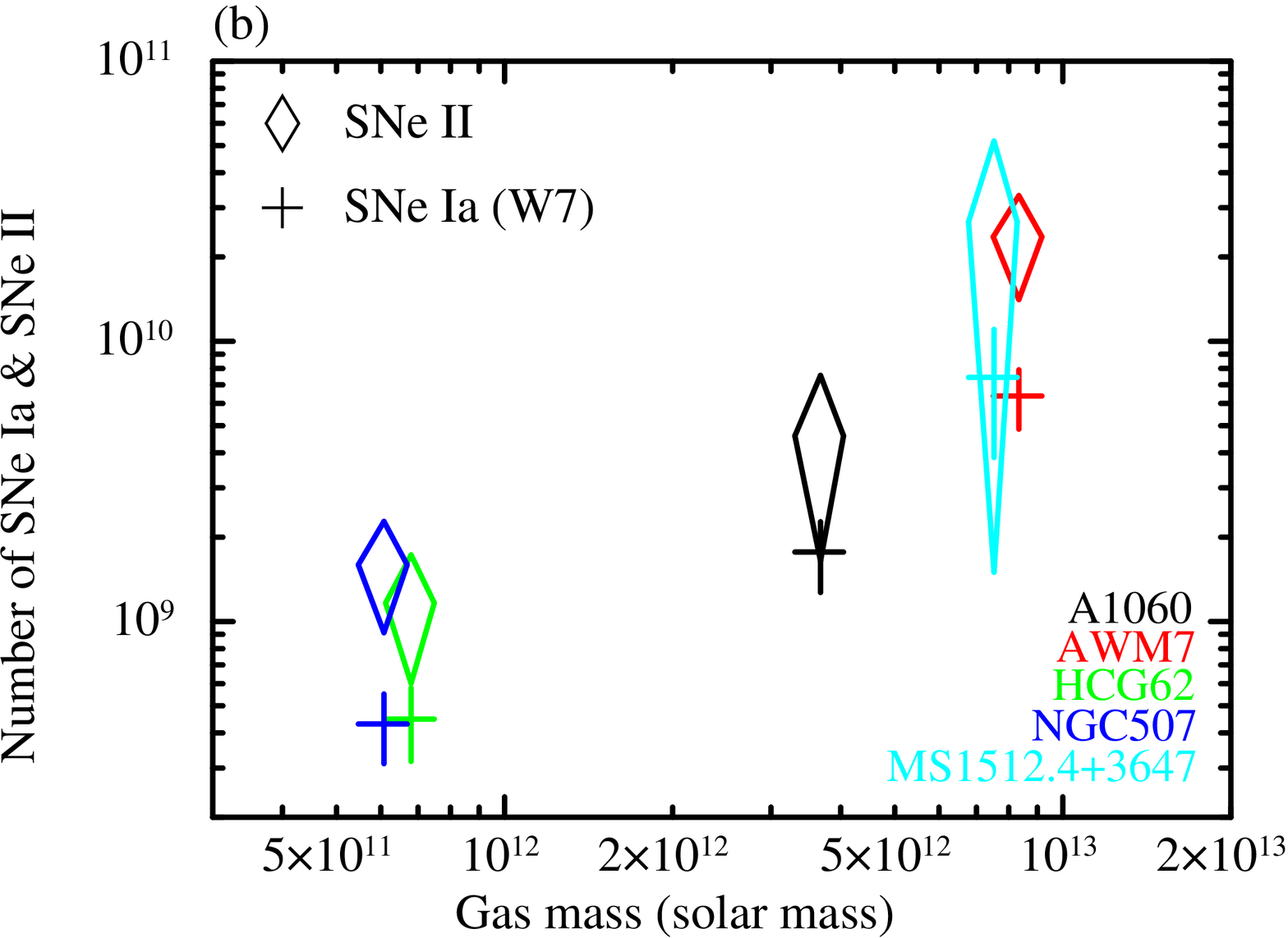}
  \end{center}
  \end{minipage}
  \caption{(a) Number ratios of SNe~II to SNe~Ia plotted against
  redshift. Dashed gray line indicates expected model, where SNe~II and
  SNe~Ia model are taken from equation (5) and figure~14 of Strolger et
  al. (2004), respectively. (b) Integrated numbers of SN~Ia (cross) and
  SNe~II (diamond) plotted against the gas mass within
  $0.3r_{200}$. Systematic errors described in section 5.1 are included
  in this figure. In both (a) and (b), the results of nearby clusters
  are also plotted.}\label{f9}
 \end{figure*}

If we assume the metals synthesized in clusters have been kept within
the cluster system, the ratio of cumulative number of SN~II explosions
to that of SN~Ia explosions in each cluster can be estimated from the
abundance pattern of metals contained in clusters, because the two types
of supernova give significantly different metal yields.
Most of Ne and Mg are synthesized by SN~II, while Fe and Ni are mostly
produced by SN~Ia.
The abundance pattern was fitted by the sum of the expected abundance
patterns of SN~Ia and SN~II (black solid line in figure~\ref{f6}(b)).
Free parameters in the fit are two normalizations of the abundance
patterns of the SN~Ia and SN~II metal yields.
In the actual fit, however, free parameters were chosen to be the
normalizations of SNe~Ia ($N_{1}$) and the number ratio of SNe~II to
SNe~Ia ($N_{2}/N_{1}$), because $N_{1}$ could be constrained well with
the relatively small error in the Fe abundance.
The derived parameters are
$N_{2}/N_{1}=3.6\pm2.9$.
Sato et al. (2007b) reported the supernova ratio, $N_{2}/N_{1}$, for nearby
four clusters to be $\sim3.5$.
de Plaa et al. (2007) also showed that the number ratio of
core-collapsed SNe (SNe~II+Ib+Ic) to SNe~Ia is $\sim3.5$, based on
XMM-Newton observations.
In figure~\ref{f9}(a), the $N_{2}/N_{1}$ ratio of MS~1512.4+3647 is
compared with those of nearby clusters and groups, which are NGC~5044,
Abell~262, and four other clusters studied by Sato et al. (2007b).
The value of MS~1512.4+3647 has a rather large error and is clearly
consistent with those of nearby clusters.

\begin{table*}[htbp]
\begin{center}
\caption{The metal abundance ratio to Fe.}\label{t4}
\begin{tabular}{lccccc}
\hline
\hline
Model&Ne/Fe&Mg/Fe&Si/Fe&S/Fe&Ni/Fe\\
&(solar)&(solar)&(solar)&(solar)&(solar)\\
\hline
1T&$<4.28$&$<2.09$&$0.99^{+0.42}_{-0.38}$&$0.66^{+0.45}_{-0.45}$&$3.03^{+2.16}_{-2.06}$\\
2T&---&$1.18^{+1.10}_{-1.03}$&$1.09^{+0.45}_{-0.40}$&$0.79^{+0.54}_{-0.51}$&$3.25^{+2.48}_{-2.20}$\\
CXB$-$6\%&$<4.55$&$<2.08$&$0.98^{+0.42}_{-0.39}$&$0.65^{+0.46}_{-0.45}$&$3.11^{+2.23}_{-2.05}$\\
CXB$+$6\%&$<4.27$&$<2.11$&$1.00^{+0.42}_{-0.39}$&$0.66^{+0.46}_{-0.45}$&$3.04^{+2.26}_{-2.05}$\\
NXB$\pm$3\%&$<4.23$&$<2.09$&$0.99^{+0.42}_{-0.38}$&$0.66^{+0.45}_{-0.45}$&$3.03^{+2.24}_{-2.06}$\\
CONTAMI$-$10\%&$<3.86$&$1.27^{+1.21}_{-1.08}$&$1.16^{+0.39}_{-0.41}$&$0.80^{+0.49}_{-0.48}$&$3.11^{+2.29}_{-2.12}$\\
CONTAMI$+$10\%&$<4.74$&$<1.82$&$0.85^{+0.39}_{-0.36}$&$0.54^{+0.42}_{-0.42}$&$2.99^{+2.15}_{-1.99}$\\
\hline
\end{tabular}
\end{center}
\end{table*}

\section{Discussion}
\subsection{Contributions from SNe Ia and SNe II}

Figure~\ref{f9}(a) also shows a curve based on a simple model.
For the model, we adopted the star formation history in equation~(5) of
Strolger et al. (2004), which is based on field galaxy data, as the time
evolution of the SN~II frequency, and the gaussian model for the SN~Ia
time evolution (see figure~14 of Strolger et al. 2004).
The present result gives little constraint to the model of the supernova
history, but the improvement of sensitivity planned in future missions
will give a meaningful result.

In figure~\ref{f9}(b), the numbers of SNe~Ia and SNe~II explosions
estimated within $0.3r_{200}$ are plotted against the gas mass.
The SN numbers of MS~1512.4+3647 are compared with those of four nearby
clusters by Sato et al. (2007b), which are AWM~7, A1060, NGC~507, and
HCG~62.
The radius of $0.3r_{200}$ is adopted by following reasons:
(1) the systematic and statistical errors of abundance profiles become
large toward larger radii,
(2) the effect of cool core becomes relatively larger for smaller radii,
and
(3) $\sim90$\% of the detected photons within $0.6r_{200}$ actually
originated from within $0.3r_{200}$, simulated with the spectrum of
MS~1512.4+3647.
We evaluated systematic error due to the scaling at several radii and
concluded that it is minimized at $\sim0.3r_{200}$.

The integrated mass of the ICM in MS~1512.4+3647 within $0.3r_{200}$ is
estimated as follows.
Because the normalization of the $vapec$ model was scaled to the value
of the input XIS image size for the A$^{\rm I}$ in the spectral
analysis, we derived the true normalization within $0.3r_{200}$ to
multiply a factor, $({\rm counts\ within}\ 0.3r_{200})/({\rm counts\ of\
whole\ region})$, to the normalization of the $vapec$ model.
The energy flux within $0.3r_{200}$ using this method is consistent with
the one of the Chandra results.
By comparing the obtained normalization to the volume integration
of squared density profile (3-dimensional $\beta$-model; $\rho_{\rm
gas}(r)=\rho_{0}(1+(r/r_{\rm core}))^{-3\beta/2}$), we derived the
normalization of 3-dimensional $\beta$-model as
$\rho_{0}=6.4\times10^{-2}$.
Where $\rho$ and $r$ are gas density and radius from cluster center,
respectively.
Finally, the integrated mass of the ICM within $0.3r_{200}$ is estimated
as $7.53\times10^{12}\ M_{\solar}$ by integrating the 3-dimensional
$\beta$-model.
The cumulative numbers of SN~Ia and SN~II explosions can be estimated
from the mass of metals contained in clusters.
Based on the metal abundances and the estimated mass of the ICM, we
calculated the mass of individual metals.
From the metal mass and relative numbers of SN~Ia and SN~II shown in
figure~\ref{f6}(b), the cumulative numbers of SN~Ia and SN~II within
$0.3r_{200}$ are derived as
$N_{\rm SN\ Ia}=(7.4\pm2.9)\times10^{9}$ 
and 
$N_{\rm SN\ II}=(2.7\pm2.4)\times10^{10}$.

For nearby four clusters, the SN numbers and gas mass are scaled to the
values within $0.3r_{200}$.
The gas density profiles and metal abundances of these nearby clusters
are spatially resolved up to
$0.36r_{200}$,
$0.26r_{200}$,
$0.25r_{200}$,
$0.22r_{200}$
for AWM~7, A1060,
NGC~507,
and HCG~62,
respectively.
Each gas mass is scaled using 3-dimensional $\beta$-model.
The $\beta$ values and core radii were taken from the literature (Sato
et al. 2008; Sato et al. 2007a; Sato et al. 2009a; Tokoi et al. 2008)
, and normalizations were derived by fitting gas mass profiles shown
in the literature (Sato et al. 2008; Sato et al. 2007a; Sato et
al. 2009a; Tokoi et al. 2008).
We assumed that the numbers of SN~Ia and SN~II are proportional to the
metal mass of Fe and Mg, respectively.
Namely, $N_{\rm SN\ Ia}\propto M_{\rm Fe}$ and $N_{\rm SN\ II}\propto
M_{\rm Mg}$.
Using above assumption, the metal mass of Fe and Mg are derived from
the metal mass profile $\rho_{\rm metal}(r)=Z_{\rm metal}(r)\times
\rho_{\rm gas}(r)$ integrated up to $0.3r_{200}$, where we adopted a
$\beta$-model for the metal abundance profile, $Z_{\rm
metal}(r)=Z_{0}(1+(r/r_{\rm core}))^{-\alpha}$ (De~Grandi et al. 2004).
Lastly, we multiplied scaling factor, $f=M_{\rm
metal}(0.3r_{200})/M_{\rm metal}(r_{\rm obs})$, to the numbers of SN~Ia
and SN~II within the individual observed radii.

There are several systematic errors due to the the scaling method.
We roughly estimated the systematic errors for the cumulative numbers
estimations of MS~1512.4+3647, and the gas mass scaling of nearby
clusters as 30\%, and 20\%, respectively (detailed in appendix~\ref{apenb}).
These systematic errors are taken into account in figure~\ref{f9}(b).
Figure~\ref{f9}(b) shows that both the SN~Ia and SN~II numbers are in good
positive correlation with the gas mass, including MS~1512.4+3647 which
is at a medium redshift, though the SN~Ia number of MS~1512.4+3647 is
slightly above the correlation.

\subsection{Implications for Metal Enrichment History}

The supernova numbers in MS~1512.4+3647 lie on the correlation between
the gas mass and the numbers of SN~Ia and SN~II as shown in
figure~\ref{f9}(b).
Figure~\ref{f9} indicates that the integrated number of SNe ($N_{\rm SN\
Ia}$ and $N_{\rm SN\ II}$) range
$4\times10^{8}$--$8\times10^{9}$ and
$1\times10^{9}$--$2\times10^{10}$, respectively, in the range of
two figures of gas mass.
This suggests that this cluster has already experienced as many
supernovae as nearby clusters.
The past history of SN~II explosions is correlated with the star
formation history (SFH), IMF, and the
lifetime function (e.g. Borgani et al. 2002).
Since massive stars above $\sim$8 $M_{\solar}$ cause a core collapse with
a short lifetime, the history of SN~II should follow the curve of SFH,
which is relatively flat at $z>2$, peaked at $1<z<2$, and then decreases
by about one order of magnitude toward $z\sim0$ (e.g. Strolger et
al. 2004).
This scenario is consistent with the SN~II contribution in
MS~1512.4+3647 (figure~\ref{f9}(b)) which is similar to that in AWM~7.
For example, in the SFH models (M1 and M2) in Strolger et al. (2004), the
number of star formation per unit volume at $0<z<0.4$ is only $\sim10$\%
of the integrated value in $0<z<6$.
This suggests that the number of SN~II in MS~1512.4+3647 would be lower
than the nearby values by about $\sim10$\%, which is within the error as
seen in figure~\ref{f9}(b). 

The history of the SN~Ia rate is more complicated, because the lifetime
function of binary systems needs to be convolved with the star formation
rate (e.g. Borgani et al. 2008).
In general, the SN~Ia rate is peaked later in time (at smaller $z$) and
prolonged longer than the SN~II rate history.
In fact, a mild decrease of the SN~Ia rate in the cluster environment is
found from $z\sim1$ to $z\sim0$ by a factor of $2\sim10$ (Gal-Yam et
al. 2002).
Our result in figure~\ref{f9}(b) suggests a low SN~Ia rate in
$0<z<0.37$, since the SN~Ia numbers normalized by the gas mass are the
same between MS~1512.2+3647 and nearby clusters.
If one assumes the evolution of the SN~Ia rate implied from the field
galaxy data (for example, model 1N2.3 in solid line in figure~10b of
Loewenstein 2006), the number of SN~Ia would increase by $\sim40$\% from
$z=0.37$ to $z=0$, namely in about 4 Gy.
In this case, the SN~Ia point of MS~1512.4+3647 in figure~\ref{f9}(b)
would exceed the level of AWM~7 when 4~Gy passed, causing some deviation
(from $0.7\sigma$ to $1.8\sigma$) from the present linear relationship
in this figure.
Therefore, models of the SN~Ia evolution with steeper gradient and/or
more contribution at larger redshifts (for example, model 2H1.05WxSt1
in solid line in figure~11b of Loewenstein 2006) would be favored.
Same results are reported from several observations of SNe~Ia (Mannucci,
Della Valle \& Panagia 2006; Maoz, Mannucci \& Brandt 2012).


\section{Conclusion}

Based on the Suzaku observation of MS~1512.4+3647, we derived metal
abundances of Ne, Mg, Si, S, Fe and Ni in the ICM.
All the elements show the similar abundance values around $\sim0.5$
solar, and the abundance ratios relative to the Fe value are
approximately 1 solar except for Ni.
The number ratios of Ne, Mg, Si, S, and Ni to Fe in MS~1512.4+3647 are
consistent with those for nearby clusters within the errors and lie
between the expected values of SN~II and SN~Ia yields.
Both SN~Ia and SN~II products are considered to have enriched the ICM of
MS~1512.4+3647 to the same amount as in nearby clusters.

The integrated number of SNe~Ia ($N_{\rm SN\ Ia}$) and the number ratio
of SNe~II to SNe~Ia ($N_{\rm SN\ II}/N_{\rm SN\ Ia}$) were derived to be
$N_{\rm SN\ Ia}=(7.4\pm2.9)\times10^{9}$
and
$N_{\rm SN\ II}/N_{\rm SN\ Ia}=3.6\pm2.9$
, respectively.
The number ratio of SNe~II to SNe~Ia in MS~1512.4+3647 is consistent
with those in nearby clusters within the errors.
The integrated numbers of both SNe~Ia and SNe~II explosions are
comparable to those in nearby clusters of galaxies when normalized by
the gas mass.
This similarity indicates that MS~1512.4+3647 has already experienced as
many SNe~Ia and SNe~II supernovae as in nearby clusters.

Massive stars heavier than $\sim$8 $M_{\solar}$ cause a core collapse in a
short lifetime, and the past SN~II rate is expected to approximately follow
the SFH.
The observed feature of SNe~II in MS~1512.4+3647 is consistent with this
standard scenario of the SN~II history.
The similarity in the number of SNe~Ia to nearby clusters suggests an
SN~Ia enrichment scenario that the SN~Ia rate steeply declines from
$z=0.37$ to $z=0$ and/or a dominant number of SN~Ia explosions occured
in higher redshifts.


\bigskip
This research was supported, in part, by a grant from the Hayakawa Satio
Fund awarded by the Astronomical Society of Japan.
YS is supported by Grant-in-Aid for JSPS Fellows.
MK acknowledges support by the Grant-in-Aid for Scientific Research,
No. 22740132.


\appendix
\section{Spectral Analysis of Point Sources}\label{apena}
\begin{figure*}[htbp]
\begin{minipage}{0.2\hsize}
\begin{center}
\FigureFile(45mm,40mm){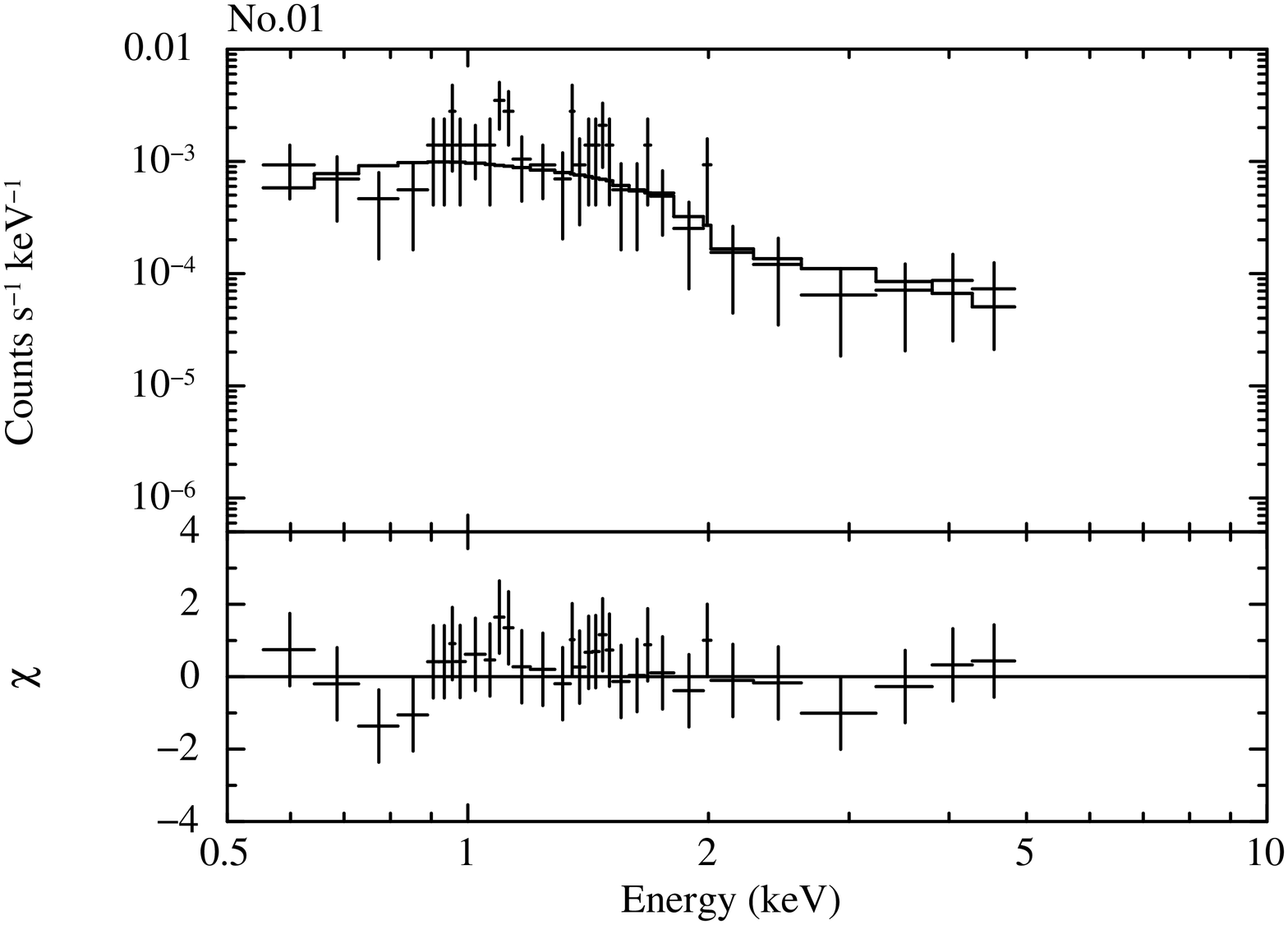}
\end{center}
\end{minipage}
\begin{minipage}{0.2\hsize}
\begin{center}
\FigureFile(45mm,40mm){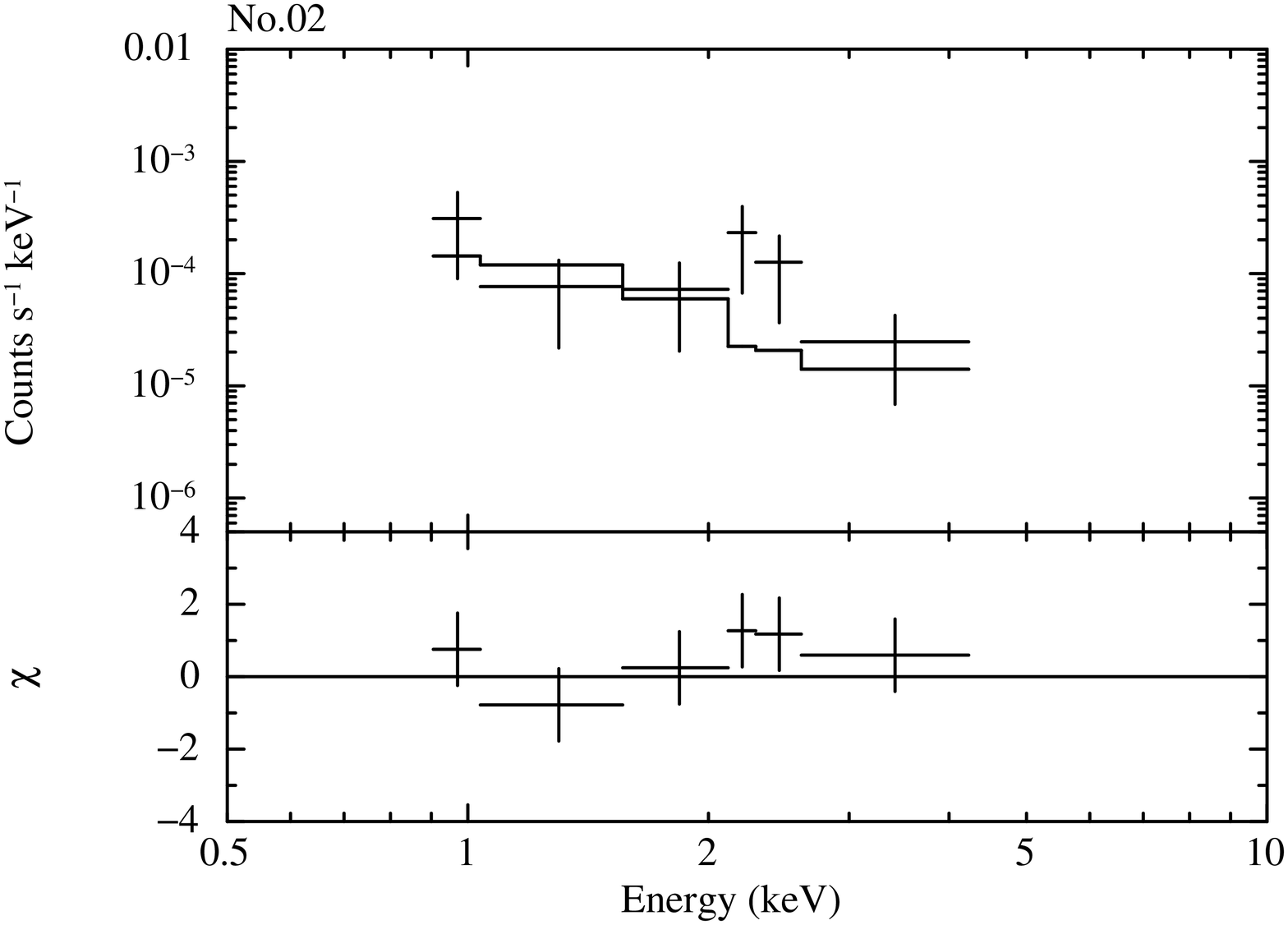}
\end{center}
\end{minipage}
\begin{minipage}{0.2\hsize}
\begin{center}
\FigureFile(45mm,40mm){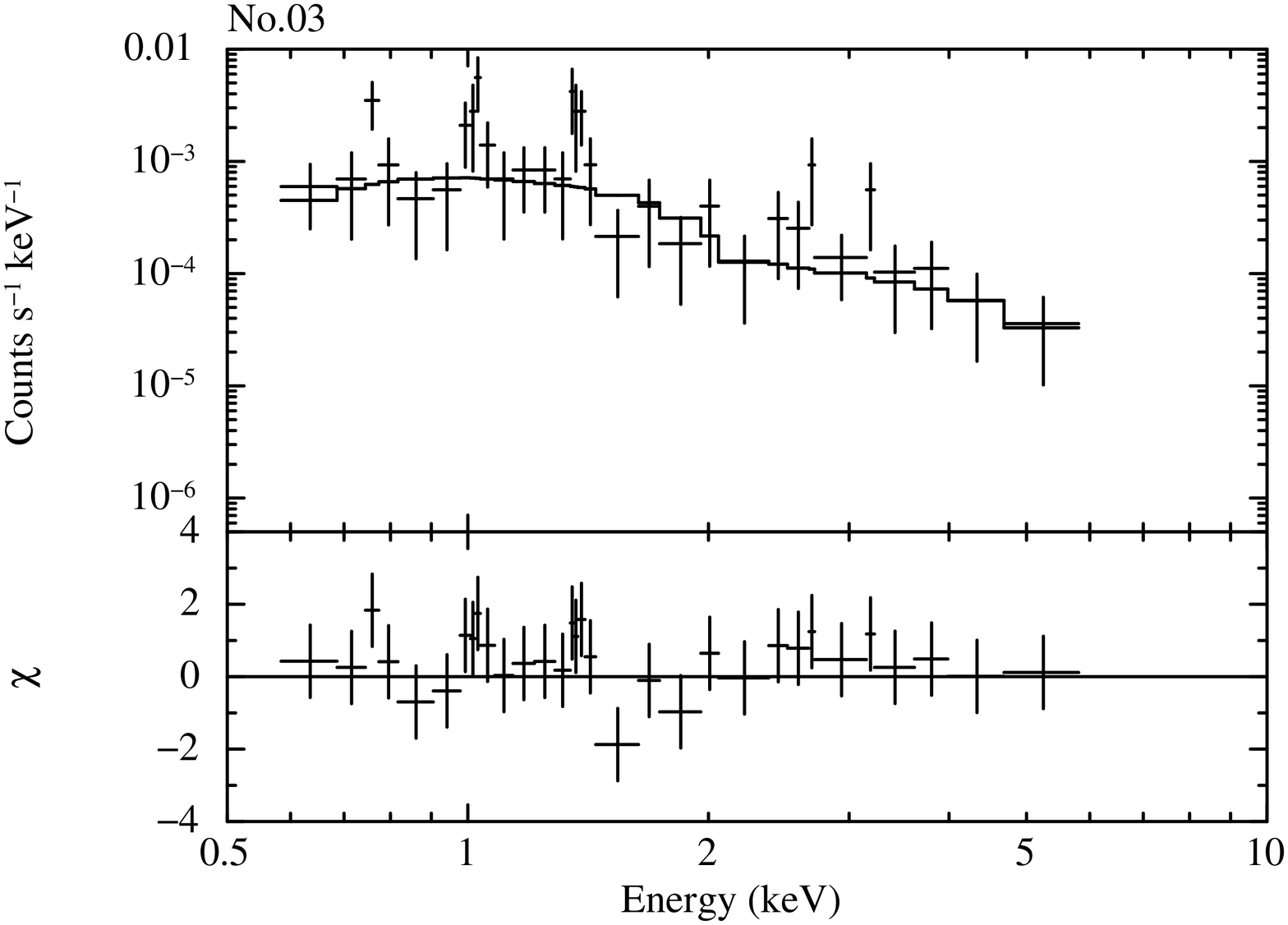}
\end{center}
\end{minipage}
\begin{minipage}{0.2\hsize}
\begin{center}
\FigureFile(45mm,40mm){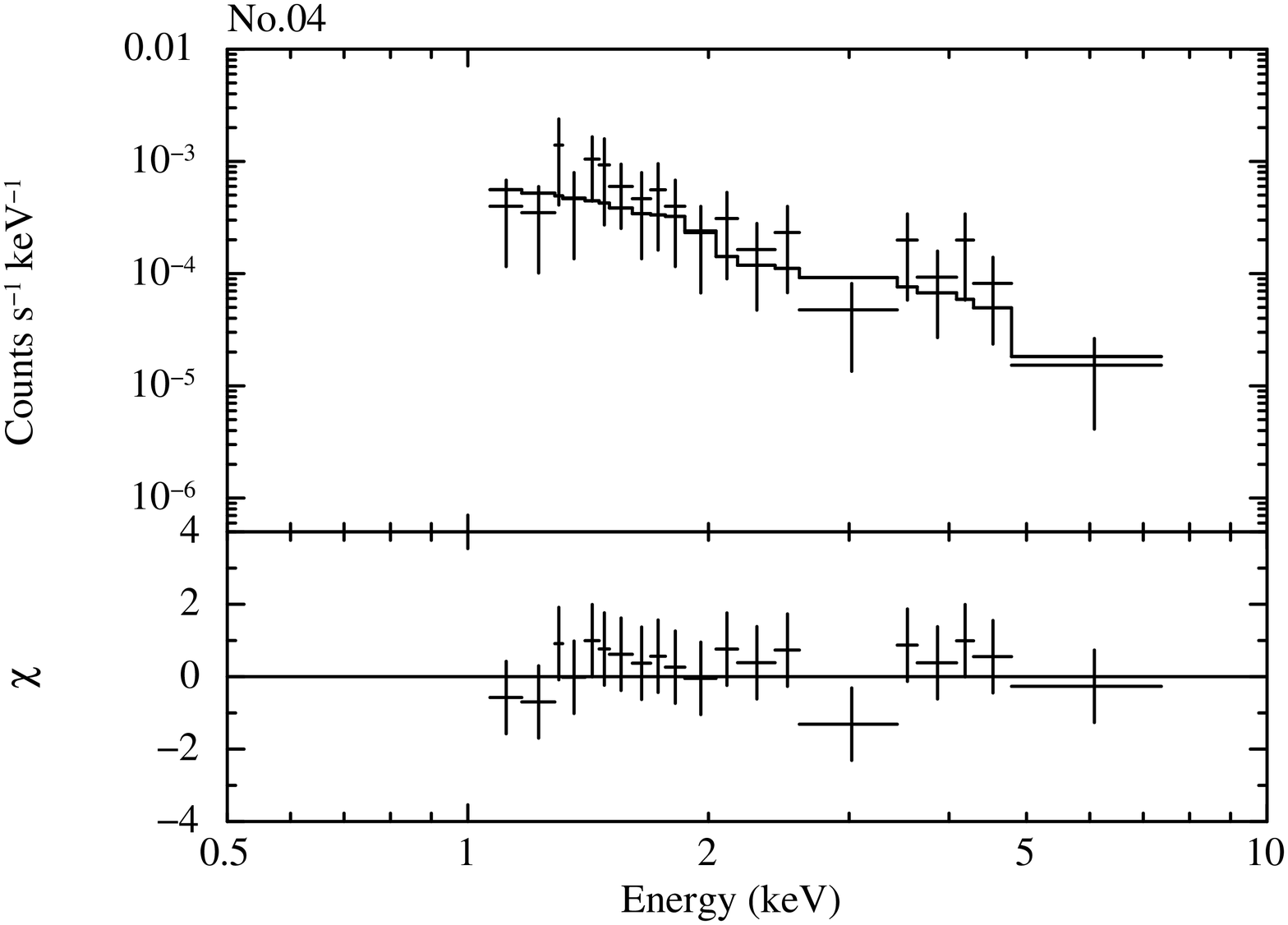}
\end{center}
\end{minipage}
\begin{minipage}{0.2\hsize}
\begin{center}
\FigureFile(45mm,40mm){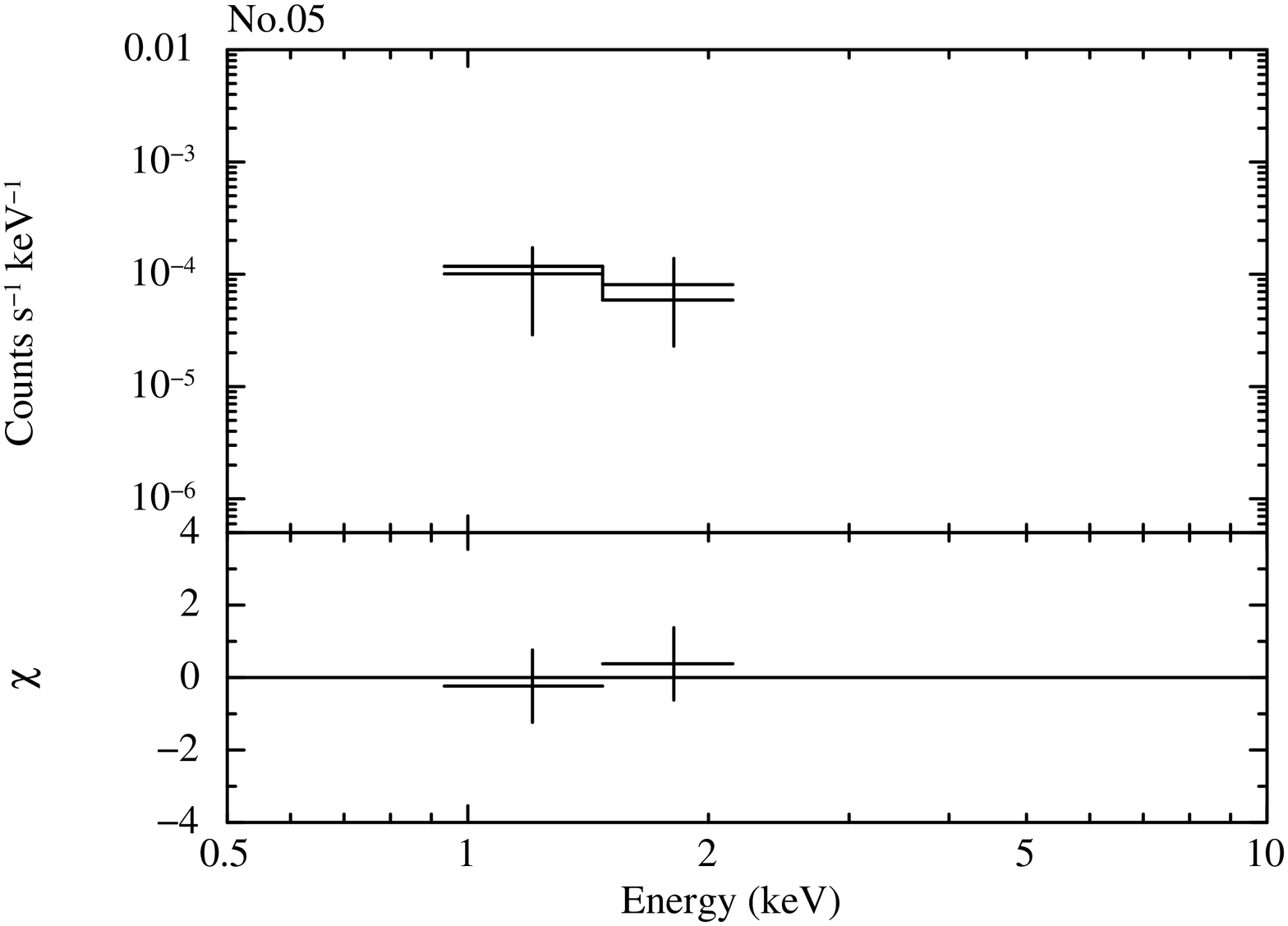}
\end{center}
\end{minipage}
\begin{minipage}{0.2\hsize}
\begin{center}
\FigureFile(45mm,40mm){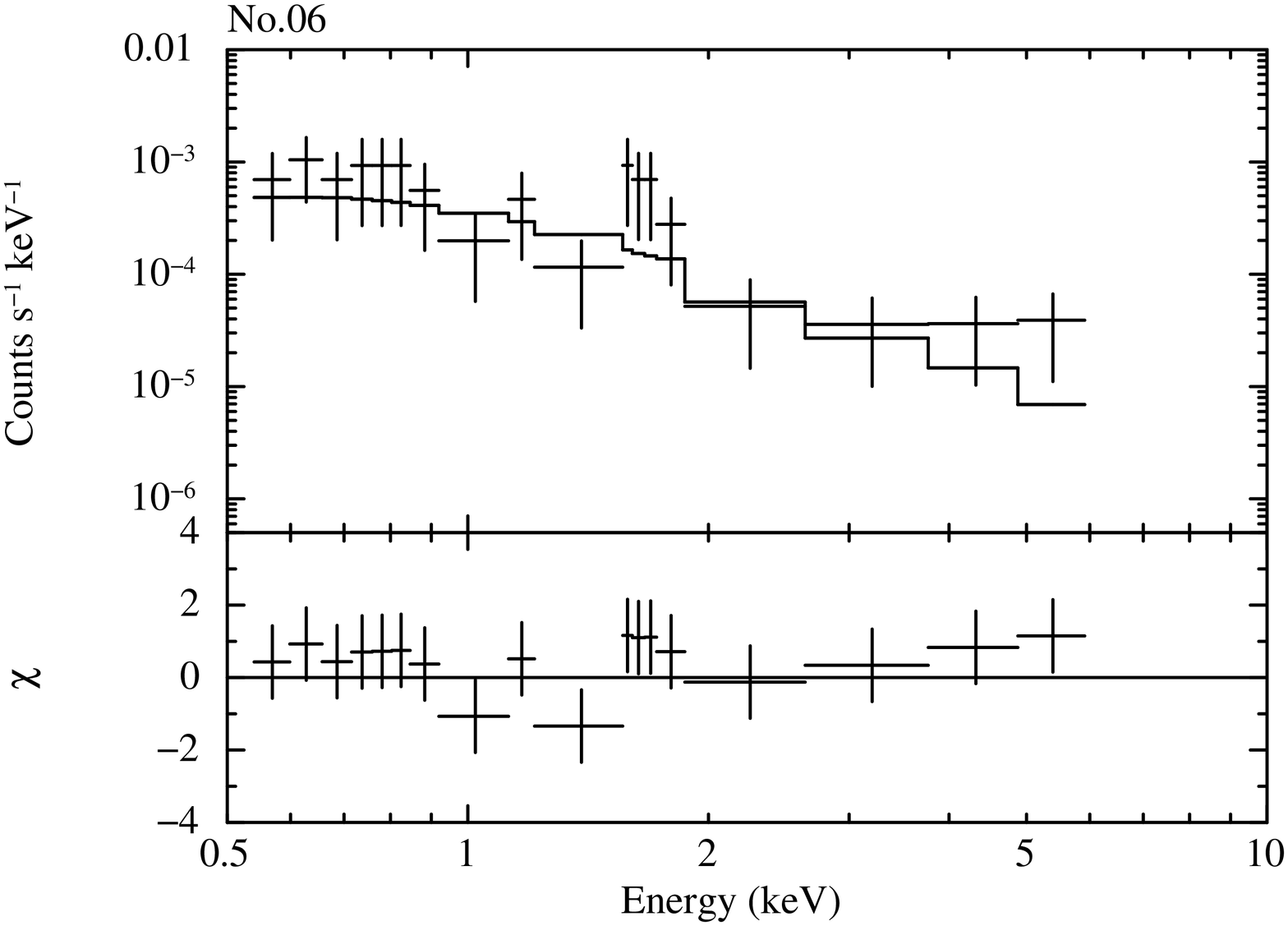}
\end{center}
\end{minipage}
\begin{minipage}{0.2\hsize}
\begin{center}
\FigureFile(45mm,40mm){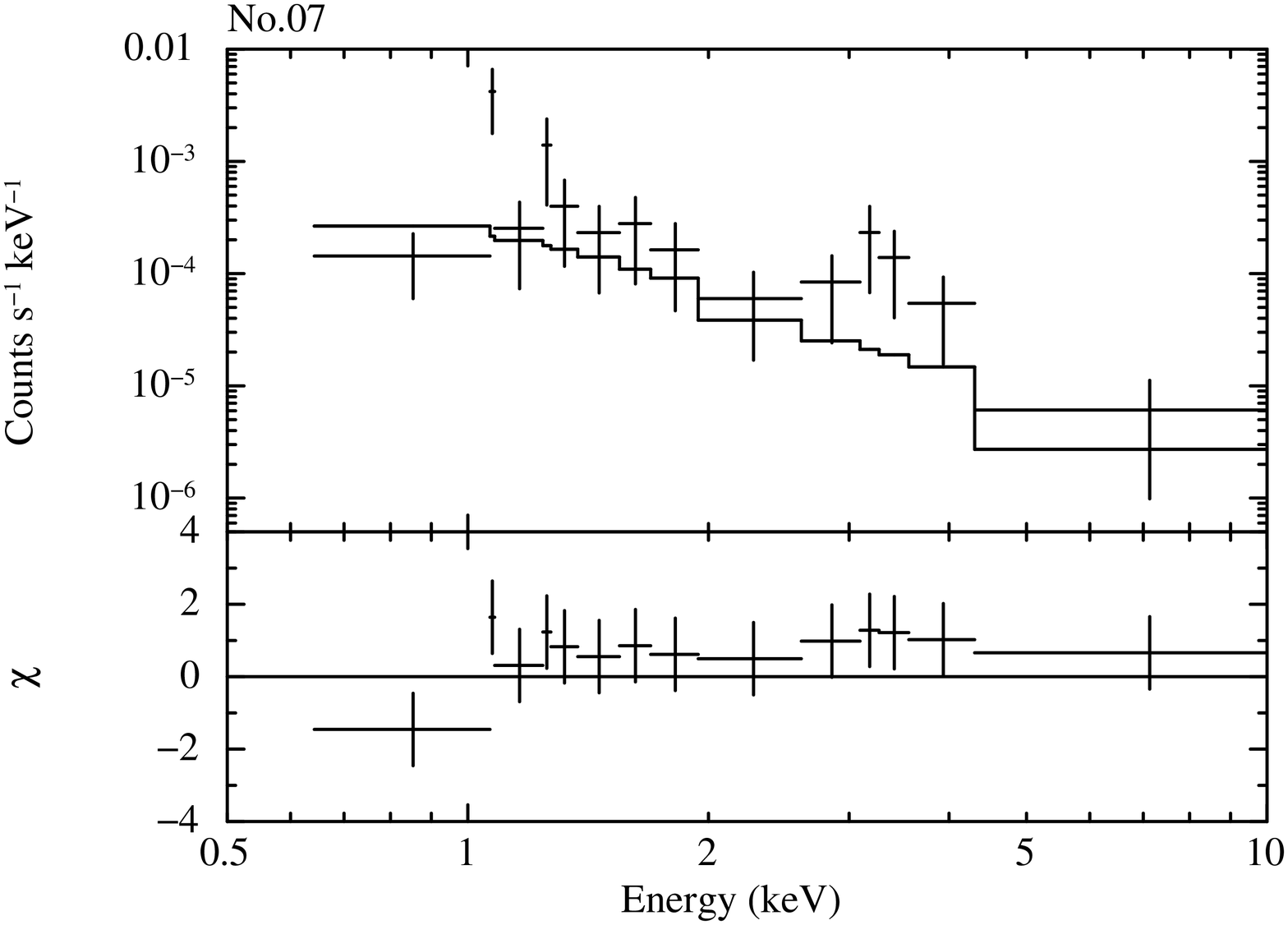}
\end{center}
\end{minipage}
\begin{minipage}{0.2\hsize}
\begin{center}
\FigureFile(45mm,40mm){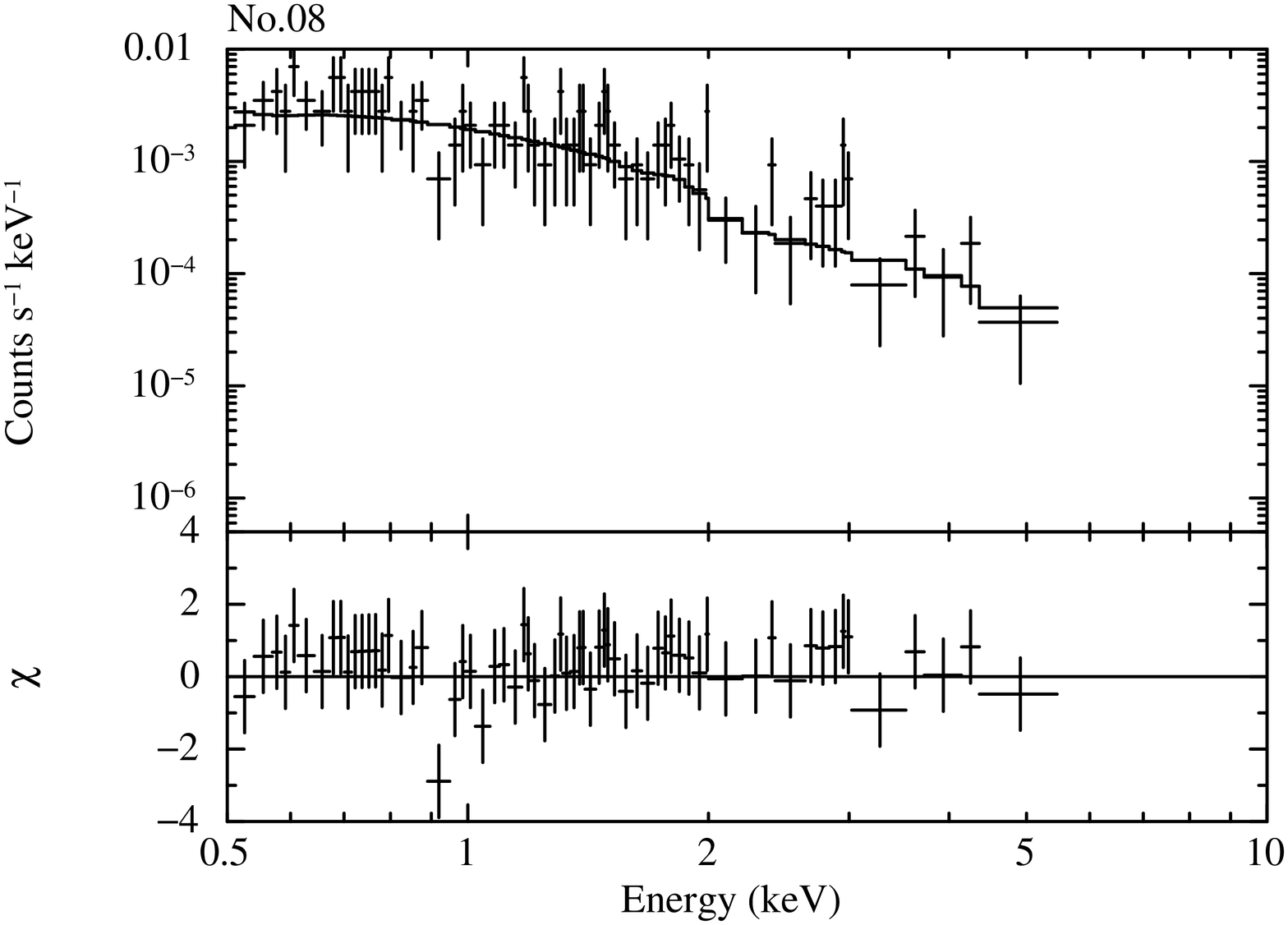}
\end{center}
\end{minipage}
\begin{minipage}{0.2\hsize}
\begin{center}
\FigureFile(45mm,40mm){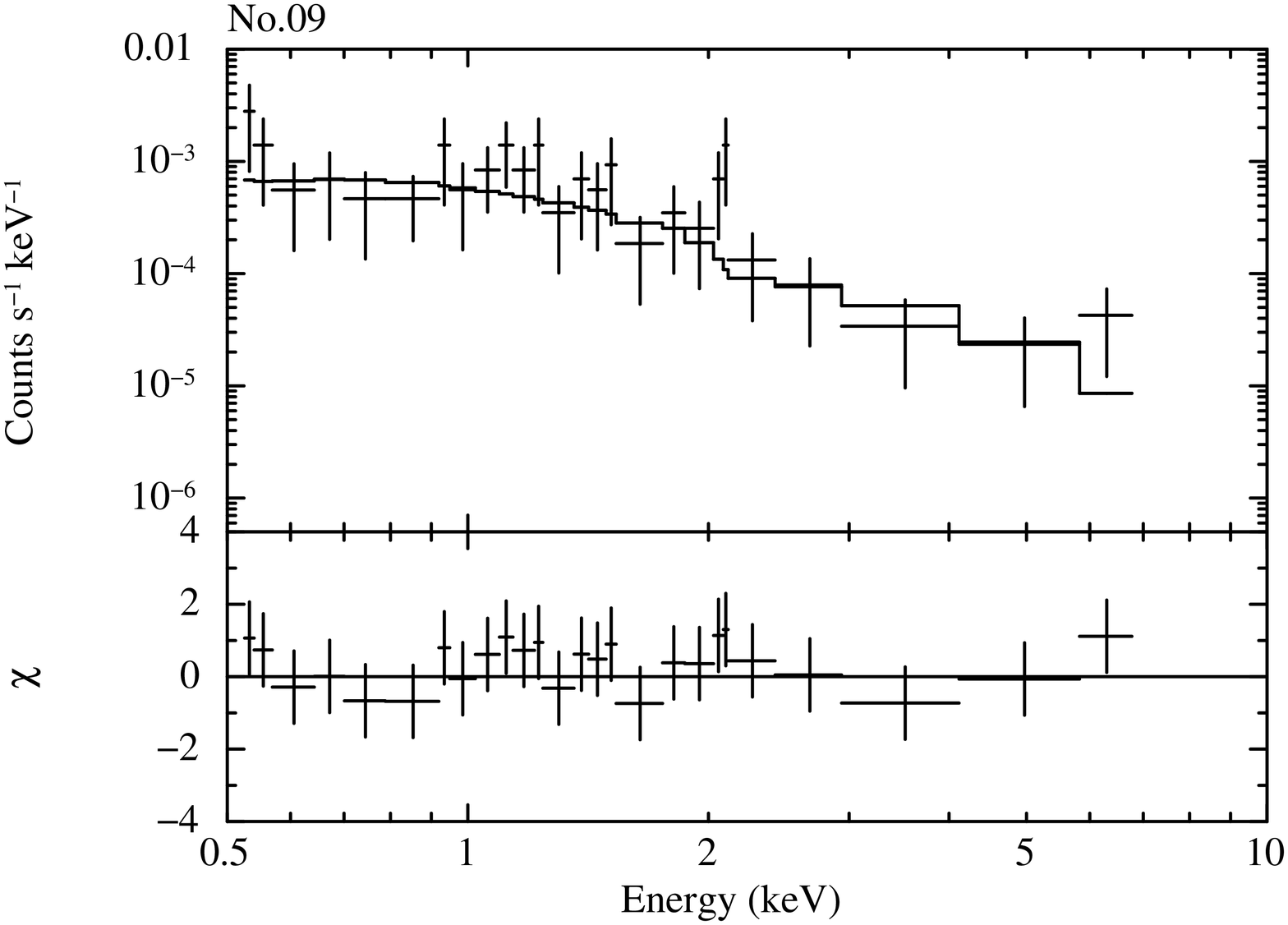}
\end{center}
\end{minipage}
\begin{minipage}{0.2\hsize}
\begin{center}
\FigureFile(45mm,40mm){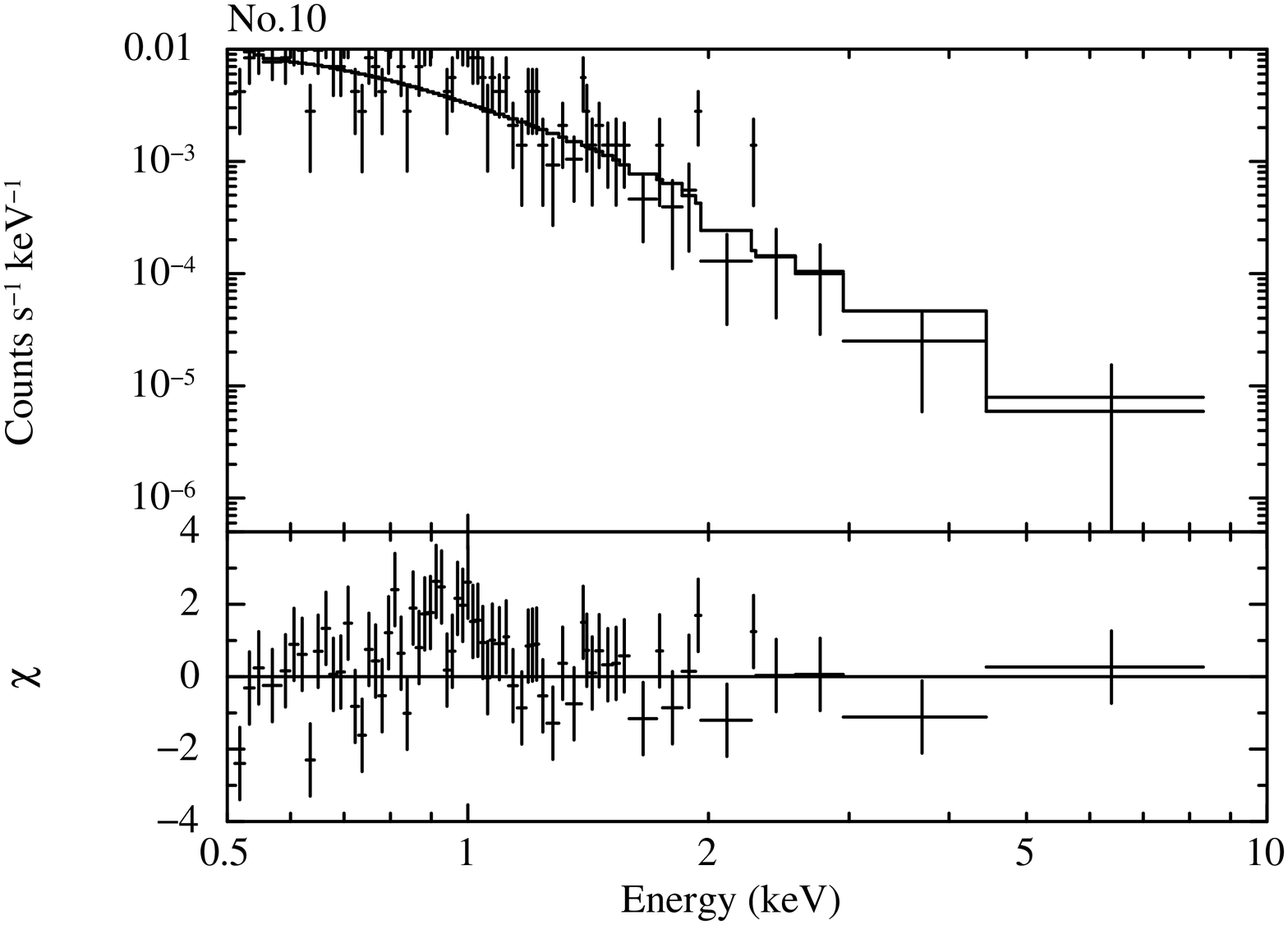}
\end{center}
\end{minipage}
\begin{minipage}{0.2\hsize}
\begin{center}
\FigureFile(45mm,40mm){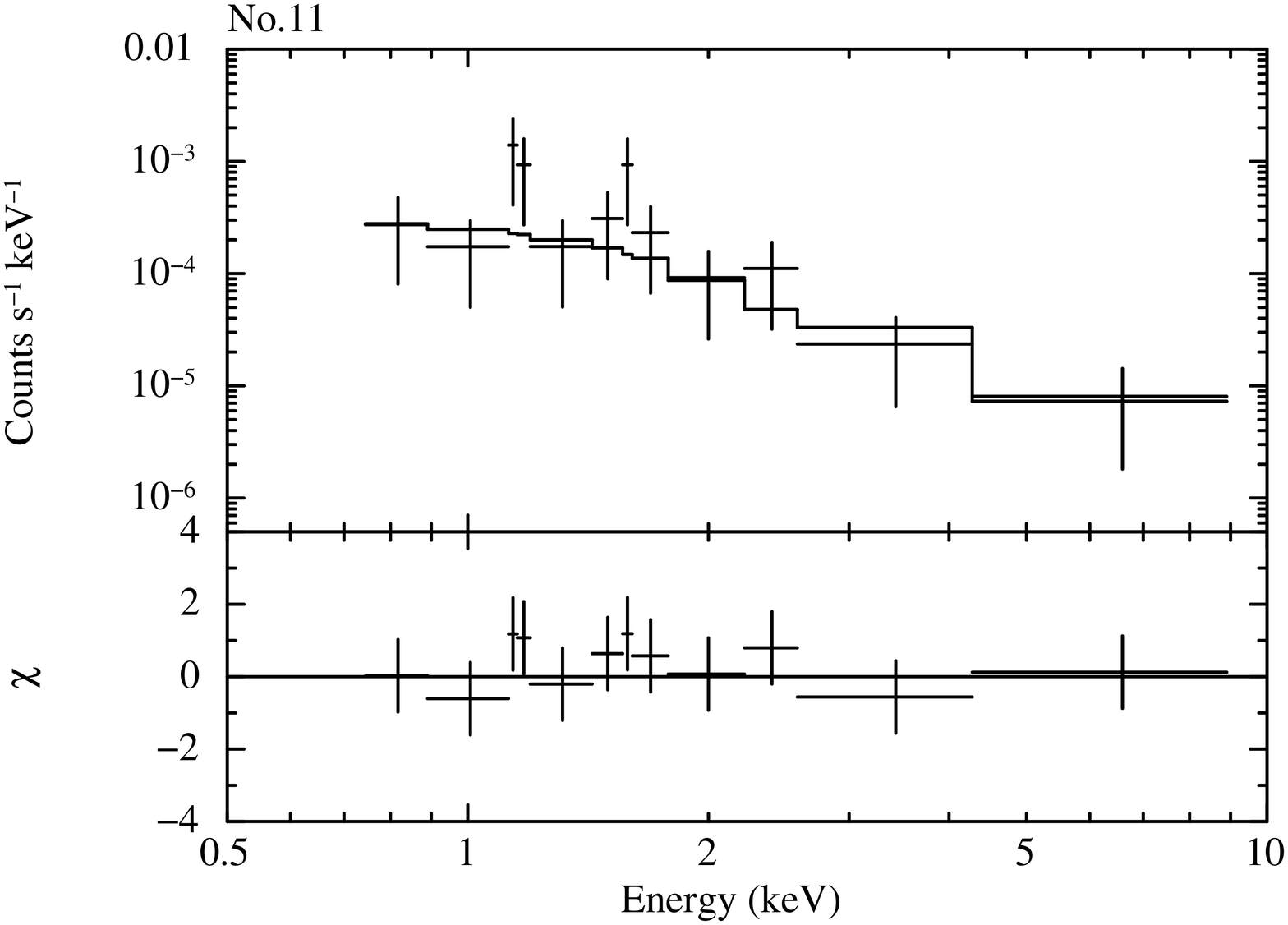}
\end{center}
\end{minipage}
\begin{minipage}{0.2\hsize}
\begin{center}
\FigureFile(45mm,40mm){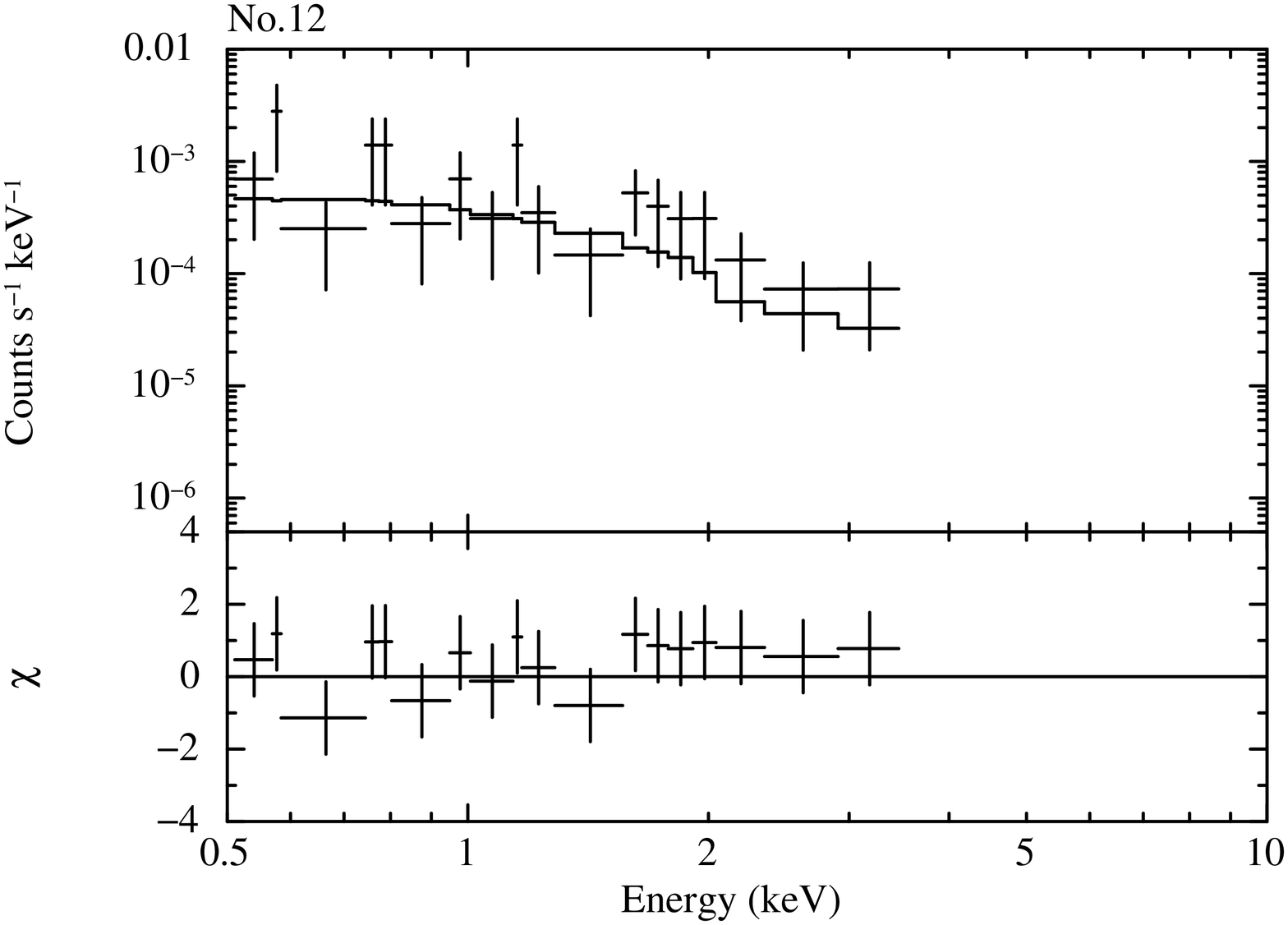}
\end{center}
\end{minipage}
\begin{minipage}{0.2\hsize}
\begin{center}
\FigureFile(45mm,40mm){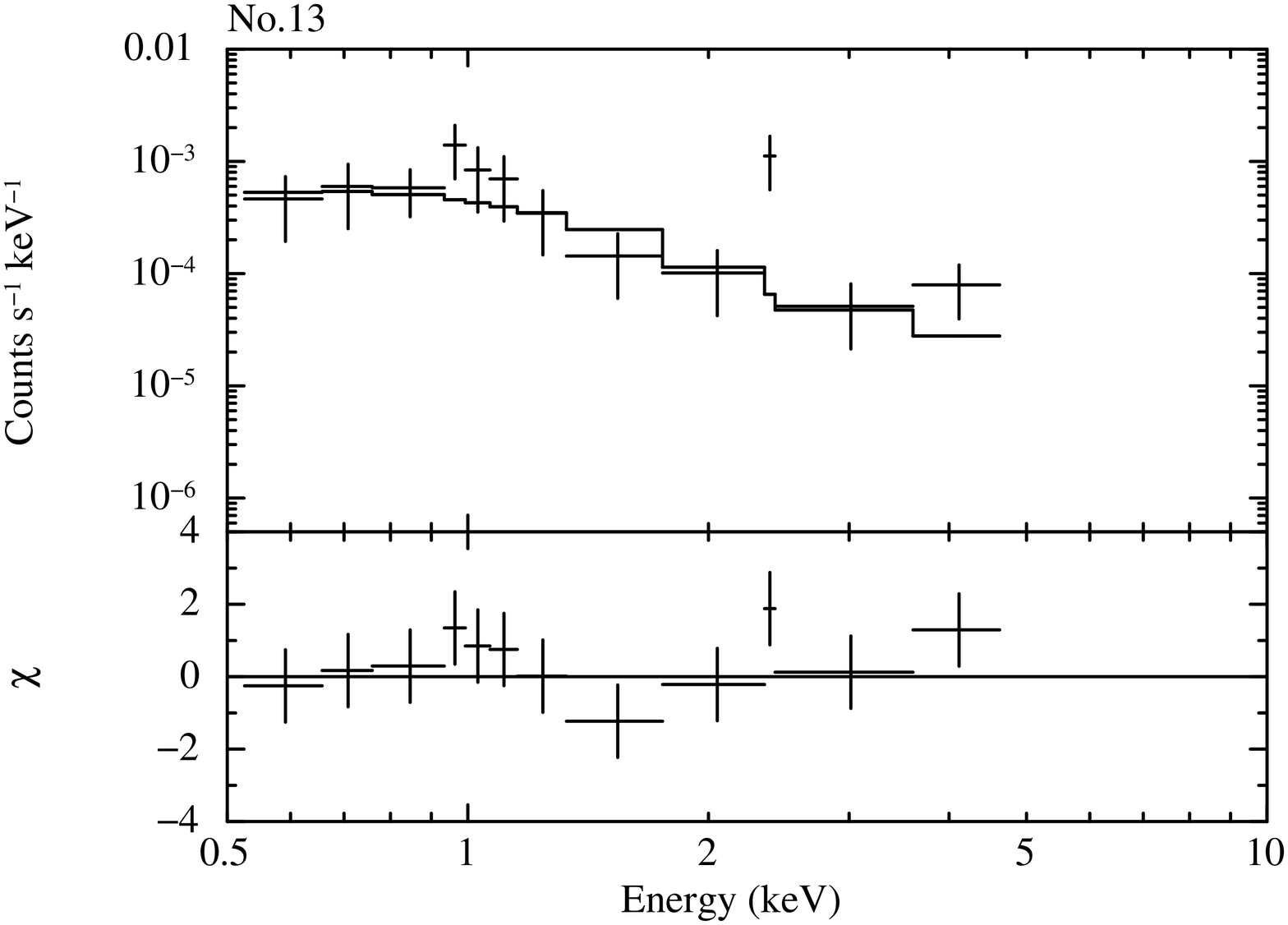}
\end{center}
\end{minipage}
\begin{minipage}{0.2\hsize}
\begin{center}
\FigureFile(45mm,40mm){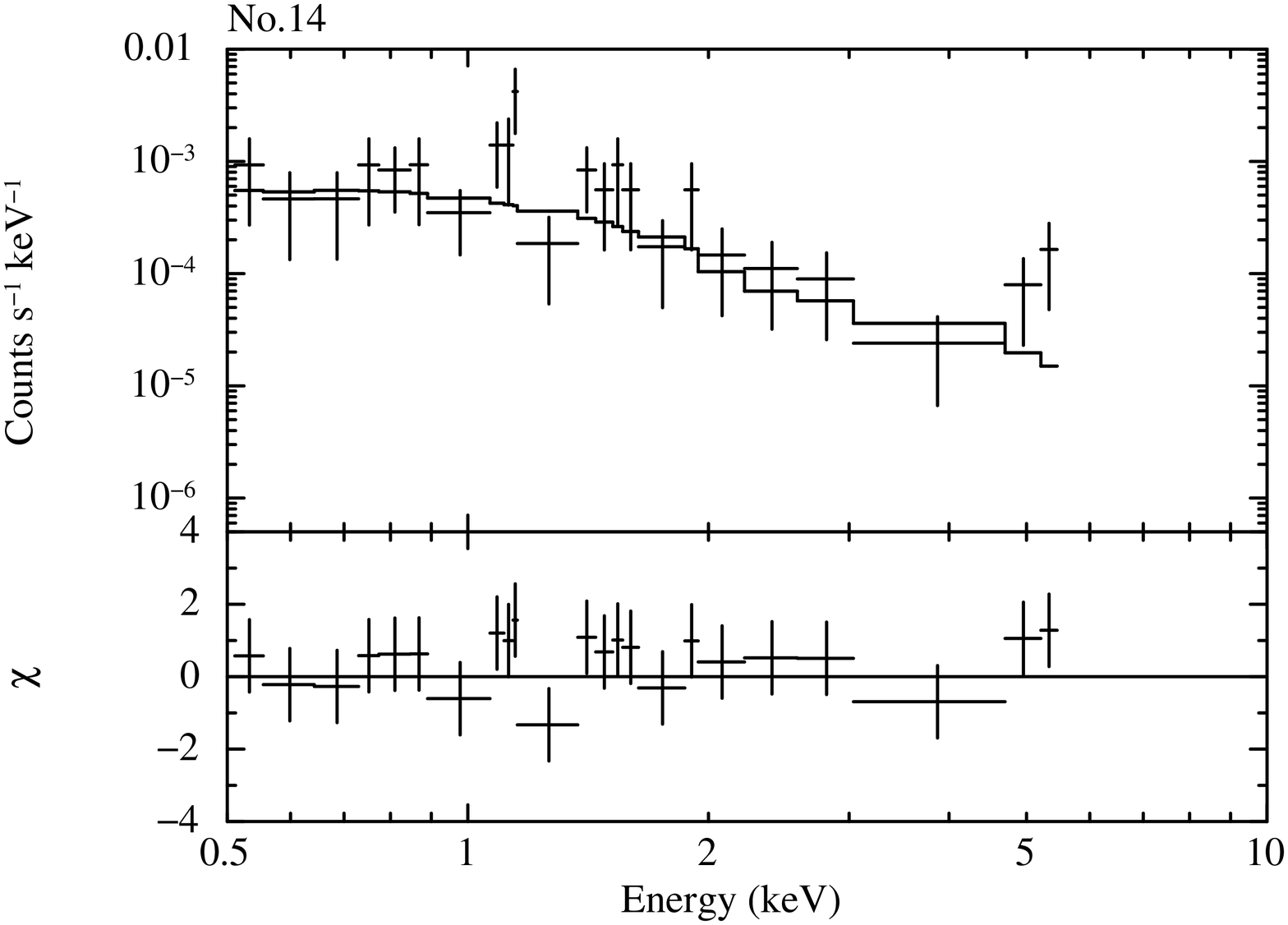}
\end{center}
\end{minipage}
\begin{minipage}{0.2\hsize}
\begin{center}
\FigureFile(45mm,40mm){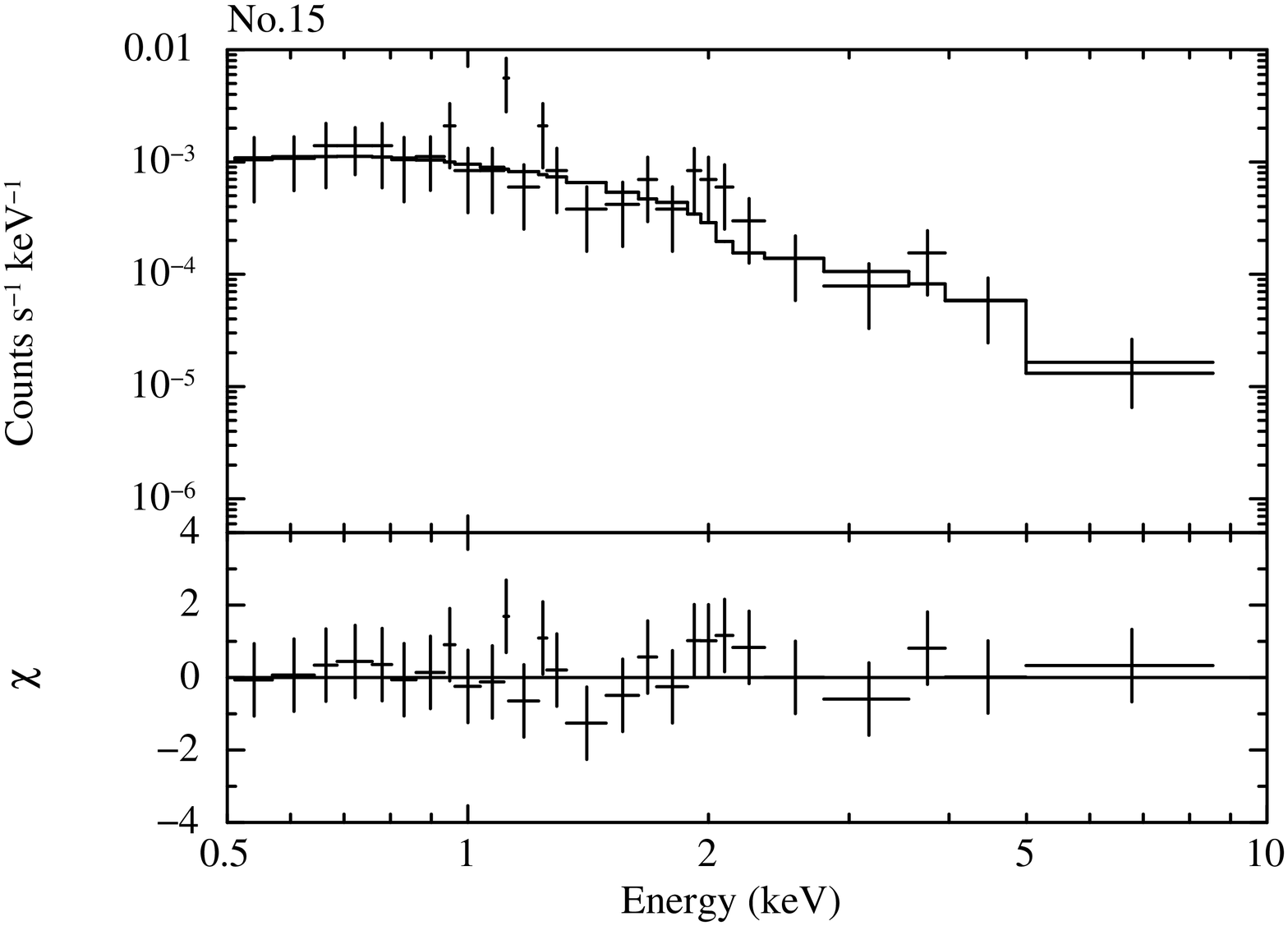}
\end{center}
\end{minipage}
\begin{minipage}{0.2\hsize}
\begin{center}
\FigureFile(45mm,40mm){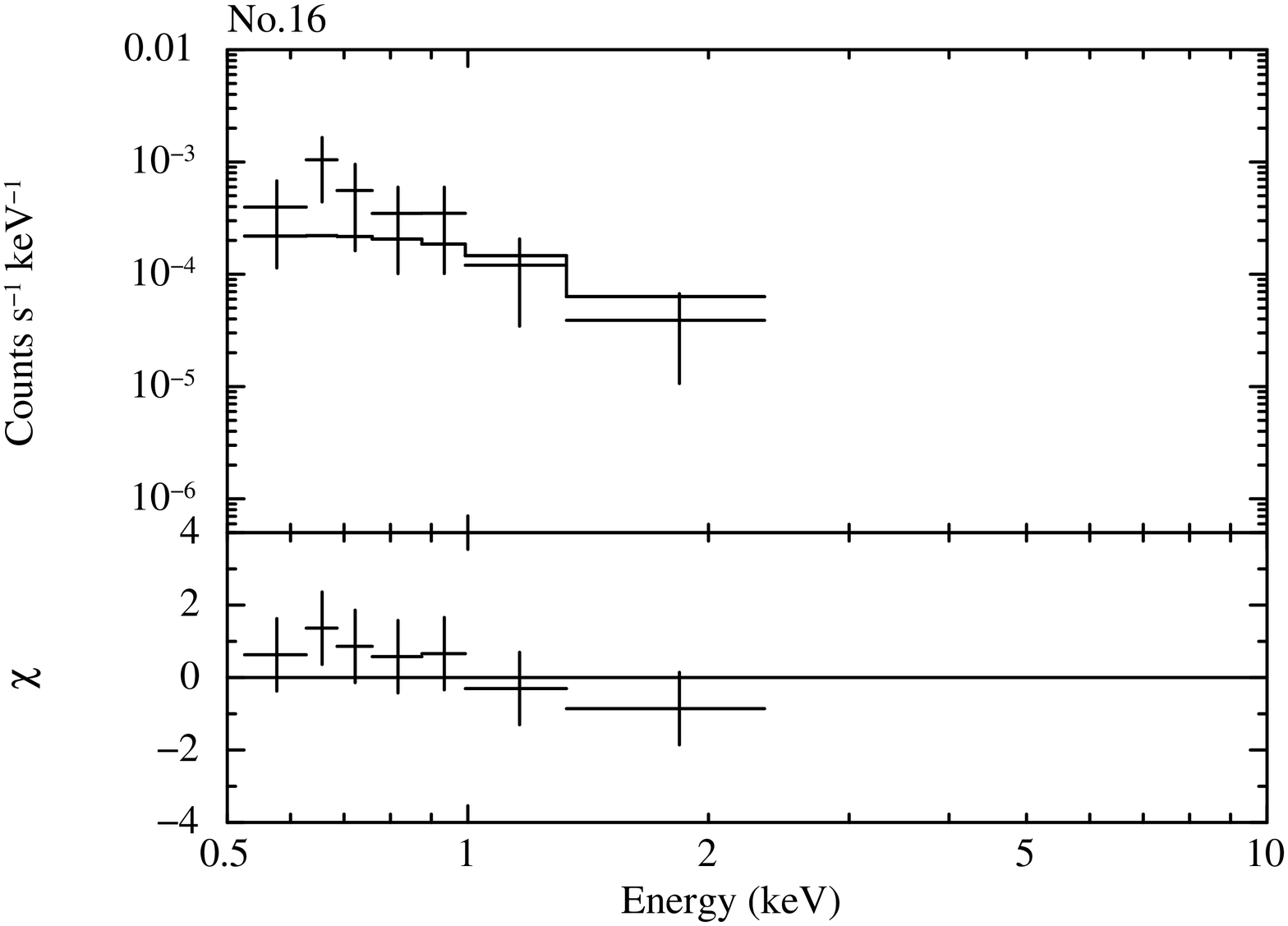}
\end{center}
\end{minipage}
\begin{minipage}{0.2\hsize}
\begin{center}
\FigureFile(45mm,40mm){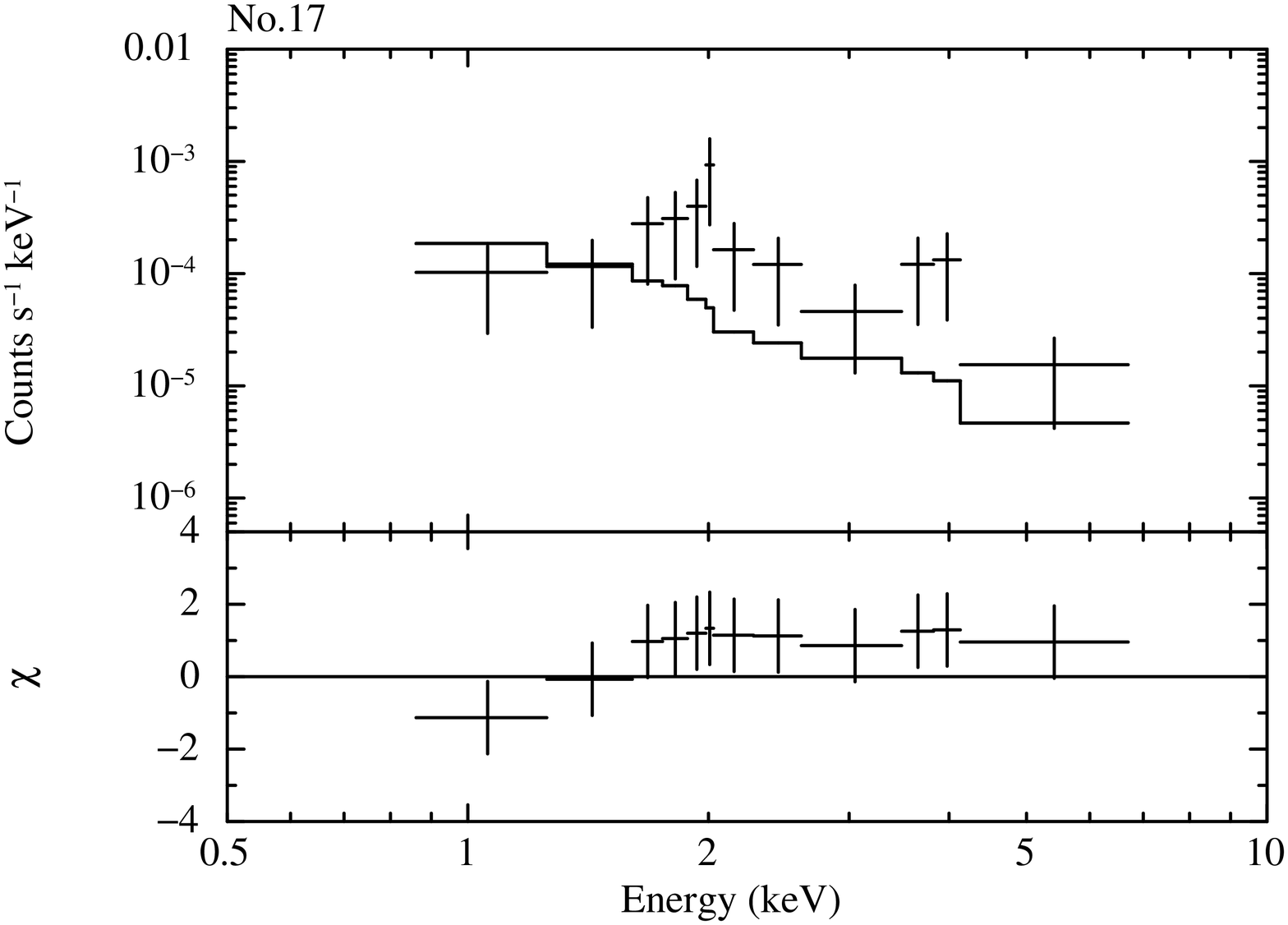}
\end{center}
\end{minipage}
\begin{minipage}{0.2\hsize}
\begin{center}
\FigureFile(45mm,40mm){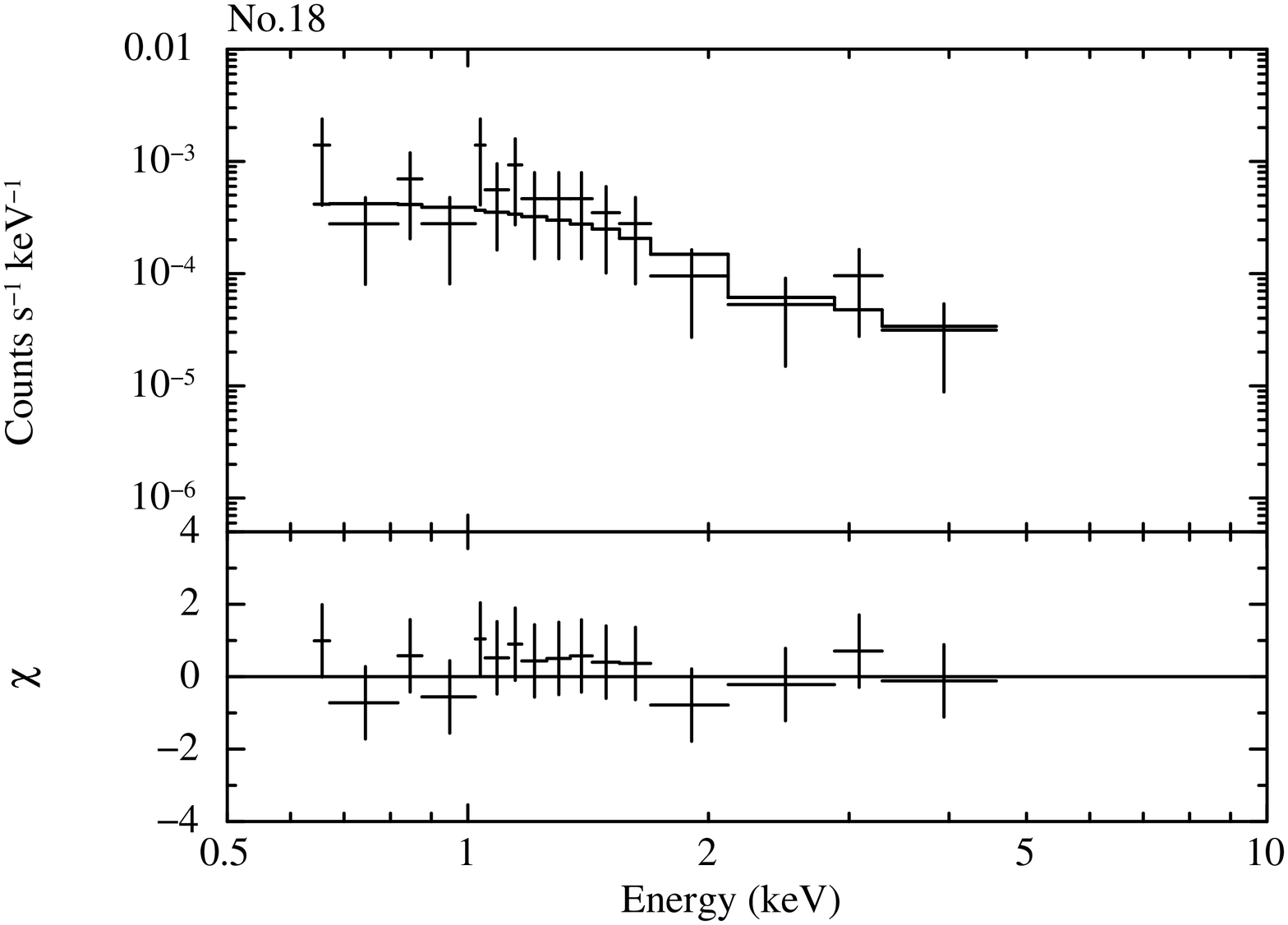}
\end{center}
\end{minipage}
\begin{minipage}{0.2\hsize}
\begin{center}
\FigureFile(45mm,40mm){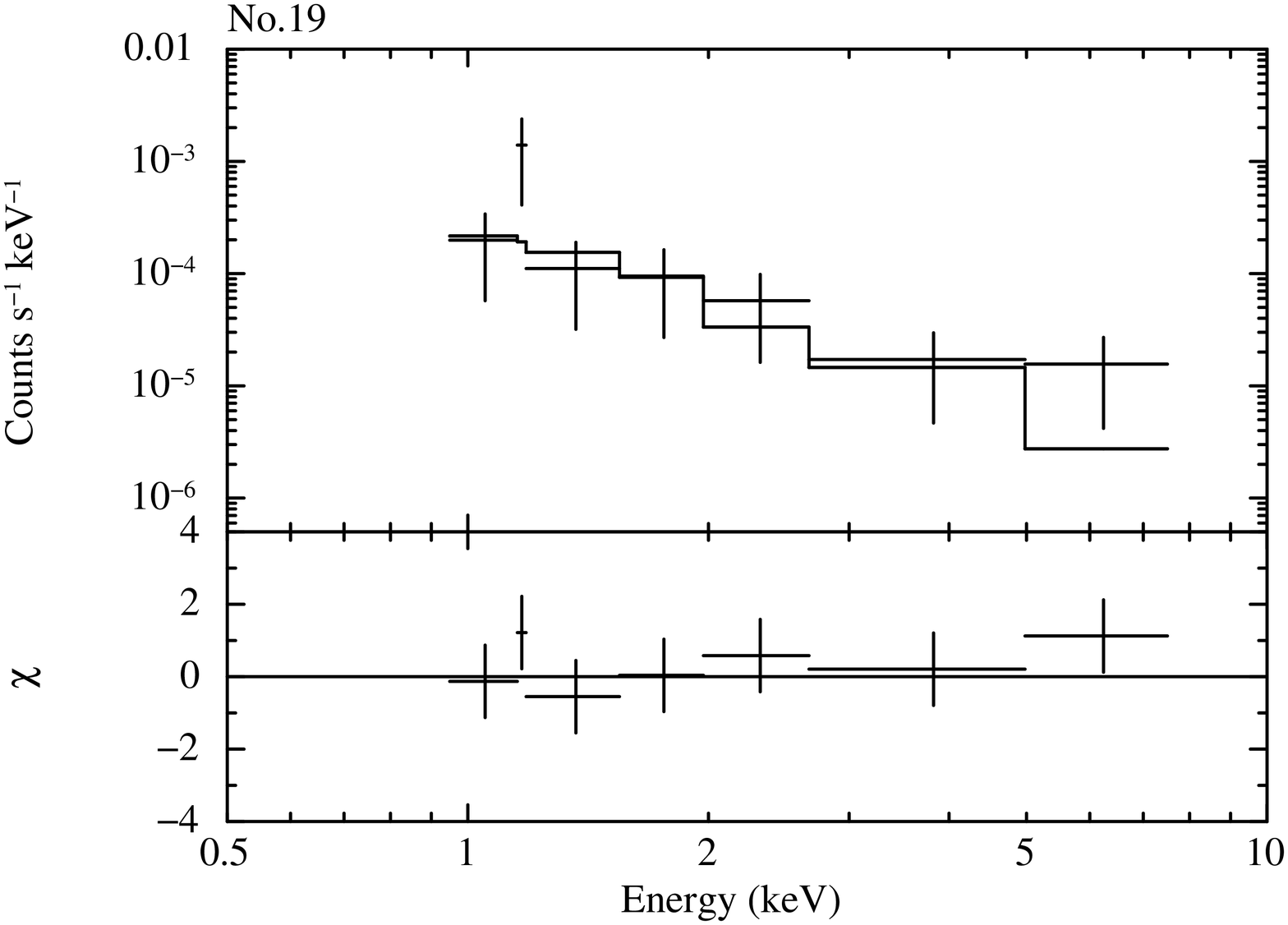}
\end{center}
\end{minipage}
\begin{minipage}{0.2\hsize}
\begin{center}
\FigureFile(45mm,40mm){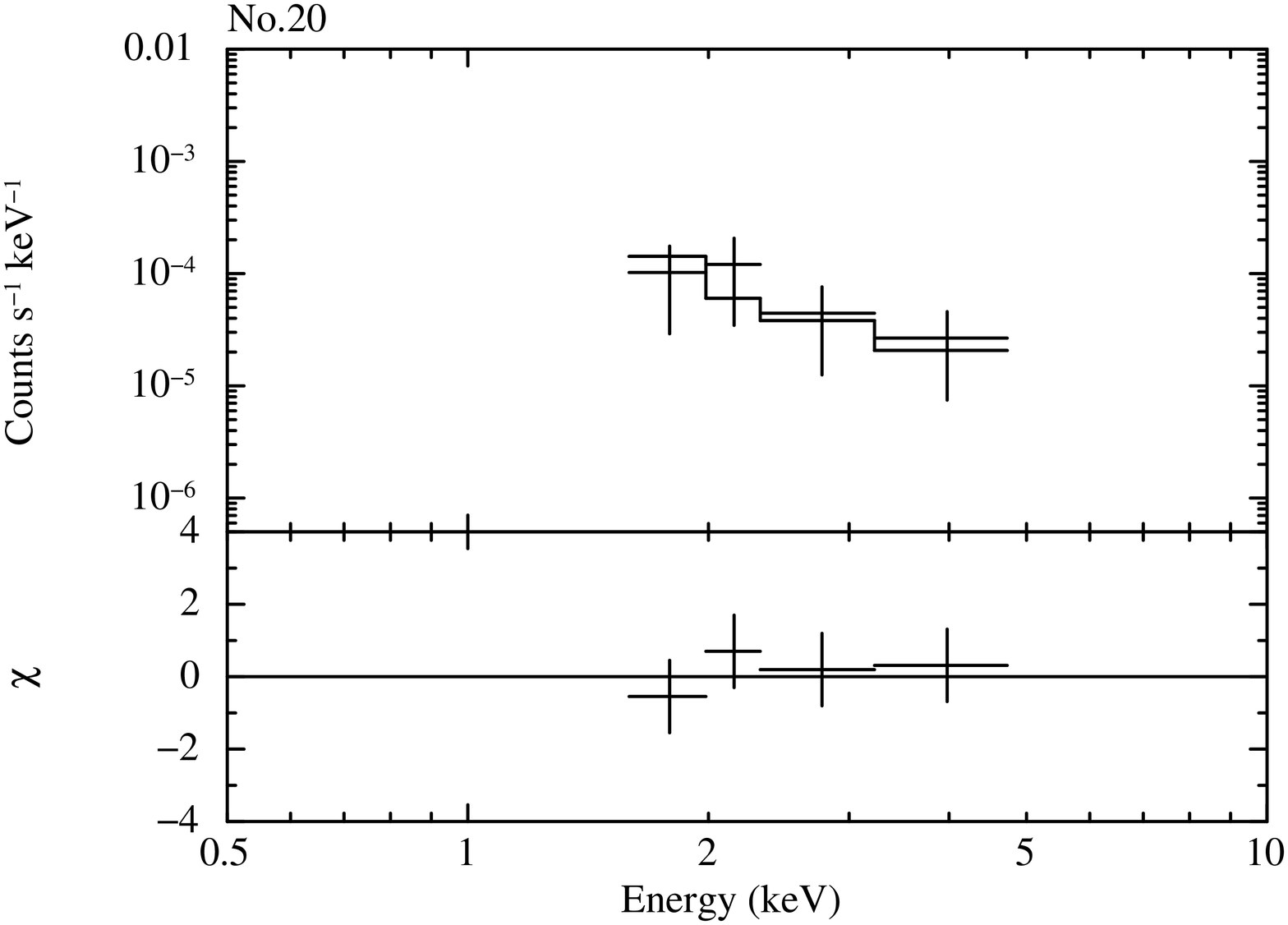}
\end{center}
\end{minipage}
\begin{minipage}{0.2\hsize}
\begin{center}
\FigureFile(45mm,40mm){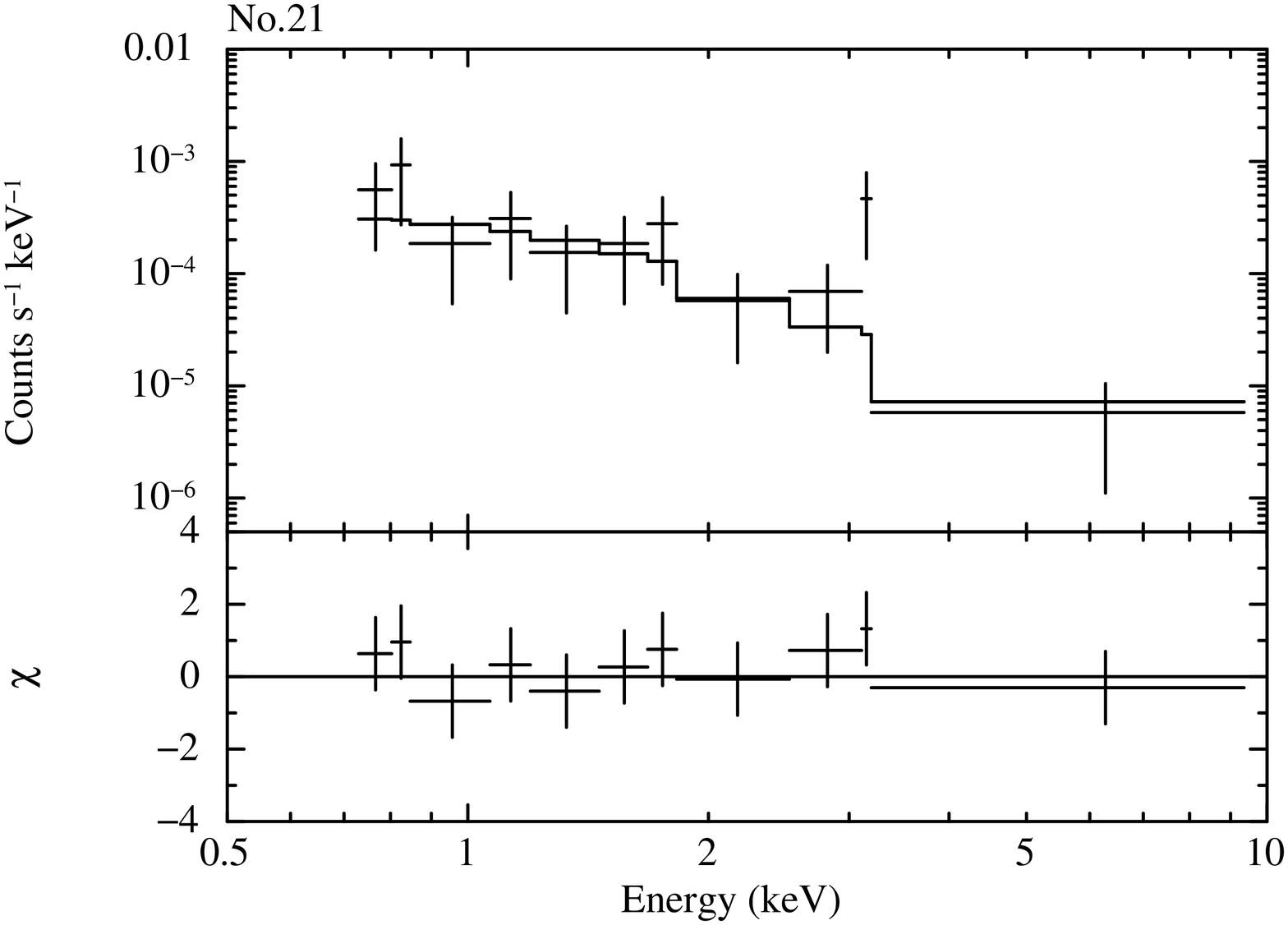}
\end{center}
\end{minipage}
\hspace{8.5mm}
\begin{minipage}{0.2\hsize}
\begin{center}
\FigureFile(45mm,40mm){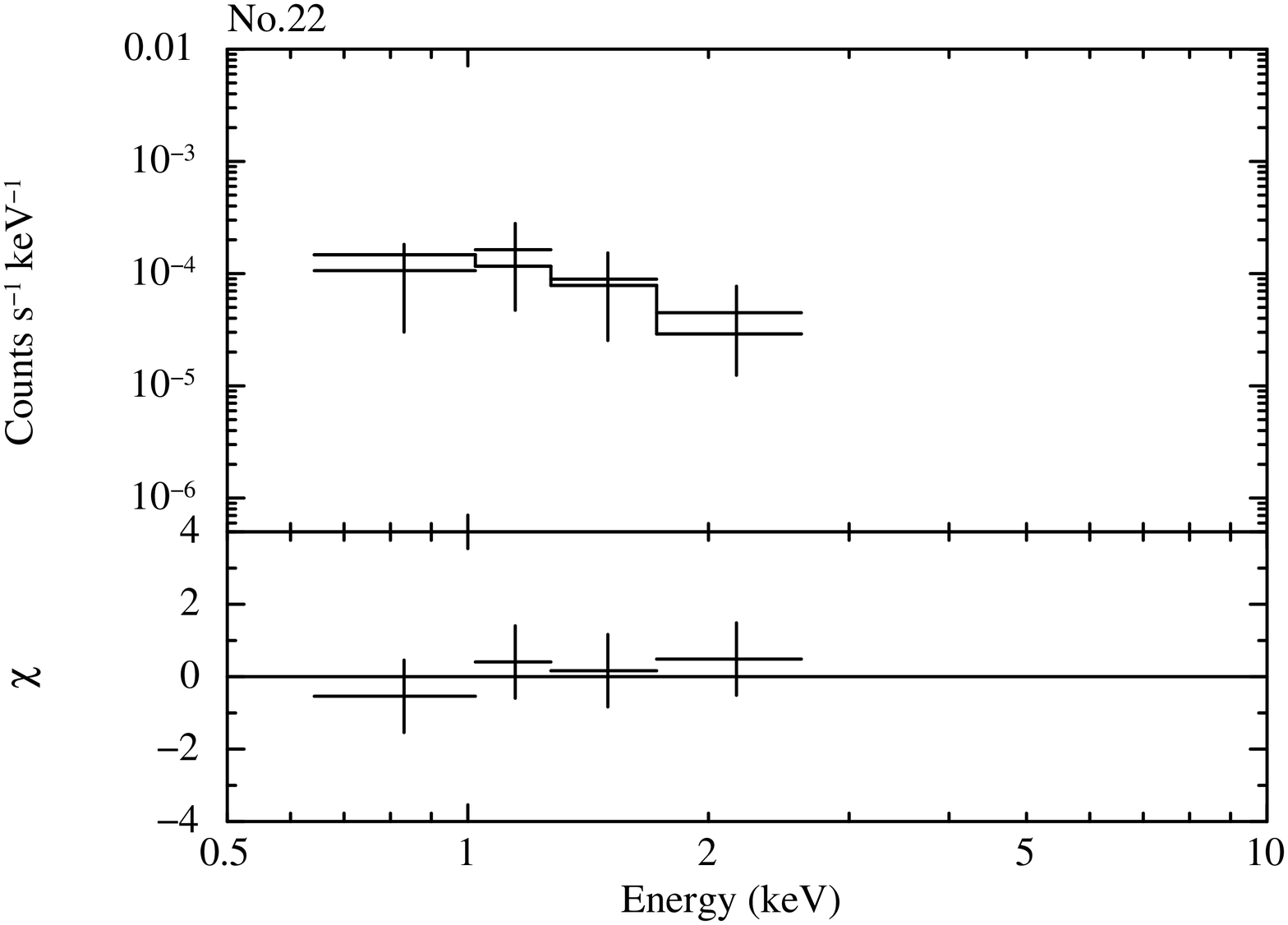}
\end{center}
\end{minipage}
\hspace{8.5mm}
\begin{minipage}{0.2\hsize}
\begin{center}
\FigureFile(45mm,40mm){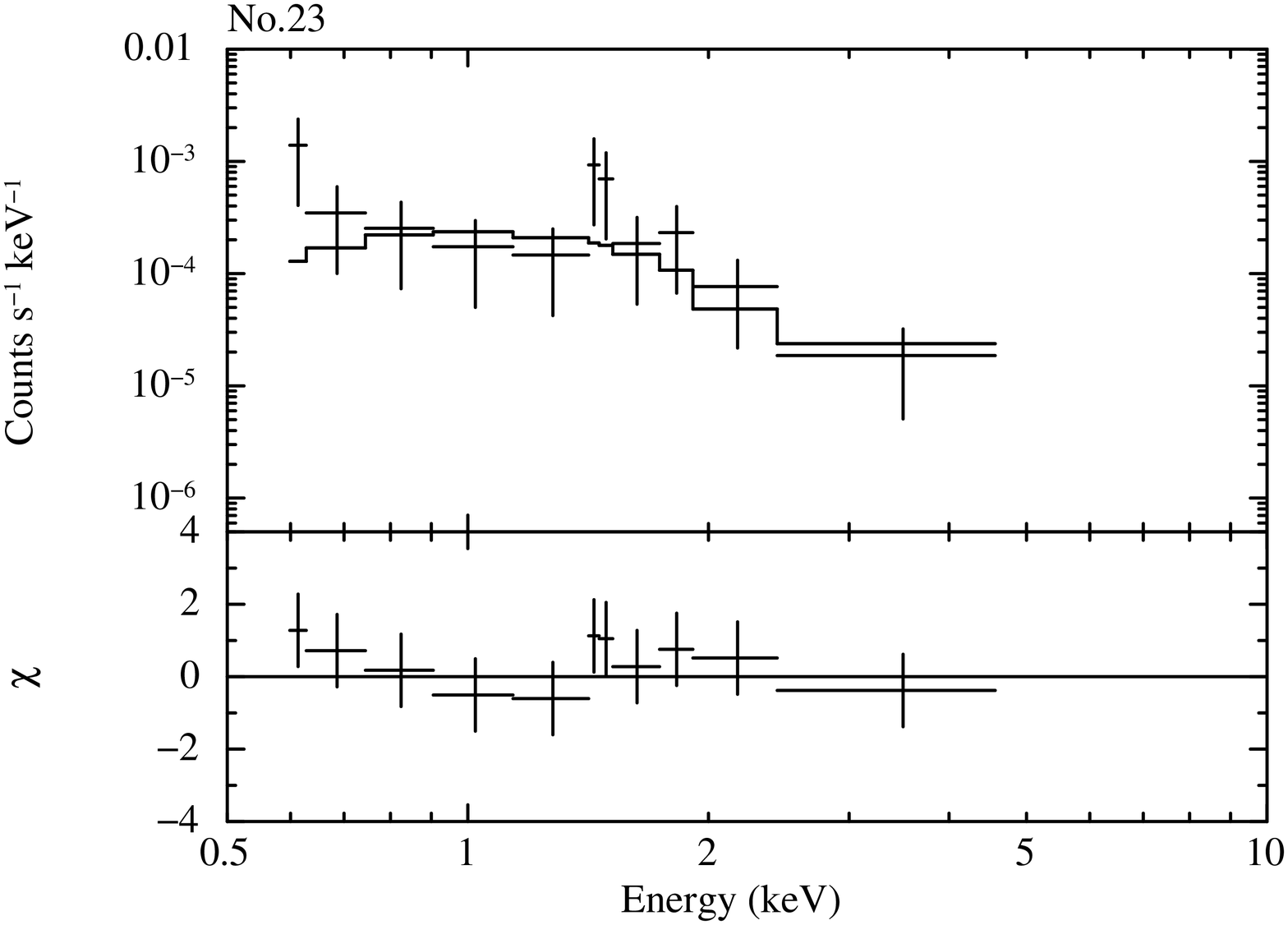}
\end{center}
\end{minipage}
\hspace{8.5mm}
\begin{minipage}{0.2\hsize}
\begin{center}
\FigureFile(45mm,40mm){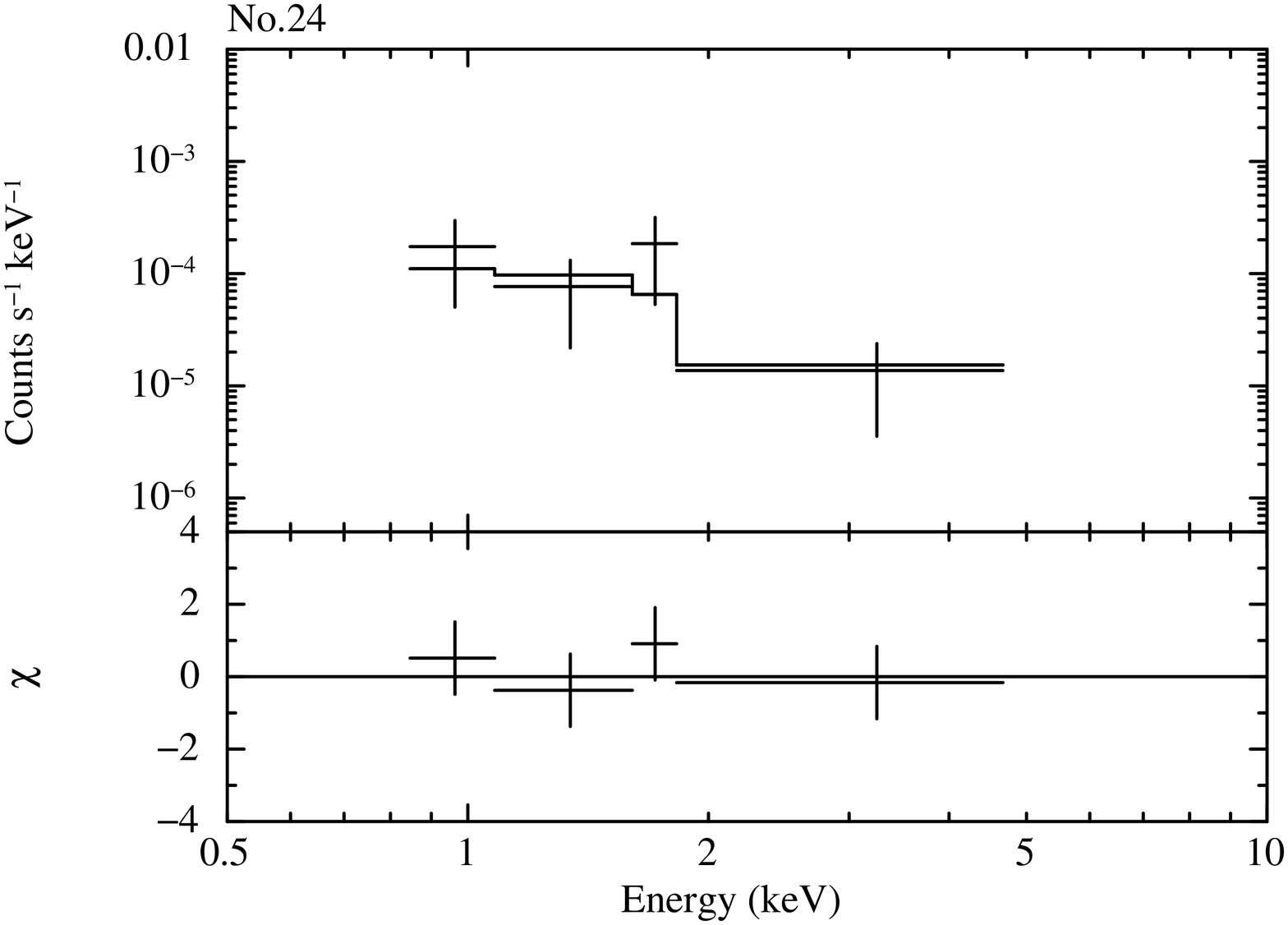}
\end{center}
\end{minipage}
\caption{ACIS spectra of detected point sources. The spectra of OBS1
 and OBS2 are added.}
\label{f2}
\end{figure*}

As for point sources subtraction, we summarized position, photon index,
and energy flux for point sources in table~\ref{t1}. In the case of
\texttt{wavdetect} point sources, the fitting results to energy spectra
are shown in figure~\ref{f2}. The detailed criteria of point source
subtraction is summarized in section~\ref{point}.

\begin{table*}[htbp]
\begin{center}
\caption{Summary of point sources.}
\label{t1}
\scalebox{0.9}{
\begin{tabular}{ccccccc}
\hline
\hline
No.&$\alpha$&$\delta$&$\Gamma$&Flux$^\ast$&Flux$^\dagger$&exclude$^\ddagger$\\
\hline
\multicolumn{7}{l}{wavdetect}\\
\hline
1 &228.551&36.538&$1.77^{+0.40}_{-0.37}$&$4.48^{+1.08}_{-1.08}$&$3.30^{+0.79}_{-0.80}$&$\circ$\\
2 &228.470&36.546&1.7 (fix)&$2.05^{+1.26}_{-1.26}$&$1.60^{+0.98}_{-0.98}$&$\circ$\\
3&228.537&36.546&$1.56^{+0.58}_{-0.48}$&$3.65^{+1.17}_{-1.14}$&$3.17^{+1.01}_{-1.00}$&$\circ$\\
4 &228.603&36.551&$1.28^{+0.62}_{-0.55}$&$2.30^{+1.25}_{-1.04}$&$2.40^{+1.31}_{-1.09}$&$\circ$\\
5 &228.438&36.577&1.7 (fix)&$3.02^{+2.58}_{-2.58}$&$2.36^{+2.01}_{-2.32}$&$\circ$\\
6 &228.596&36.580&$1.87^{+1.59}_{-1.06}$&$0.46^{+0.17}_{-0.21}$&$0.31^{+0.11}_{-0.14}$&---\\
7 &228.594&36.582&1.7 (fix)&$0.33^{+0.14}_{-0.14}$&$0.26^{+0.11}_{-0.11}$&---\\
8 &228.642&36.584&$1.86^{+0.30}_{-0.26}$&$3.97^{+0.59}_{-0.60}$&$2.72^{+0.41}_{-0.41}$&$\circ$\\
9 &228.562&36.585&$1.56^{+0.46}_{-0.40}$&$0.91^{+0.25}_{-0.26}$&$0.79^{+0.22}_{-0.23}$&---\\
10&228.599&36.587&$2.99^{+0.22}_{-0.20}$&$1.86^{+0.21}_{-0.21}$&$0.41^{+0.04}_{-0.05}$&---\\
11&228.596&36.590&$1.29^{+0.80}_{-0.66}$&$0.52^{+0.28}_{-0.27}$&$0.54^{+0.29}_{-0.28}$&---\\
12&228.577&36.602&1.7 (fix)&$0.50^{+0.17}_{-0.17}$&$0.39^{+0.13}_{-0.13}$&---\\
13&228.617&36.605&$1.61^{+0.71}_{-0.71}$&$0.71^{+0.23}_{-0.25}$&$0.59^{+0.20}_{-0.21}$&---\\
14&228.580&36.614&$1.54^{+0.59}_{-0.49}$&$0.71^{+0.22}_{-0.23}$&$0.62^{+0.20}_{-0.20}$&---\\
15&228.618&36.629&$1.48^{+0.38}_{-0.33}$&$2.41^{+0.53}_{-0.54}$&$2.21^{+0.49}_{-0.50}$&$\circ$\\
16&228.613&36.634&1.7 (fix)&$0.33^{+0.17}_{-0.17}$&$0.26^{+0.13}_{-0.13}$&---\\
17&228.628&36.638&1.7 (fix)&$0.42^{+0.22}_{-0.23}$&$0.32^{+0.18}_{-0.17}$&---\\
18&228.587&36.647&$1.39^{+0.70}_{-0.62}$&$0.87^{+0.35}_{-0.34}$&$0.85^{+0.33}_{-0.34}$&---\\
19&228.607&36.648&1.7 (fix)&$0.47^{+0.24}_{-0.24}$&$0.36^{+0.19}_{-0.18}$&---\\
20&228.598&36.657&1.7 (fix)&$0.95^{+0.59}_{-0.60}$&$0.74^{+0.46}_{-0.46}$&---\\
21&228.656&36.678&$1.47^{+1.09}_{-0.71}$&$2.61^{+1.39}_{-1.32}$&$2.39^{+1.28}_{-1.21}$&$\circ$\\
22&228.592&36.680&1.7 (fix)&$0.59^{+0.37}_{-0.36}$&$0.46^{+0.29}_{-0.28}$&---\\
23&228.696&36.697&1.7 (fix)&$9.21^{+3.86}_{-3.36}$&$7.18^{+3.00}_{-3.01}$&$\circ$\\
24&228.490&36.704&1.7 (fix)&$2.87^{+1.86}_{-1.86}$&$2.24^{+1.45}_{-1.45}$&$\circ$\\
\hline
\multicolumn{7}{l}{CSC}\\
\hline
25&228.450&36.548&1.7 (fix)&$1.44^{+0.38}_{-0.38}$&$1.16^{+0.32}_{-0.32}$&---\\
26&228.496&36.563&1.7 (fix)&$0.95^{+0.25}_{-0.25}$&$0.77^{+0.21}_{-0.21}$&---\\
27&228.400&36.568&1.7 (fix)&$1.55^{+0.66}_{-0.67}$&$1.25^{+0.54}_{-0.54}$&---\\
28&228.727&36.659&1.7 (fix)&$1.61^{+0.38}_{-0.38}$&$1.30^{+0.31}_{-0.31}$&---\\
29&228.799&36.675&1.7 (fix)&$2.50^{+0.80}_{-0.81}$&$2.02^{+0.66}_{-0.66}$&$\circ$\\
30&228.604&36.694&1.7 (fix)&$1.17^{+0.60}_{-0.60}$&$0.94^{+0.48}_{-0.48}$&---\\
31&228.685&36.710&1.7 (fix)&$11.60^{+0.86}_{-0.87}$&$9.37^{+0.74}_{-0.74}$&$\circ$\\
32&228.732&36.721&1.7 (fix)&$1.67^{+0.34}_{-0.33}$&$1.35^{+0.27}_{-0.27}$&---\\
33&228.724&36.732&1.7 (fix)&$1.18^{+0.29}_{-0.29}$&$0.95^{+0.23}_{-0.23}$&---\\
34&228.638&36.774&1.7 (fix)&$1.27^{+0.32}_{-0.33}$&$1.03^{+0.26}_{-0.26}$&---\\
\hline
\multicolumn{7}{l}{2XMMi}\\
\hline
35&228.600&36.621&1.7 (fix)&$8.73$&$7.53$&$\circ$\\
36&228.487&36.665&1.7 (fix)&$1.99$&$1.71$&$\circ$\\
37&228.487&36.673&1.7 (fix)&$4.43$&$3.80$&$\circ$\\
38&228.543&36.696&1.7 (fix)&$6.67$&$5.72$&$\circ$\\
\hline
\end{tabular}
}
\begin{description}
\item[$^\ast$] 0.5--7.0 keV energy flux in units of $10^{-14}$ erg
	   s$^{-1}$ cm$^{-2}$. 
\item[$^\dagger$] 2.0--10.0 keV energy flux in units of $10^{-14}$
	   erg s$^{-1}$ cm$^{-2}$.
\item[$^\ddagger$] The mark of $\circ$ means the excluded point source from
	   the spectral analysis.
\end{description}
\end{center}
\end{table*}

\section{Systematic Errors of Scaling Method}\label{apenb}
In the derivation of cumulative numbers, we assumed the uniform
abundance profile of MS~1512.4+3647.
We calculated the systematic error due to this assumption reffering the
average Fe abundance of several clusters in Matsushita et al. 2011.
By comparing the amount of Fe integrated up to $0.3r_{200}$ between the
case of uniform Fe abundance and the case of taking account of radial
profile, the difference between two cases became roughly $\sim30$ \%.
We took account this error into the derivation of cumulative numbers
for MS~1512.4+3647.
In the gas mass scaling of nearby clusters, the systematic error due to
the accuracy of the abundance profile is considered.
From several studies of spatially resolved nearby clusters, the amount
of Fe integrated up to $0.3r_{200}$ have the statistical error of
typically $\sim20$ \% (e.g. Sato et al. 2007a).
We took account this error into the gas mass scaling for the nearby
clusters.


\end{document}